\documentclass[twocolumn]{aastex631}
\usepackage{tablefootnote}
\usepackage{footmisc}
\usepackage{ulem} 

\newcommand\omc{$\omega$\,Cen}

\defcitealias{2010ApJ...710.1032A}{AvdM10}

\received{March 26, 2024}
\revised{May 3, 2024}
\accepted{May 5, 2024}
\published{July 31, 2024}

\begin{document}

\title{oMEGACat II - Photometry and proper motions for 1.4 million stars in Omega Centauri and its rotation in the plane of the sky}

\correspondingauthor{Maximilian Häberle}
\email{haeberle@mpia.de}
\author[0000-0002-5844-4443]{Maximilian Häberle}
\affiliation{Max Planck Institute for Astronomy, K\"onigstuhl 17, D-69117 Heidelberg, Germany}
\author[0000-0002-6922-2598]{N. Neumayer}
\affiliation{Max Planck Institute for Astronomy, K\"onigstuhl 17, D-69117 Heidelberg, Germany}
\author[0000-0003-3858-637X]{A. Bellini}
\affiliation{Space Telescope Science Institute, 3700 San Martin Drive, Baltimore, MD 21218, USA}
\author[0000-0001-9673-7397]{M. Libralato}
\affiliation{AURA for the European Space Agency (ESA), Space Telescope Science Institute, 3700 San Martin Drive, Baltimore, MD 21218, USA}
\affiliation{INAF, Osservatorio Astronomico di Padova, Vicolo dell’Osservatorio 5, Padova,I-35122, Italy}
\author[0009-0005-8057-0031]{C. Clontz}
\affiliation{Max Planck Institute for Astronomy, K\"onigstuhl 17, D-69117 Heidelberg, Germany}
\affiliation{Department of Physics and Astronomy, University of Utah, Salt Lake City, UT 84112, USA}
\author[0000-0003-0248-5470]{A. C. Seth}
\affiliation{Department of Physics and Astronomy, University of Utah, Salt Lake City, UT 84112, USA}
\author[0000-0002-2941-4480]{M. S. Nitschai}
\affiliation{Max Planck Institute for Astronomy, K\"onigstuhl 17, D-69117 Heidelberg, Germany}
\author[0000-0001-6604-0505]{S. Kamann}
\affiliation{Astrophysics Research Institute, Liverpool John Moores University, 146 Brownlow Hill, Liverpool L3 5RF, UK}
\author[0000-0002-1212-2844]{M. Alfaro-Cuello}
\affiliation{Facultad de Ingenier\'{i}a y Arquitectura, Universidad Central de Chile, Av. Francisco de Aguirre 0405, La Serena, Coquimbo, Chile}
\affiliation{Space Telescope Science Institute, 3700 San Martin Drive, Baltimore, MD 21218, USA}
\author[0000-0003-2861-3995]{J. Anderson}
\affiliation{Space Telescope Science Institute, 3700 San Martin Drive, Baltimore, MD 21218, USA}
\author[0000-0001-6187-5941]{S. Dreizler}
\affiliation{Institut für Astrophysik und Geophysik, Georg-August-Universität Göttingen, Friedrich-Hund-Platz 1, 37077 Göttingen, Germany}
\author[0000-0002-0160-7221]{A. Feldmeier-Krause}
\affiliation{Max Planck Institute for Astronomy, K\"onigstuhl 17, D-69117 Heidelberg, Germany}
\affiliation{Department of Astrophysics, University of Vienna, T\"urkenschanzstrasse 17, 1180 Wien, Austria}
\author[0000-0002-6072-6669]{N. Kacharov}
\affiliation{Leibniz Institute for Astrophysics, An der Sternwarte 16, 14482 Potsdam, Germany}
\author[0000-0002-7547-6180]{M. Latour}
\affiliation{Institut für Astrophysik und Geophysik, Georg-August-Universität Göttingen, Friedrich-Hund-Platz 1, 37077 Göttingen, Germany}

\author[0000-0001-7506-930X]{A. P. Milone}
\affiliation{Dipartimento di Fisica e Astronomia “Galileo Galilei,” Univ. di Padova, Vicolo dell’Osservatorio 3, Padova, I-35122, Italy}
\author[0000-0002-1670-0808]{R. Pechetti}
\affiliation{Astrophysics Research Institute, Liverpool John Moores University, 146 Brownlow Hill, Liverpool L3 5RF, UK}
\author[0000-0003-4546-7731]{G. van de Ven}
\affiliation{Department of Astrophysics, University of Vienna, T\"urkenschanzstrasse 17, 1180 Wien, Austria}
\author[0000-0001-6215-0950]{K. Voggel}
\affiliation{Universit\'{e} de Strasbourg, CNRS, Observatoire astronomique de Strasbourg, UMR 7550, F-67000 Strasbourg, France}
\begin{abstract}

Omega Centauri (\omc{}) is the most massive globular cluster of the Milky Way. It is thought to be the nucleus of an accreted dwarf galaxy because of its high mass and its complex stellar populations. To decipher its formation history and study its dynamics, we created the most comprehensive kinematic catalog for its inner region, by analyzing both archival and new \textit{Hubble Space Telescope (HST)} data. Our catalog contains 1\,395\,781 proper-motion measurements out to the half-light radius of the cluster ($\sim5.0\arcmin$) and down to $m_{\rm F625W}\approx25$\,mag. The typical baseline for our proper-motion measurements is 20 years, leading to a median 1D proper motion precision of $\sim$11~µas yr$^{-1}$ for stars with $m_{\rm F625W}\approx 18$\,mag, with even better precision ($\sim$6.6~µas yr$^{-1}$) achieved in the extensively observed centermost ($r<1.5\arcmin$) region.  In addition to our astrometric measurements, we also obtained precise \textit{HST} photometry in seven filters spanning from the ultraviolet to the near-infrared. This allows detailed color-magnitude-diagram studies and to separate the multiple stellar populations of the cluster. In this work, we describe the data reduction used to obtain both the photometric and the proper-motion measurements. We also illustrate the creation and the content of our catalog, which is made publicly available. Finally, we present measurements of the plane-of-sky rotation of \omc{} in the previously unprobed inner few arcminutes and a precise measurement of the inclination $i=(43.9\pm1.3)^\circ$.
\end{abstract}



\section{Introduction} \label{sec:intro}

\subsection{The accretion history of the Milky Way}
In recent years the formation history of the Milky Way has been unraveled, thanks to the \textit{Gaia} satellite \citep{2016A&A...595A...2G}, which provides 6-D phase space information for millions of stars, in combination with large spectroscopic surveys such as APOGEE \citep{2008AN....329.1018A, 2017AJ....154...94M}, LAMOST \citep{2012RAA....12..735D}, and GALAH \citep{2015MNRAS.449.2604D,2021MNRAS.506..150B}.
These surveys have revealed that our Galaxy has experienced a series of mergers, where smaller dwarf galaxies were accreted by the more massive Milky Way. During those mergers, the dwarf galaxy is disrupted by tidal forces \citep{2001MNRAS.323..529H, 2002MNRAS.336..119M} and its stars are scattered across the Halo of the Milky Way.

The largest of the recently discovered mergers is the \textit{Gaia Enceladus} event, a merger $\sim$10 Gyrs ago with a satellite galaxy with a stellar mass of $6\times10^8$~M$_\odot$\space \citep{2018MNRAS.478..611B,2018ApJ...863..113H}, similar to the present-day mass of the Small Magellanic Cloud. There are also signs of other smaller accretion events such as \textit{Sequoia} \citep{2019MNRAS.488.1235M},the Helmi-Streams \citep{1999Natur.402...53H} or the Pontus merger \cite{2022ApJ...926..107M,2022ApJ...930L...9M}. An example, where accretion is still ongoing, is the Sagittarius Dwarf Galaxy \citep{1997AJ....113..634I, 2018MNRAS.481..286L}. The central regions of Sagittarius remain bound, including the nuclear star cluster of the galaxy, M54 \citep{2019ApJ...886...57A,2020ApJ...892...20A,2022ApJ...939..118K}.

While the field stars of accreted galaxies are scattered across the halo of the Milky Way and can only be identified in action space and via their chemistry, the dense globular clusters of the accreted galaxy can survive the merger \citep{2009MNRAS.399.1275P} and are added to the globular cluster population of the Milky Way \citep{1978ApJ...225..357S, 2019MNRAS.486.3180K}. \cite{2019A&A...630L...4M} kinematically linked the globular clusters of the Milky Way to different known accretion events, and found that only about 40\% of the clusters are likely to have formed in-situ. In addition to globular clusters, most galaxies contain a very dense and massive nuclear star cluster in their center (e.g. \citealp{2020A&ARv..28....4N}), which remains intact during accretion. These nuclear star clusters can be fully stripped of their surrounding galaxy \citep{2013MNRAS.433.1997P} and look very similar to massive globular clusters. \cite{2019MNRAS.486.3180K} predict 6$\pm$1 stripped nuclear star clusters hiding within the Milky Way's globular cluster population.

The most promising stripped nuclear star cluster candidate is Omega Centauri (\omc), the most massive ($M\approx 3.55\times10^6$~M$_\odot$, \citealp{2018MNRAS.478.1520B}) globular cluster in the Milky Way (e.g. \citealp{1999Natur.402...55L,2003MNRAS.346L..11B}). \omc{}  is relatively close to the Sun ($d_\odot\approx 5.43$~kpc, \citealp{2021MNRAS.505.5957B}), which allows us to study it in great detail. Decades of observations have shown that \omc{} is unique among the Milky Way's globular clusters in many ways.
The first evidence for \omc's complex stellar populations was the discovery of a large scatter of the cluster's red giant branch by \cite{1973MNRAS.162..207C}, followed by spectroscopic observations that revealed a large metallicity spread \citep{1975ApJ...201L..71F}. Newer spectroscopic catalogs confirmed those early findings and include spectra of thousands \citep{2010ApJ...722.1373J} or most recently even hundreds of thousands of stars \citep{2018MNRAS.473.5591K, 2023ApJ...958....8N}. These studies found a spread in iron abundance of almost 2~dex, ranging from [Fe/H]$\sim-2.2$ to $-0.5$, a much larger spread than for other Milky Way globular clusters. In addition to these spectroscopic findings, precise \textit{Hubble Space Telescope (HST)} photometry played a crucial role in highlighting the complexity of the stellar populations for a much larger sample, including fainter stars \citep{1997PhDT.........8A,2004ApJ...603L..81F,2010AJ....140..631B}. The detailed color-magnitude diagrams (CMDs) show an amazing complexity of several split sequences and subpopulations. Based on studying various ultra-violet (UV) CMDs, \cite{2017ApJ...844..167B} were able to distinguish at least 15 subpopulations along the main sequence. Another very powerful tool to photometrically disentangle the different subpopulations are the so-called chromosome maps based on UV filters \citep{2017MNRAS.464.3636M}. 
The different subpopulations in \omc{}  are also believed to have different ages \citep{2004A&A...422L...9H,2014ApJ...791..107V}, although the exact duration of the star formation is still controversial and estimates range from less than 0.5~Gyrs \citep{2016MNRAS.457.4525T}, 1-2~Gyrs \citep{2013ApJ...762...36J} to 4-5~Gyrs \citep{2007ApJ...663..296V}. The determination of relative ages is complicated by differences between the abundances of light elements for the different subpopulations \citep{2012ApJ...746...14M}. 

Besides these peculiar stellar populations, there is also kinematic evidence supporting the stripped nucleus scenario: \cite{2006A&A...445..513V} found evidence for the presence of a central stellar disk and a preference for tangential orbits in the outer parts. More recently, both kinematic and chemical associations with stellar streams such as the \textit{Fimbulthul} stream have been found in e.g. \cite{2012ApJ...747L..37M}, \cite{2019NatAs...3..667I}, and \cite{2022ApJ...935..109L}. Another approach is taken in \cite{2022A&A...659A..96M}, in which a connection between the low retrograde binary fraction in the Milky Way and the star formation conditions in \omc{}'s progenitor is studied. Both the Sequoia and the Gaia-Enceladus/Sausage progenitors have been discussed as potential former host-galaxies of \omc{} \citep{2019MNRAS.488.1235M,2019A&A...630L...4M,2020MNRAS.493..847F,2021MNRAS.500.2514P}.

To summarize, \omc{}  is most likely an accreted nuclear star cluster and therefore, both the closest galactic nucleus (even closer than the Galactic Center) and a remnant of an important accretion event in the history of the Milky Way. Studying its formation can reveal both details of the Milky Way's assembly history and nuclear star clusters.

\subsection{Project overview}
The \textit{oMEGACat} project aims to decipher the formation history and dynamics of \omc{} by assembling the largest spectroscopic, photometric, and astrometric data set out to the cluster's half-light radius. The spectroscopic part of this dataset is an extensive study performed with the Very Large Telescope Multi-Unit Spectroscopic Explorer (MUSE, \citealt{2010SPIE.7735E..08B}) integral field spectrograph. This spectroscopic catalog has recently been published \citep{2023ApJ...958....8N} and contains line-of-sight velocities and metallicity measurements for more than 300\,000 stars.

In this paper, we describe the creation of the second part of the dataset, a complementary astro-photometric catalog, based on archival and new \textit{HST} observations.

Both the spectroscopic and the astro-photometric catalogs are made public and therefore provide a legacy dataset for the community.

\subsection{Outline of this work}
In Section \ref{sec:review} we review other published ground- and space-based astrometric and photometric catalogs for \omc{}. In Section~\ref{sec:datasets} we give a brief overview of the dataset that has been used to create our catalog. In Section~\ref{sec:data_reduction} we describe our data reduction, which yields individual astro-photometric measurements from the \textit{HST} images. We explain how we determine proper motions based on those individual data points in Section~\ref{sec:proper_motions}. In Section~\ref{sec:photometry} we describe how we create a uniform photometric catalog based on the individual measurements. In Section~\ref{sec:catalog_validation} we perform several crosschecks and comparisons to other catalogs, which we use to test the quality of our dataset. In  Section~\ref{sec:rotation} we present the first science results based on our catalog: a redetermination of the plane-of-sky rotation of \omc{}. We conclude the work with a description of the published data products (Section~\ref{sec:data_products}) and our conclusions in Section~\ref{sec:conclusions}.

\section{Review of other astrometric and photometric catalogs for \omc{}}
\label{sec:review}
\subsection{Other proper motion catalogs}
The study of the proper motions within \omc{} has a long history, including several ground- and space-based catalogs. In Table \ref{tab:pm_catalogs} we provide a complete overview of all these catalogs along with information on their coverage, depth, and astrometric precision.

Astrometric studies of \omc{} started with photographic plate measurements by \cite{1965RGOB..100...81M} and \cite{1966ROAn....2....1W} at the Royal Greenwich Observatory. Another large plate-based effort was taken by \cite{2000AA...360..472V} and reached the impressive precision of 0.1\,mas\,yr$^{-1}$ thanks to the long baseline of more than 50 years. Other plate-based (or hybrid plate / CCD) studies with the goal of constraining the absolute motion of \omc{} were published in \cite{1999AJ....117..277D} and \cite{2002ASPC..265..399G}, although no proper motion catalog was made public along these works.
A more recent remarkable wide-field ground-based study was done by \cite{2009AA...493..959B} with the CCD-imager WFI@ESO/MPG2.2m.

\begin{table*}[]
\caption{List of all published proper-motion catalogs for Omega Centauri}
\label{tab:pm_catalogs}
\scriptsize

\begin{flushleft}
\begin{tabular}{@{}p{2.75cm}p{2.2cm}p{3.5cm}p{2.2cm}p{1.5cm}p{1.5cm}p{2.0cm}@{}}
\hline
Catalog                                         & Instrument       & Covered Area                      & Limiting                      & Number of   & Maximum   & Bright Star                 \\
                                                &                  &                                   & Magnitude                     & entries     & Baseline  & Proper Motion Error              \\ \hline
\cite{1965RGOB..100...81M,1966ROAn....2....1W}  & Royal Greenwhich Observatory&                        & $B < 16.8$                           &  $\sim$4000           & 56 yrs    & 2.0~mas~yr$^{-1}$                  \\\hline
Hipparcos \citep{1997AA...323L..49P}                                      & Hipparcos        & Allsky                            &                               & 3\footnote{\label{note1}See \cite{2001ASPC..228...43F}}      & 3.5\,yrs          &                             \\ \hline
Tycho 2 \citep{2000AA...355L..27H}                                       & Hipparcos        & Allsky                            &                               & 53\footref{note1}     & 3.5\,yrs          &                             \\ \hline
\cite{2000AA...360..472V}                       & Yale-Columbia 66cm refractor   & $r\leq29\farcm5$                  & $B < 16.0-16.5$    & 9847        & 52 yrs    & 0.1~mas~yr$^{-1}$                  \\ \hline
\cite{2009AA...493..959B}                       & MPG~2.2m  WFI   & $33\arcmin\times33\arcmin$        & $B < 20$                        & 360\,000     & 4 yrs     & 1.1~mas~yr$^{-1}$                  \\ \hline
\cite{2010ApJ...710.1032A}                      & HST              & \mbox{Central Field ($r\leq2\arcmin$)} \mbox{Major Axis field ($r\approx4\arcmin$)}   & $m_{\rm F625W} < 23$ \mbox{ $m_{\rm F625W} < 22.5$}& 108\,507 61\,293    & 4.07\,yrs 2.5\,yrs & 0.1~mas~yr$^{-1}$ 0.2~mas~yr$^{-1}$ \\ \hline
\cite{2017ApJ...842....6B}                      & HST              & $r\leq2\farcm5$                   & $m_{\rm F606W} < 24$          & 279\,909     & 10.6 yrs  & 0.025~mas~yr$^{-1}$                \\ \hline
Gaia (E-)DR3 \citep{2021AA...649A...1G}                                        & Gaia             & Allsky                            & \textit{Gaia G} $<$17 (center)                              &  321\,698 (within $r\leq0.8^\circ$)              & 2.8 yrs          &  0.02~mas~yr$^{-1}$                           \\\hline
\cite{2018ApJ...853...86B}                      & HST              & 1 field at 3.5$ r_{HL}\approx17\arcmin$  & $m_{\rm F606W} < 27$          & 5\,153        & 15 yrs    & 0.01~mas~yr$^{-1}$                 \\ \hline
\cite{2021MNRAS.505.3549S}                      & HST              & 2 fields at 2.5$ r_{HL}\approx12\arcmin$ & $m_{\rm F606W} < 27$          & 27\,885      & 2 yrs     & 0.07~mas~yr$^{-1}$                 \\ \hline
Gaia FPR \citep{
2023AA...680A..35G}            & Gaia             & $r\leq0.8^\circ$                                   &  \textit{Gaia G} $<$20.5                             & 526\,587              &     5 yrs      &  0.3~mas~yr$^{-1}$                           \\ \hline
oMEGACat \mbox{(this work)}                                         & HST              & $10\arcmin\times10\arcmin$                          & $m_{\rm F625W} < 25$                              & 1\,399\,455   & 20.89 yrs & \footnote{We reach a precision of 0.007~mas~yr$^{-1}$ in the well-covered center, and a precision of 0.012~mas~yr$^{-1}$ over the full field}0.007~mas~yr$^{-1}$  0.012~mas~yr$^{-1}$     \\ \hline
\end{tabular}
\end{flushleft}

\end{table*}

The era of space astrometry was initiated with the Hipparcos Satellite \citep{1997AA...323L..49P}, but due to the limited depth, neither the HIPPARCOS catalog, nor the hybrid Tycho-2 catalog \citep{2000AA...355L..27H} allowed to study the internal kinematics of  \omc{}. \cite{2001ASPC..228...43F} report only 3 Hipparcos stars and 53 Tycho-2 stars in common with \cite{2000AA...360..472V}.

In comparison, the \textit{Hubble Space Telescope} proved to be the perfect tool for crowded field astrometry. Its high resolution and well-characterized, stable point spread function allow individual astrometric measurements with 0.4~mas precision \citep{2006acs..rept....1A, 2011PASP..123..622B}. Thanks to its high sensitivity, very faint stars can also be studied. For \omc, the main limitation is the comparatively small field of view of its main imaging instruments ($3\farcm3\times3\farcm3$ for ACS/WFC, $2\farcm7\times2\farcm7$ for WFC3/UVIS); the existing \textit{HST} proper-motion catalogs cover an area of only one or two \textit{HST} pointings.
The first \textit{HST} proper motion study of \omc's innermost region was done in \cite{2010ApJ...710.1032A}, with an additional field South-East of the center. The measurements in this area were significantly improved in \cite{2014ApJ...797..115B} and published along an extensive photometric catalog in \cite{2017ApJ...842....6B}. This most recent public catalog covers the core region out to a radius of $\sim$2$\farcm$7 and has a maximum temporal baseline of 12~years.
One other notable work based on \textit{HST} observations of the center of \omc{} was the detection of astrometric acceleration by dark companions presented in \cite{2023arXiv231216186P}, however the astrometric catalog has not been made public.
Other \textit{HST} fields at larger radii have been analyzed in \cite{2018ApJ...853...86B} ($r\sim 17\arcmin \sim 3.5r_{\rm HL}$) and \cite{2021MNRAS.505.3549S} ($r\sim 12\arcmin \sim 2.5r_{\rm HL}$). Due to the lower stellar density at these radii and the long exposure times, those fields mark the deepest observations of \omc{} at the time of writing, reaching  magnitudes of $m_{\rm F606W}\sim27$.

In addition to \textit{HST}, the \textit{Gaia} astrometry satellite \citep{2016A&A...595A...1G} has measured hundreds of thousands of absolute proper motions in the outer regions of \omc. However, even in the most recent general data release DR3 \citep{2023A&A...674A...1G} (whose astrometric component was already published as Early Data Release 3; \citealt{2021AA...649A...1G,2021A&A...649A...2L}), the \textit{Gaia} measurements are both limited in depth and precision in the center due to the high crowding and the limited resolution of the satellite. One main challenge is the \textit{Gaia} readout window strategy, which runs into processing and downlink limitations for extremely crowded fields such as \omc{}. For this reason, the \textit{Gaia} collaboration has taken dedicated engineering images, the so called  Service Interface Function (SIF) images. An extension of the regular \textit{Gaia} catalog for \omc{} has been made public during the \textit{Gaia Focused Product Release (FPR)} \citep{2023AA...680A..35G}. Using the engineering images and dedicated on-the-ground data processing, measurements for 526\,587 additional stars have been added for a region with a radius of around $\sim0.8^\circ$ around the center of \omc{}. Especially for the central few arcminutes, this leads to much better completeness than in \textit{Gaia} DR3, however with relatively large astrometric errors (due to the binned nature of the SIF images). In Section~\ref{sec:catalog_validation} we present a detailed comparison between the different astrometric datasets. 

The new catalog presented in this work represents a significant improvement over previous astrometric catalogs in several ways: In comparison with earlier \textit{HST} catalogs, we cover a much larger field of view out to the half-light radius, with a significantly longer and highly uniform baseline as well as new, state-of-the-art, photometry tools. Our catalog is complementary to the recent \textit{Gaia FPR}: while the strength of the \textit{Gaia FPR} is the uniform completeness out to very large radii and its anchoring in an absolute reference frame, we tackle the crowded inner regions with a higher sensitivity and resolution and much longer temporal baseline, resulting in significantly lower astrometric errors and measurements for fainter stars. Within the half-light radius, the new \textit{HST} measurements probe around 3 magnitudes deeper than the \textit{Gaia FPR} data and have proper motion errors at least one order of magnitude better. In addition, the proper-motion catalog presented in this work is complemented by the uniform 6-band photometry we publish along with it.

\subsection{Other \textit{HST} photometry catalogs of \omc}
In the past years, several photometric catalogs based on \textit{HST} imaging have been published for \omc{} with various science goals. Some of them were published together with the astrometric catalogs already mentioned in the previous section. The first \textit{HST} based photometric catalog of \omc{} was created as part of the ``The ACS Survey of Globular Clusters" and is described in \cite{2008AJ....135.2055A}. A much larger catalog was then published by \cite{2010ApJ...710.1032A}, covering a grid of $3\times3$ ACS/WFC pointings, giving an on-sky extent of $10\arcmin\times10\arcmin$ and containing deep photometry for 2$\times10^6$ stars in the F435W and F625W filters. The data used for this study - observed in 2002 - marks the first epoch for most of our proper-motion measurements.
With the installation of the WFC3/UVIS instrument, a new range of UV filters became available. They have been used to study multiple populations, for example in \cite{2010AJ....140..631B} and \cite{2013ApJ...769L..32B}. 
The deepest photometry for \omc{} has been obtained for a Large Project \citep{2017MNRAS.469..800M} with the goal of studying the stars at the faint end of the main sequence. Several astrometric and photometric catalogs based on this data have been published \citep{2018ApJ...853...86B,2018ApJ...854...45L,2021MNRAS.505.3549S,2022ApJ...930...24G}. The most recent catalog for the core of \omc{} was published by \cite{2017ApJ...842....6B}, containing the most comprehensive set of filters (18 WFC3/UVIS filters and 8 WFC3/IR filters) and the same state-of-the-art photometry software as in this work. This catalog is limited to the centermost region of \omc{} with $r\leq2\farcm5$.

All the mentioned photometric \textit{HST} catalogs have excellent photometric quality and some of them reach even deeper than the data presented in this work or have a larger set of filters. The unique feature of our catalog is the large field which is uniformly covered with deep photometry in 6 filters while at the same time also adding high-precision astrometry.

\section{Dataset}
\label{sec:datasets}
\omc{} is one of the individual objects with the largest number of \textit{HST} observations. This is in part due to its interesting properties which have sparked many science programs, but also because it provides an almost ideal calibration target for high-resolution imaging instruments, due to its high and fairly uniform central stellar density. For this reason, it was chosen as the astrometric calibration field for the WFC3/UVIS instrument and is repeatedly observed to monitor the astrometric stability \citep[see e.g.][]{2015wfc..rept....2K}.

For our study, we used imaging data obtained with the Advanced Camera for Surveys (ACS) Wide Field Channel and the Wide Field Camera 3 UVIS Channel. The data from both of these instruments are similar: both instruments have a mosaic of two 2048$\times$4096 pixel CCD detectors with a narrow chip gap, giving approximately a square footprint on the sky. The ACS/WFC, installed during Service Mission 3B, has a nominal pixel scale of around 50\,mas\,pixel$^{-1}$, giving a field of view of  $3\farcm3\times3\farcm3$. WFC3/UVIS was installed during Service Mission 4 and has a slightly higher resolution with a pixel scale of around 40\,mas\,pixel$^{-1}$, resulting in a field of view of $2\farcm7\times2\farcm7$. 

\begin{figure*}
  \centering
    \includegraphics[width=1.0\textwidth]{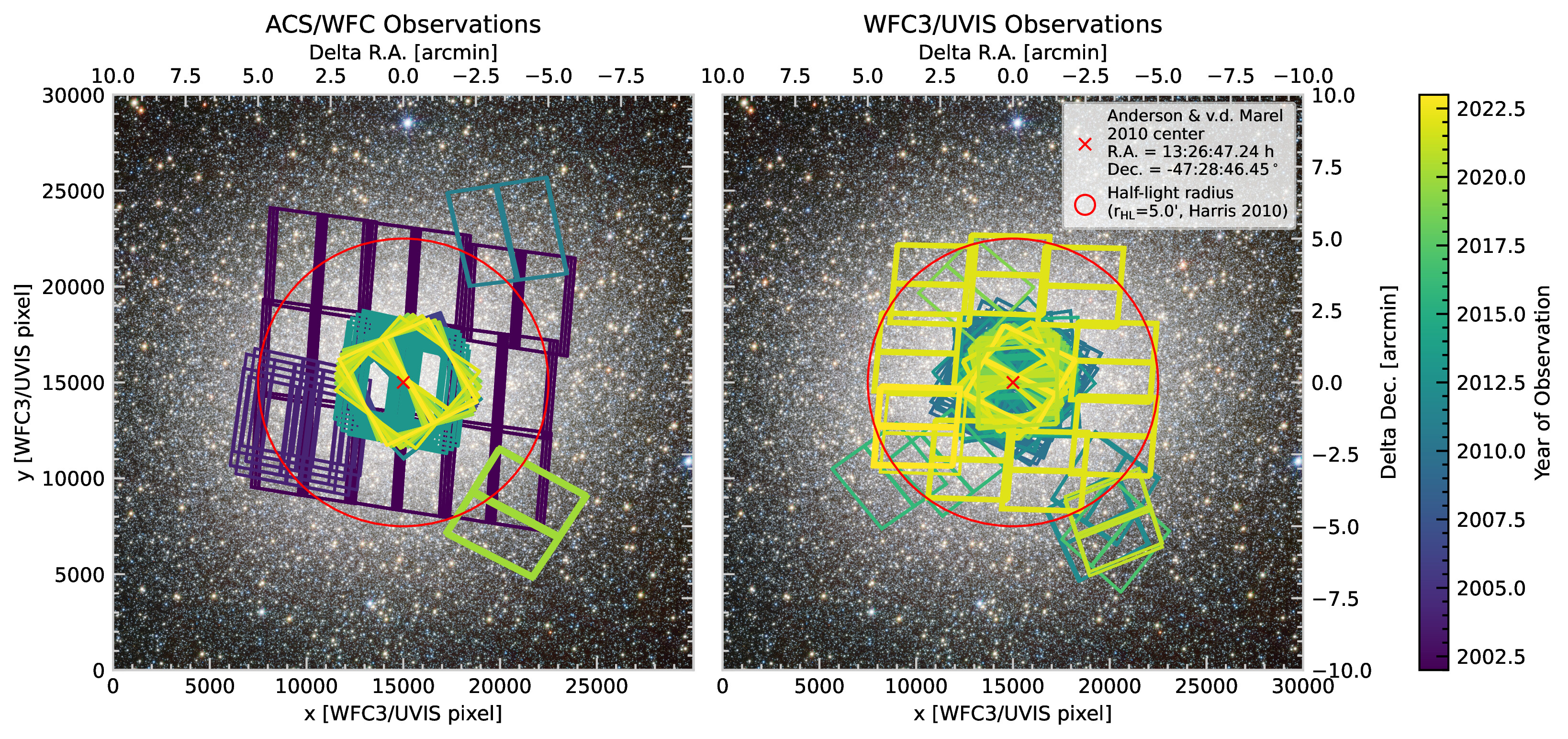}
  \caption{The footprint of the \textit{HST} observations used to measure photometry and determine proper motions, color-coded according to the year of observation. The plots are in both our pixel-based coordinate system and in relative angular units. The left panel shows observations with the ACS/WFC instrument, the right panel with the WFC3/UVIS instrument. The background shows a wide field image of \omc{} taken with the ESO/VST telescope (Image Credit: ESO/INAF-VST/OmegaCAM. Acknowledgement: A. Grado, L. Limatola/INAF-Capodimonte Observatory, \url{https://www.eso.org/public/images/eso1119b/})}
  \label{fig:finding_chart}
\end{figure*}

In total, we reduced 236 images taken with the Advanced Camera for Surveys (ACS) Wide Field Channel, and 561 images taken with the Wide Field Camera 3 UVIS Channel, including both archival data and data from a new, dedicated program (GO-16777, PI: A. Seth). However, not all filters are suitable for high-precision astrometry, due to the unavailability of dedicated high-precision geometric distortion corrections. Proper motions are, therefore, based on a subset of the data, including 196 ACS/WFC and 476 WFC3/UVIS exposures. For the photometric catalogs, we restricted ourselves to the six filters with the widest field coverage (WFC3/UVIS F275W, F336W, F814W; ACS/WFC F435W, F625W, F658N). These 6 filters fill the half-light radius with only minimal gaps. In addition, we also included the WFC3/UVIS F606W filter which has only been used in the central region of \omc{}. Due to the large number of calibration observations ($N_{used}=184$) in this filter it provides excellent photometric quality out to $r\sim2.5\arcmin$. Footprints of the utilized observations can be found in Figure~\ref{fig:finding_chart}. Table~\ref{tab:filter_counts} lists all filters and the number of images used for the creation of our catalogs. We note that, while the ACS High Resolution Channel is principally suitable for high-precision astrometry, there are no usable observations within the field covered in this study. Although available in the archive, we also did not make use of any WFC3/IR images, as they are less useful for high-precision astrometry due to their relatively large pixel size (130\,mas\,pixel$^{-1}$). A state-of-the-art reduction of the WFC3/IR data can be found in \cite{2017ApJ...842....6B}.
A detailed list of all program IDs, filters, and exposure times used for our analysis is shown in Tables \ref{tab:uvisdata} and \ref{tab:acsdata} in Appendix~\ref{sec:appendixdataset}.
In addition, all the {\it HST} data used in this paper can be found under the following DOI in the Mikulski Archive for Space Telescopes (MAST): \dataset[10.17909/26qj-g090]{http://dx.doi.org/10.17909/26qj-g090}.

\begin{table}[]
\caption{List of all filters used for the creation of our astro-photometric catalog. In the fourth column we state whether a filter has been used only for photometry (phot.), astrometry (astro.), or both.}
\label{tab:filter_counts}
\begin{tabular}{llll}
\hline
Instrument & Filter & N$_{\rm exp.}$ & Usage  \\ \hline
ACS/WFC    & F435W  & 69  & astro. \& phot. \\
ACS/WFC    & F475W  & 7   & astro. only \\
ACS/WFC    & F555W  & 4   & astro. only \\
ACS/WFC    & F606W  & 35  & astro. only \\
ACS/WFC    & F625W  & 40  & astro. \& phot.\\
ACS/WFC    & F658N  & 39  & phot. only \\
ACS/WFC    & F775W  & 8   & astro. only \\
ACS/WFC    & F814W  & 33  & astro. only \\ \hline
WFC3/UVIS  & F275W & 85 & phot. only \\
WFC3/UVIS  & F336W  & 106 & astro. \& phot. \\
WFC3/UVIS  & F390W  & 15  & astro. only \\
WFC3/UVIS  & F438W  & 49  & astro. only \\
WFC3/UVIS  & F555W  & 25  & astro. only \\
WFC3/UVIS  & F606W  & 184 & astro. \& phot. \\
WFC3/UVIS  & F775W  & 18  & astro. only \\
WFC3/UVIS  & F814W  & 79  & astro. \& phot.\\ \hline
\end{tabular}
\end{table}

\section{Astrophotometric data reduction}
\label{sec:data_reduction}
In general, our data reduction follows the procedures described in \cite{2017ApJ...842....6B}, \cite{2018ApJ...861...99L}, and \cite{2022ApJ...934..150L}. However, we had to adapt the procedures due to the larger field and the large number of epochs.
\subsection{First photometry iteration with \texttt{hst1pass}}
\label{subsec:firstpass}
As a first step in our data reduction, we ran the point spread function (PSF) photometry code \texttt{hst1pass} \citep{2022wfc..rept....5A} on all individual exposures in our dataset. For all data reduction steps, we used \texttt{*\_flc.fits} images that are flat-field and charge-transfer-efficiency corrected, but not resampled. This preserves the original astrometric signal we aim to study. 

\texttt{hst1pass} uses the library effective PSF (ePSF) models described in \citet{2006acs..rept....1A}. When available, we used the state-of-the-art focus-diverse ePSF models described in \cite{2018acs..rept....8B} for ACS/WFC and \cite{2018wfc..rept...14A} for WFC3/UVIS. \texttt{hst1pass} further improves the library ePSF to better match those of each individual image. This is necessary due to the time variability of the \textit{HST} PSF caused by telescope breathing and focus variations. For typical observations, the central values of the PSF change by around 5\% (RMS over the full field), although in some rare cases, this change can be up to 25\%.
We saved these perturbed ePSF models for each image, as we need them later for the second photometry stage.

We used the geometric distortion corrections from \cite{2006acs..rept....1A,2006hstc.conf...11A} for ACS/WFC\footnote{For the proper-motion measurements we also applied look-up table corrections to post SM4-\textit{HST} observations, see Section~\ref{subsec:pmiter}}, and from \cite{2009PASP..121.1419B,2011PASP..123..622B} for WFC3/UVIS to correct stellar positions in each \texttt{\_flc} exposure.
For WFC3/UVIS filters with no dedicated high-precision correction available, we used the F606W correction. These filters were not used for the proper motion determination but are only used for photometry.

\subsection{Grouping the data into epochs}
After we obtained single-image catalogs for each exposure, we grouped all these individual exposures in 1-year bins, from 2002 to 2023. If there were multiple sets of non-overlapping exposures (e.g. a set of observations of the center and another one of a different region), we created separate bins for them. In total, this leaves us with 26 groups that we reduced separately. For each of them, we created astrometric master frames (see Section~\ref{subsec:ref_frame}), performed an initial photometric registration (see Section~\ref{subsec:photometric_ref}) and finally run the second iteration of the photometry (see Section~\ref{subsec:ks2}). This is a compromise between creating very deep image stacks to improve the completeness and having image stacks based on short-time scales to facilitate the detection of fast-moving stars.

\subsection{Reference frame and astrometric image registration}
\label{subsec:ref_frame}
For the second iteration of photometry and the proper-motion determination, we need to set up a common reference frame in which we can precisely anchor each \textit{HST} image. To do so, we use a hybrid \textit{Gaia-HST} reference frame, which is created in the following way:

First, we queried the \textit{Gaia} EDR3 \citep{2021AA...649A...1G, 2021A&A...649A...2L} with a search-radius of 10~arcmin around the center of \omc. This initial spatial selection gives us 100\,170 sources. To them, we applied strict quality selections:
\begin{itemize}
    \item Successfully measured photometry in both the \texttt{RP} and the \texttt{BP} bands
    \item Renormalized Unit Weight Error \texttt{RUWE < 1.5}
    \item Total Position Error \texttt{< 1 mas}
    \item Total proper-motion error \texttt{< 0.3 mas/yr}
\end{itemize}
Many \textit{Gaia} measurements in the center of \omc{} suffer from crowding, and therefore, only 13\,520 stars pass our combined selection criteria. These well-measured stars are typically very bright (at $r = 0\arcmin$,  \textit{Gaia G} mag $< 12$; at $r = 2.5\arcmin$, \textit{Gaia G} mag $< 16$; and at $r = 5.0\arcmin$, \textit{Gaia G} mag $ < 18$). We note that at the time of making this work the \textit{Gaia FPR}, which addresses some of the crowding issues, was not available yet. Regardless, the astrometric precision of the bright stars from \textit{Gaia DR3} used as absolute astrometric reference is higher than the precision of the sources in \textit{Gaia FPR} (see also Section \ref{sec:catalog_validation}); therefore, the inclusion of the \textit{Gaia FPR} data would not significantly improve the astrometric registration.

The \textit{Gaia} proper motions of these reference stars were used to extrapolate their positions from the \textit{Gaia} reference epoch (2016.0) to the epoch of the GO-9442 \textit{HST} observations ($\sim$2002.5), to allow for a more precise astrometric match to the oldest \textit{HST} data.
Then, the angular \textit{Gaia} coordinates were converted to a convenient, pixel-based coordinate system using a tangent-plane projection. Our reference frame is defined with North up, East to the left, a pixel scale of 40~mas~pixel$^{-1}$ (similar to the UVIS instrument) and the cluster center at $(x,y) = (15000, 15000)$. We used \texttt{R.A. = 13:26:47.24h Dec. = -47:28:46.45$^\circ$} for the cluster center as found by \cite{2010ApJ...710.1032A}; these are also the central coordinates provided in the \cite{2010arXiv1012.3224H} catalog. 

We then crossmatched this \textit{Gaia}-based reference frame with the single-image \textit{HST} catalogs by determining the ideal linear six-parameter transformations to convert our image-based coordinates to the reference frame. At this step we encounter a fundamental challenge: While the well-measured \textit{Gaia} stars are typically very bright, they are saturated in the deep \textit{HST} exposures and therefore unusable for high-precision astrometry.

Therefore, we applied a two-step procedure. In the first step, we only crossmatched the short exposure time ACS observations (12~s-F435W, 8~s-F625W) from the 2002 epoch with the \textit{Gaia} reference stars. For these exposures a sufficiently high number of unsaturated stars was available and we could reliably determine the linear transformations. All the transformed short-exposures were combined to a first short-exposure \textit{HST} master frame.
In the second step, we crossmatched all other (long) \textit{HST} exposures with the short-exposure master frame and created, by combining positions measured from all 2002 exposures, our second \textit{HST} astrometric master frame.

For all epochs post-2002 we distinguished between the center and the off-center observations:
\begin{itemize}
\item In the central region (where \textit{Gaia} stars are sparse), for each epoch we incrementally crossmatched all exposures with the astrometric master frame of the previous epoch and determined the optimal linear transformations onto this preceding master frame. Then, we averaged the individual measured positions of the new data to create a new master frame (using only filters with a dedicated high-precision geometric distortion correction). This approach is reasonable because the time difference between the central epochs is low (typically just one year) and therefore the spatial displacements between epochs are expected to be small.

\item For the non-central regions (where the temporal gaps between \textit{HST} epochs are longer and there are more \textit{Gaia} stars), we updated the \textit{Gaia} based reference frames by propagating the stars to the correct epoch and then using the same hybrid-approach as described above for the 2002 epoch (but also only using exposures with a dedicated geometric distortion correction). When propagating the \textit{Gaia} positions to the correct epoch, we corrected for the absolute motion of \omc, as we want all our frames registered to the same cluster-based reference system.
\end{itemize}

\subsection{Known-offsets and motion of the center}
As described above, our astrometric reference system is based on positions from \textit{Gaia} (E-)DR3 and using the center estimate from \cite{2010ApJ...710.1032A} (in the following AvdM10). Although both the \citetalias{2010ApJ...710.1032A} center coordinates and the \textit{Gaia} positions are given in the International Celestial Reference System (ICRS), we have to note two caveats here: first of all, the \citetalias{2010ApJ...710.1032A} was anchored on the \textit{HST} catalog from \cite{2008AJ....135.2055A}, which itself was anchored on a small number of bright stars from the 2MASS catalog. The 2MASS astrometric reference frame has an absolute astrometric accuracy of 15~mas \citep{2006AJ....131.1163S}, but the errors of individual sources can be significantly larger. Indeed we observe an astrometric offset between the \citetalias{2010ApJ...710.1032A} catalog and our new (Gaia-based) absolute astrometry of around 100~mas. Although noticeable, this is still significantly smaller than the uncertainty of $1\arcsec$ given for the position of the center in \citetalias{2010ApJ...710.1032A}, therefore, we refrain from correcting the center estimate.

In addition, one has to take into account the absolute proper motion of \omc{}, which leads to a movement of the center over time. As our proper motions are determined in a reference frame co-moving with \omc{}, the center stays fixed at its 2002.5 position in our reference frame. This will lead to a time-dependent offset with respect to other astrometric catalogs.
From 2002 (the epoch of our first observations) to 2023 (the epoch of the last observations) the center will have moved by 68~mas (R.A. direction) and 141~mas (Dec. direction). 
This has to be taken into account when comparing our data with other catalogs (with absolute astrometry) such as the \textit{Gaia} FPR (see also Figure~\ref{fig:gaia_footprint}).

\subsection{Initial photometric registration and creation of a list of bright stars}
\label{subsec:brightlist}
As there are small photometric zero point variations even between exposures with the same integration time, in this step we determine the relative zero points between them. For each epoch, we then combine all single exposure measurements with the same filter and similar exposure time to a ``photometric masterframe". We start by searching for the best linear transformations and relative photometric zero points to crossmatch individual catalogs of a similar exposure time. Then the master frame is created by combining multiple measurements of each single star by calculating the averaged position and photometric measurement. The zero point estimates are iteratively improved, by crossmatching the individual exposure catalogs with the masterframe and then updating the masterframe until convergence.
After that, we combine the different exposure master frames of a single filter using the following rules: If a star was measured in multiple master frames, we use the measurement from the longest exposure time master frame in which it was not saturated. If there was no unsaturated measurement, we used the saturated measurement with the shortest exposure time as our best available estimate.

We compile a list of all stars that have an instrumental magnitude\footnote{We define instrumental magnitudes as ${m_{\rm inst.} = -2.5\log_{10}(N_{e^-})}$ with $N_{e^-}$ being the number of electrons fit with the PSF model.} brighter than \texttt{-9}. This list of bright stars is then used in the next step to mask PSF artifacts around bright stars. The list also contains saturated stars; these are not remeasured in the second photometry iteration and thus the \texttt{hst1pass} photometry is the best photometry available for these stars.

\subsection{Second photometry iteration with \texttt{KS2}}
\label{subsec:ks2}
To obtain our final astro-photometric measurements we used the \texttt{KS2} software written by Jay Anderson \cite[for more details, see][]{2017ApJ...842....6B}. \texttt{KS2} uses the ePSFs tailored to each image in Section~\ref{subsec:firstpass}, the transformations determined in Section~\ref{subsec:ref_frame}, and the list of bright stars compiled in Section~\ref{subsec:brightlist}.

\begin{figure*}
  \centering
    \includegraphics[width=1.0\textwidth]{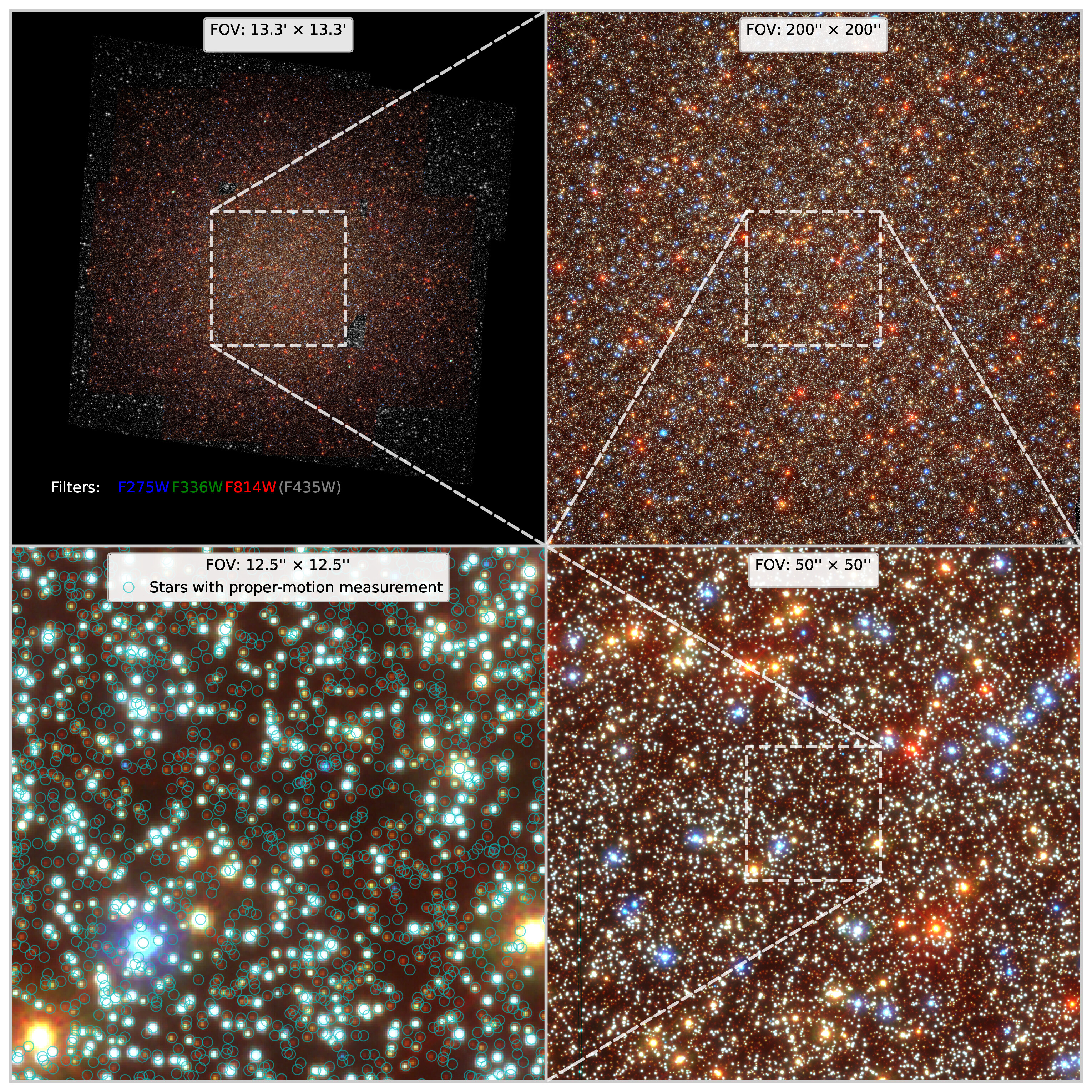}
  \caption{Zoom into a three-color composite image based on our stacked images. The red channel is WFC3/UVIS F814W, the green channel is WFC3/UVIS F336W and the blue channel is WFC3/UVIS F275W. Where no three filter coverage was available we used ACS/WFC F435W in gray-scale. Due to the wide color spread blue horizontal branch stars and red giant stars show strong colors and can be easily identified. In the highest magnification panel (lower left) we mark all stars with a successful proper motion measurement with a light-blue circle to demonstrate the depth and completeness of our catalog.}
  \label{fig:rgb_stack}
\end{figure*}

The program goes through several iterations of source finding, PSF fitting, and source subtraction. The source detection is based on peak maps from multiple images, which enables the detection of faint sources that do not produce a significant peak in each individual image. A byproduct of this process is the creation of deep stacked images, which we used to create a high-resolution 3-color composite image (see Figure~\ref{fig:rgb_stack}).

After the finding stage, the program performs photometry on the individual exposures. Before each star is measured, the flux of neighboring stars is subtracted using the ePSF model. KS2 measures photometry with three different methods that are more appropriate for different signal-to-noise regimes. Method~1 fits the ePSF to a 5x5 pixel aperture of the individual exposures with the flux and the $(x,y)$ position being free parameters. This only works if the star is bright enough to produce a significant peak in individual exposures. Method~2 takes the position determined from the peak map in the finding stage and only fits the flux within a 3x3 aperture. Finally, method~3 uses only the 4 brightest pixels and weights them according to their expected flux (based on the ePSF model). For the astrometric measurements, we rely on the method~1 measurements, the only method where position measurements are obtained in each individual image. We still keep the method~2 and method~3 photometry, as they might be useful for some science cases, e.g. when studying the photometry of stars on the faint end of the main sequence or along the white dwarf cooling sequence (e.g., \citealp{2013ApJ...769L..32B}).

\section{Proper Motions}
\label{sec:proper_motions}
\subsection{Inter-Epoch Crossmatch}
\label{subsec:crossmatch}
The goal of this step is to identify all stars that appear in multiple epochs, which are the stars for which a proper-motion measurement is possible.

We start by crossmatching each epoch with each of the other epochs. As stars move between epochs (due to their proper motions), to limit the number of miss-identifications we run the crossmatch with increasing matching radii, starting with a search radius of 0.1~UVIS~pixel, removing all stars that have been found from both catalogs, and then continuing with increasingly larger search radii up to a maximum radius of 5.0~UVIS~pixel. The individual search radii are \small{[0.1, 0.2, 0.3,..., 1.9, 2.0, 2.25, 2.5, 2.75, 3.0, 3.5, 4.0, 4.5, 5.0]}~UVIS~pixels.

\begin{figure}
  \centering
    \includegraphics[width=0.49\textwidth]{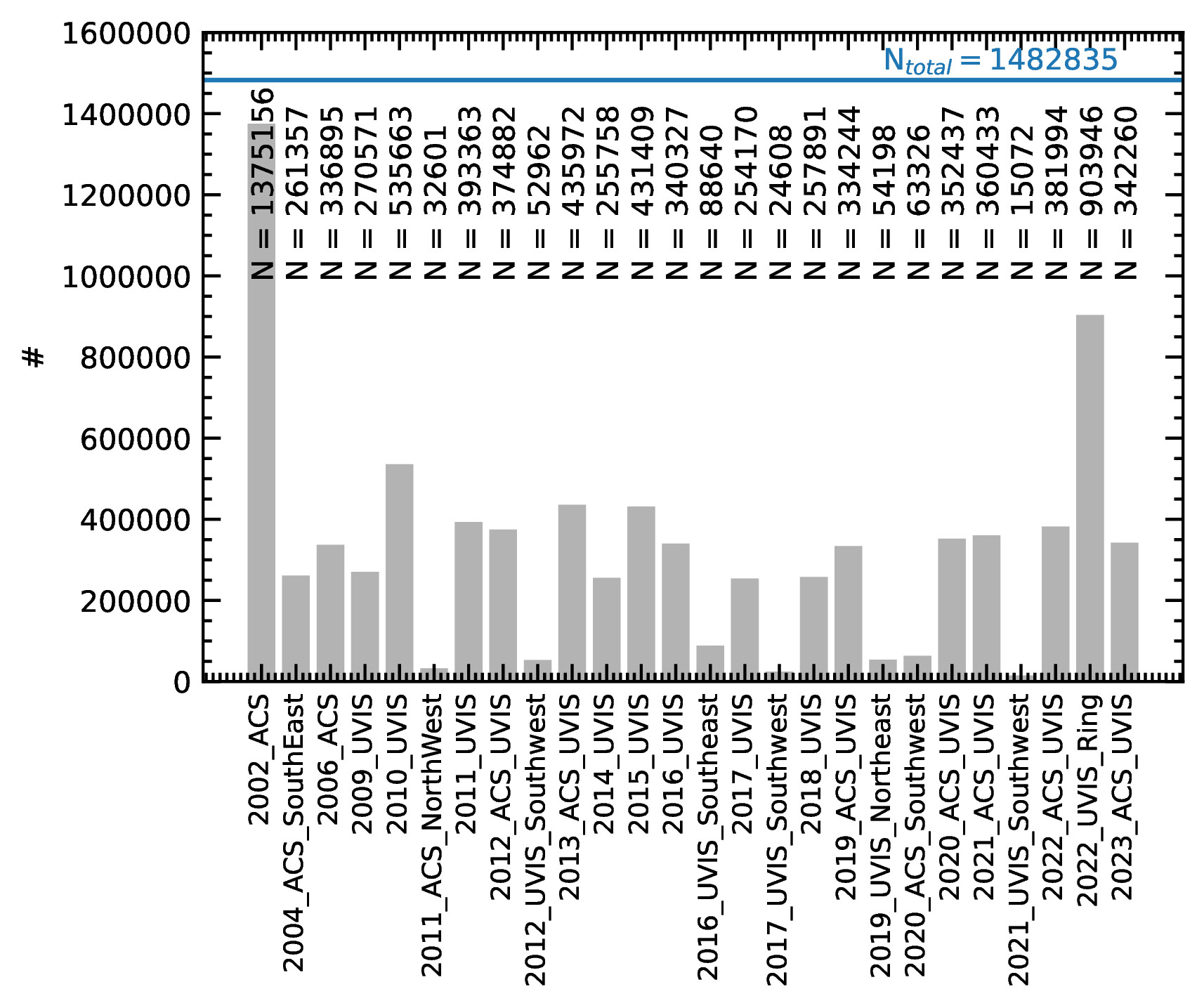}
  \caption{Number of stars of each epoch that could be crossmatched with at least one other epoch.}
  \label{fig:crossmatch_result}
\end{figure}

After all individual epoch pairs have been crossmatched, we combine the results into a single large table. This final table contains 1\,482\,835 stars measured in at least two epochs, all other (one-epoch-only) detections were discarded from the further analysis.
In Figure~\ref{fig:crossmatch_result} we show how many stars were contributed from each epoch. The epochs with the highest number of contributed stars are, as expected, the ones with the widest field coverage, i.e. the 3$\times$3 2002 ACS mosaic (GO-9442) with 1\,375\,156 measurements, and the newly observed ring of 10 UVIS fields contained in the half-light radius (GO-16777) with 903\,946. The central epochs contributed typically between 250\,000 and 500\,000 stars depending on the depth and the dither pattern.

\subsection{Iterative Proper Motion Determination}
\label{subsec:pmiter}
Proper motions are measured using the method developed in \cite{2014ApJ...797..115B} and improved in \cite{2018ApJ...853...86B, 2018ApJ...861...99L}, and we refer to those publications for detailed descriptions of the procedure. Proper motions are measured relative to a subset of well-measured cluster stars. The set of reference stars is iteratively improved. In the first iteration, it is based on photometric-quality indicators and the stars' position in the color-magnitude diagram alone. Once proper motions become available, non-members are also removed based on their position in the vector-point diagram (Figure~\ref{fig:vpd}) or if they have a spurious proper motion measurement. 

The program treats each individual image catalog as a stand-alone epoch. As a first step, all these geometric-distortion-corrected\footnote{Our software uses the library geometric distortion corrections from \cite{2003hstc.conf...13A} (ACS), \cite{2009PASP..121.1419B}, and \cite{2011PASP..123..622B} (UVIS), with an additional look-up table for ACS observations post \textit{HST}-SM4 (2009). We noticed that the ACS/WFC distortion has worsened over time. For this reason, we made additional, time-depend table-of-residuals corrections for the latest observations  (epoch $>$ 2018) with the ACS/WFC detector following the prescriptions in \cite{2011PASP..123..622B}.} individual-image catalogs are transformed to the reference frame using linear six-parameter transformations. These transformations are determined individually for the 4 amplifiers used to read out the images (i.e. the single image catalogs are split into 4 quadrants, corresponding to the 4 amplifiers reading out the detectors of the instruments) to mitigate potential amplifier-based systematic effects. In the final iteration, the transformations are determined individually for each star using its 100 closest reference stars. The transformed positions of each star are fitted with a straight line in both the $x$ and $y$ directions, to directly fit the two proper-motion components. The fit takes into account the magnitude-dependent astrometric errors and has several stages of outlier rejection.

In another iterative loop, the crossmatch and the transformations to the master frame positions are improved by using the proper motions to propagate the reference stars to the same epoch as the individual image catalogs. In total, 103\,616\,339 individual position measurements were used for the proper-motion measurements, making this one of the largest astrometric datasets of all times.

Some detections that appeared as two separate sources during the crossmatch could be reassigned to a single source using the proper motions. Therefore, the final number of entries in our catalog is 1\,475\,096, slightly lower than the 1\,482\,835 sources measured at least twice during the crossmatch.

In total, 1\,395\,781 individual sources pass the iterative process and have a high-precision proper-motion measurement.

\begin{figure*}
  \centering
    \includegraphics[width=1.0\textwidth]{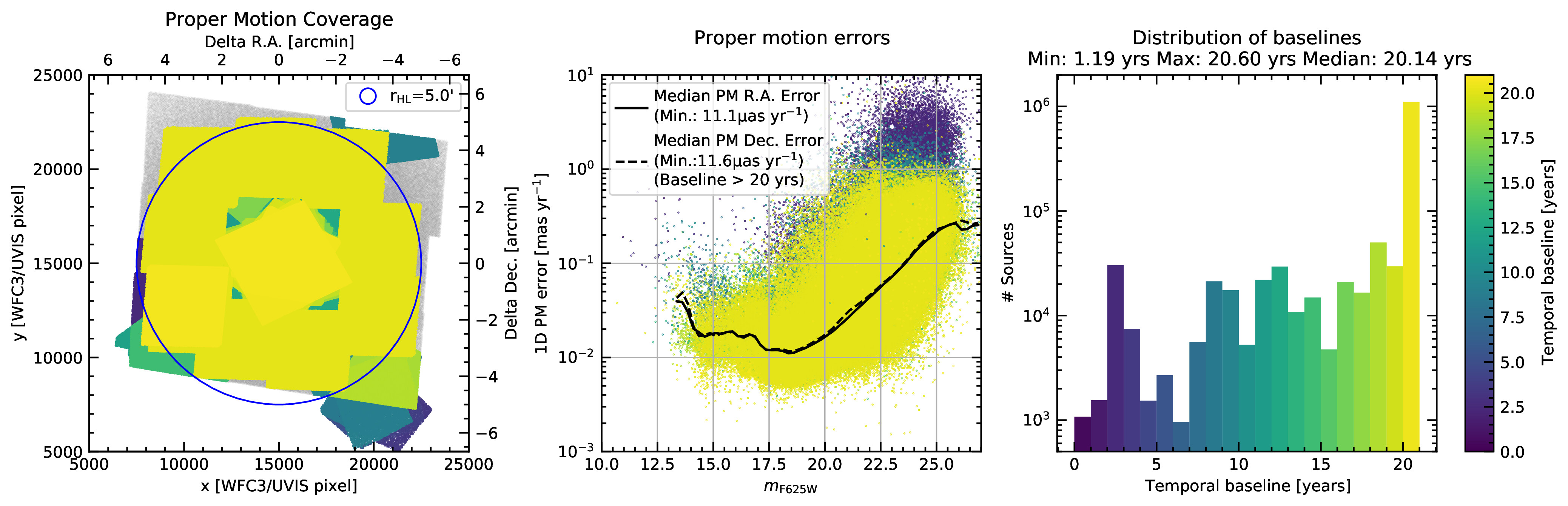}
  \caption{This figure shows how the temporal baseline of the astrometric data depends on the position within the observed field and how this affects the proper-motion error. \textbf{Left panel:} The colored areas show the parts of the field, where we were able to measure proper motions. The color-coding indicates the maximum temporal baseline used for the measurements. We achieve a highly uniform baseline of typically 20.6~years across most of the field. \textbf{Middle Panel:} We plot the proper-motion error as a function of the F625W magnitude and the temporal baseline. A longer baseline leads to a lower proper-motion error. The black line indicates the magnitude-dependent median error. For brighter stars, a better proper-motion error is achieved due to better S/N. This trend is reversed when the star is saturated in an increasing number of exposures (for $m_{\rm F625W} < 17.5$). \textbf{Right Panel:} Here we show the distribution of baselines. The majority (79\%) of proper-motion measurements have a baseline longer than 20~years.}
  \label{fig:pm_baseline}
\end{figure*}

\begin{figure*}
  \centering
    \includegraphics[width=1.0\textwidth]{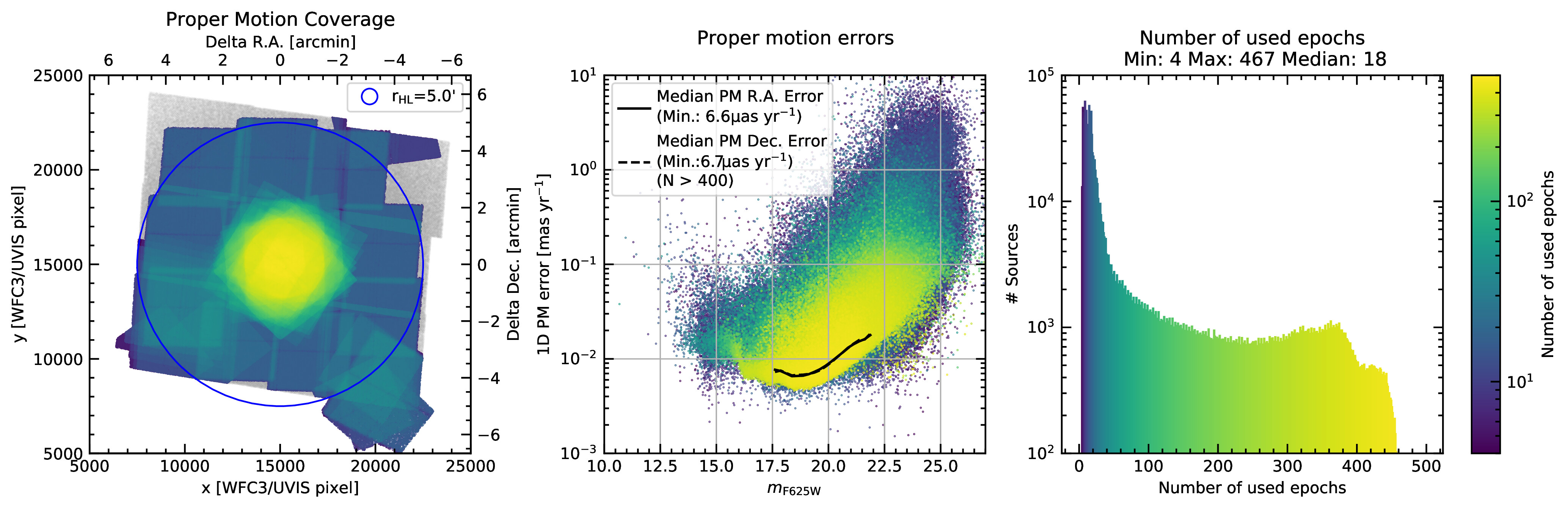}
  \caption{Similar to Figure~\ref{fig:pm_baseline} but with the number of used epochs as color coding. We can see how the number of astrometric measurements depends on the position within the observed field and how this affects the proper-motion error. \textbf{Left panel:} The colored areas show the available number of measurements at different locations in the field. In the center, a very large number of astrometric measurements (up to 467) is available due to the larger amount of data. In the outer parts of the field, there are typically around 10 measurements. \textbf{Middle Panel:} We show how the 1D proper-motion error depends on the F625W magnitude and the number of available measurements. The higher the number of measurements, the lower the proper-motion error. The black line indicates the magnitude-dependent median error for stars with more than 400 measurements. These stars are all located close to the center of the cluster and are in an ideal magnitude range from $m_{\rm F625W} = 17.5$ to $m_{\rm F625W} = 22.0$. For these stars, a median proper-motion error of only 6.6~$\mu$as yr$^{-1}$ is achieved in both directions. \textbf{Right Panel:} Here we show a histogram of the number of available measurements. The majority of stars have less than 100 measurements, with a median of 18 measurements.}
  \label{fig:pm_nused}
\end{figure*}

The vast majority of our proper motions (1\,102\,818) have a temporal baseline longer than 20 years. The median number of individual astrometric measurements used for the proper-motion determination is 17, it is lowest in the outskirts of our field of view and quickly increases towards the center. In the very center and in the best-covered magnitude range ($m_{\rm F625W} = 17.5$ to $m_{\rm F625W} = 22.0$), many stars have more than 400 individual measurements (with a maximum of 467), leading to a median proper-motion error of only 6.6~µas\,yr$^{-1}$ ($\sim$0.15\,km\,s$^{-1}$ at the distance of \omc{}) with individual stars reaching as low as 3.3~µas\,yr$^{-1}$. The field dependence of the temporal baseline and the number of available measurements and their effect on the proper-motion error are presented in Figures \ref{fig:pm_baseline} and \ref{fig:pm_nused}.

\subsection{A-posteriori corrections}
\label{subsec:corrections}
\begin{figure*}
  \centering
    \includegraphics[width=1.0\textwidth]{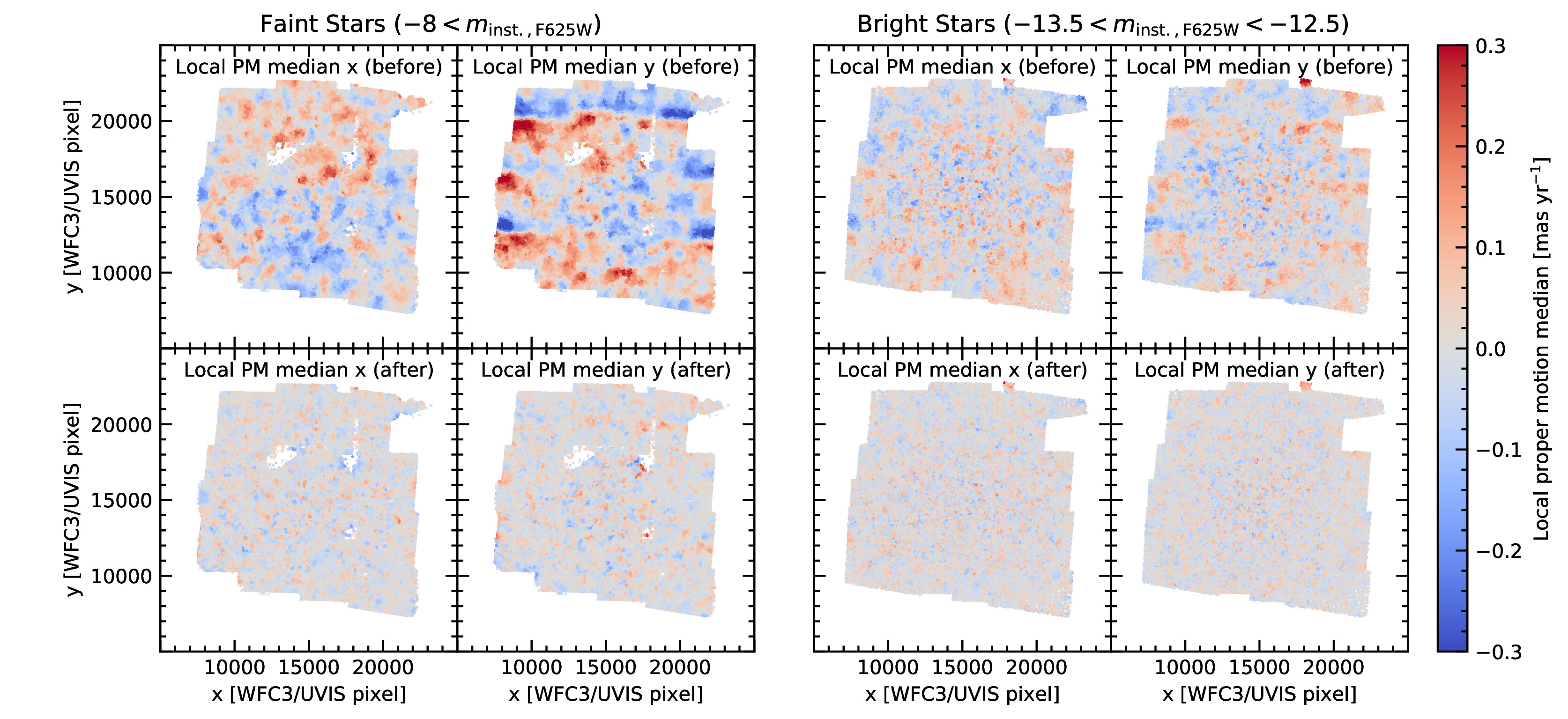}
  \caption{The top panels show both locally averaged proper-motion components for faint stars (left sub-figure) and bright stars (right sub-figure) before the a-posteriori corrections are applied. Especially for the faint stars, the imprint of uncorrected charge-transfer-efficiency effects in the $y$/declination direction becomes clearly visible. The lower panels show the local average of the proper motions after the corrections have been applied. The systematic residuals have disappeared.}
  \label{fig:pm_corrections}
\end{figure*}

The resulting (amplifier-based) proper motions are of excellent quality. However, uncorrected charge transfer efficiency effects and residual distortion can lead to small systematic trends in the proper motions that vary both spatially and with the magnitude of the stars. We correct for these with a-posteriori corrections, following the prescriptions from \cite{2014ApJ...797..115B} and \cite{2022ApJ...934..150L}. For each star, we search for neighboring cluster stars with a similar ($\Delta m < 0.5$) magnitude within a radius of 600 UVIS pixel. If there are less than 50 neighbors matching those criteria, we do not calculate a correction. This is only the case at the edges of the observed field. If there are more than 150 neighboring stars, we use the 150 closest neighbors as reference stars. 
Using the assumption that the mean motion of those neighboring cluster stars should be zero in both proper-motion directions by construction, we then calculate the 3.5 sigma clipped median of the proper motion of the neighboring stars and use this as correction value. The effectiveness of this method can be seen in Figure~\ref{fig:pm_corrections}.
Applying this correction removes systematic errors, but comes at the cost of adding an additional statistical uncertainty. For a typical 1D velocity dispersion of 0.65~mas~yr$^{-1}$ and 150 reference stars, this uncertainty is $\sigma_{\rm correction} = \frac{0.65~\text{mas~yr}^{-1}}{\sqrt{150}} = 0.053~\text{mas~yr}^{-1}$.
As there is no filter for which we have measurements for all stars, we use the following approach to obtain a correction for each star: We calculate the correction in multiple filters, then take the correction from the ACS F625W filter (i.e. the filter with the largest field coverage and the largest number of measurements). If a star has not been measured in that filter, we take the correction from other filters in the order of the number of measurements (F814W, F435W, F336W, F606W). This approach is reasonable as there is no strong dependence on the filter used for the local corrections, as we search the reference stars in a narrow magnitude interval. In total, we were able to obtain a local correction for 1\,384\,877 of 1\,395\,781 stars with a proper-motion measurement.

It is important to note that this local approach to measure proper motions also removes any signature of rotation from the proper motions (as this is a systematic effect on scales larger than the areas used to determine the transformations and local corrections). In Section~\ref{sec:rotation} we discuss how the rotation can be recovered.

\subsection{Vector-point diagram}
\begin{figure*}
  \centering
    \includegraphics[width=1.0\textwidth]{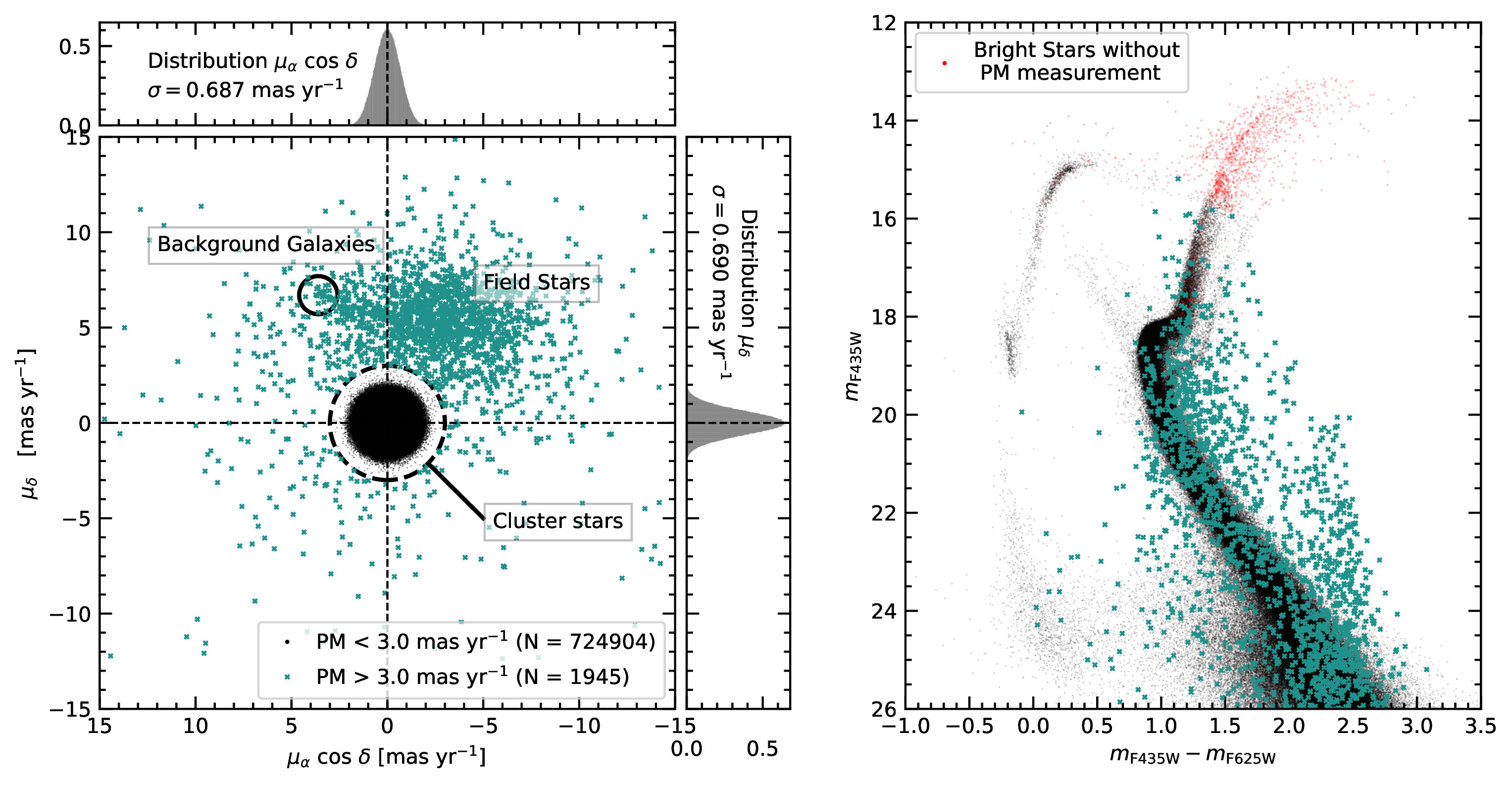}
  \caption{\textbf{Left panel: } A vector-point diagram of the relative proper motions for a subset of around 700\,000 well-measured stars in our catalog. Most stars are distributed around the origin, as expected for \omc's member stars and follow a normal distribution in both velocity components (see marginalized histograms at the edge of the plot). A small fraction of stars has a relative proper motion incompatible with the cluster's motion. Those stars lie outside of the 3.0~mas~yr$^{-1}$ radius indicated with a dashed circle (corresponding to a 4.5$\sigma$ deviation from the velocity distribution of the cluster stars) and are marked with turquoise crosses. The non-member stars show substructures that can be attributed to background galaxies (marked with a solid circle) and Galactic foreground/background stars. \textbf{Right panel:} A color-magnitude diagram of the same sample. One can see that the stars with a high relative proper motion do not follow the CMD sequences of \omc. Our proper motions cover almost the entire CMD, however the brightest stars, indicated with red dots, do not have a proper motion measurement, as they are saturated even in the shortest \textit{HST} exposures.}
  \label{fig:vpd}
\end{figure*}

As a first demonstration of the quality and size of our proper-motion catalog, we show the vector-point diagram of the proper motions in Figure~\ref{fig:vpd}. In this plot, we only show stars with both well-measured proper motion and well-measured photometry in the F435W and F625W filters, using the exemplary quality cuts described in Section \ref{sec:data_products}. These selections leave us with a subset of around 700\,000 stars from the sub-giant branch down to white dwarfs and faint main-sequence stars. As expected, most stars are concentrated around the origin $(0,0)$ and show a normal distribution with $\sigma \approx 0.66$~mas yr$^{-1}$ in both velocity components\footnote{Please note that this is not a proper measurement of the velocity dispersion of \omc{} yet. For this, we would have to account for the errors on the proper-motion measurements, split the dataset into different radial and mass bins, and perform a more careful selection of cluster stars. A detailed study of the kinematics of \omc{} will be done in a follow-up work.}. In addition to the cluster stars, there are additional over-densities visible corresponding to background galaxies and Galactic field stars in the fore- and background of \omc. As shown in the color-magnitude diagram (CMD) in Figure~\ref{fig:vpd}, a simple total proper motion cut of \texttt{PM < 3 mas yr$^{-1}$} allows an effective separation of cluster stars from field stars.
\section{Creation of the photometric catalog}
\label{sec:photometry}
After the proper-motion determination, we are left with individual image catalogs that are all matched very precisely with the crossmatched master catalog. However, the photometric information is still based on various different \texttt{KS2} runs (see Section~\ref{subsec:ks2}) and still in uncalibrated instrumental magnitudes. The goal of this step is to combine all these single-image measurements into a uniform, calibrated, photometric catalog for the 6 filters for which we have coverage over the full field (WFC3/UVIS F275W, F336W, F814W; ACS/WFC F435W, F625W, F658N). In addition, we include the WFC3/UVIS F606W filter, as this is the filter with the most uniform and extensive coverage in the center. 

\subsection{Creation of a photometric reference catalog}
\label{subsec:photometric_ref}
The goal of this step is to create calibrated, aperture-photometry-based reference catalogs for all filters that we can then crossmatch with our PSF photometry catalogs to obtain their zero points.
The reference catalogs are created similarly to the procedure described in \cite{2017ApJ...842....6B},  using the current version of the WFC3/UVIS zero points. We perform aperture photometry on a selected subset of exposures for all filters. In contrast to our PSF photometry measurements, we now use the resampled and flux-normalized (to 1~s exposure time) \texttt{*drc} type images. For the photometric reference catalog we choose exposures with a representative exposure time for the respective filter and covering the full field of view. For the WFC/ACS filters (F435W, F625W, F658N) we use the full GO-9442 data. For the WFC3/UVIS filters (F275W, F336W, F606W, F814W) we use the full GO-16777 data for the outer fields and the data from GO-11911/12094 for the central field.

We perform aperture photometry with various radii between 2.5 and 10 pixels (and a sky annulus between 12 and 16 pixels). For the ACS data we then add the respective infinite aperture correction from \cite{2016AJ....152...60B} and the date-specific VegaMag zero point from the ACS zero point calculator\footnote{\url{https://acszeropoints.stsci.edu/}}. For the WFC3/UVIS data these two steps are unified in the python package \texttt{stsynphot}\footnote{\url{https://stsynphot.readthedocs.io/en/latest/index.html}} including the most recent photometric calibrations \citep{2021wfc..rept....4C,2022AJ....164...32C}. We follow the example notebook\footnote{\url{https://github.com/spacetelescope/WFC3Library/blob/master/notebooks/zeropoints/zeropoints.ipynb}} to calculate the WFC3/UVIS zero points. Once we have the calibrated, single-exposure, aperture-based photometric catalogs we crossmatch them to our astrometric master catalog and then create a combined reference catalog for each filter. If a star has multiple measurements in the aperture photometry catalogs, we combine them using the median. We remove all stars from the catalog with a brighter neighbor within 20 pixels, as those brighter neighbors contaminate the aperture photometry. Also, we chose the aperture radius that provides the lowest scatter when compared to the PSF photometry. This is 3.5 pixels for the ACS filters (F435W, F625W, F658N), 4.0 pixels for WFC3/UVIS F606W and F814W, and 4.5 pixels for WFC3/UVIS F275W and F336W.

\subsection{Creation of an error model}
\label{subsec:error_model}
\begin{figure*}
  \centering
    \includegraphics[width=1.0\textwidth]{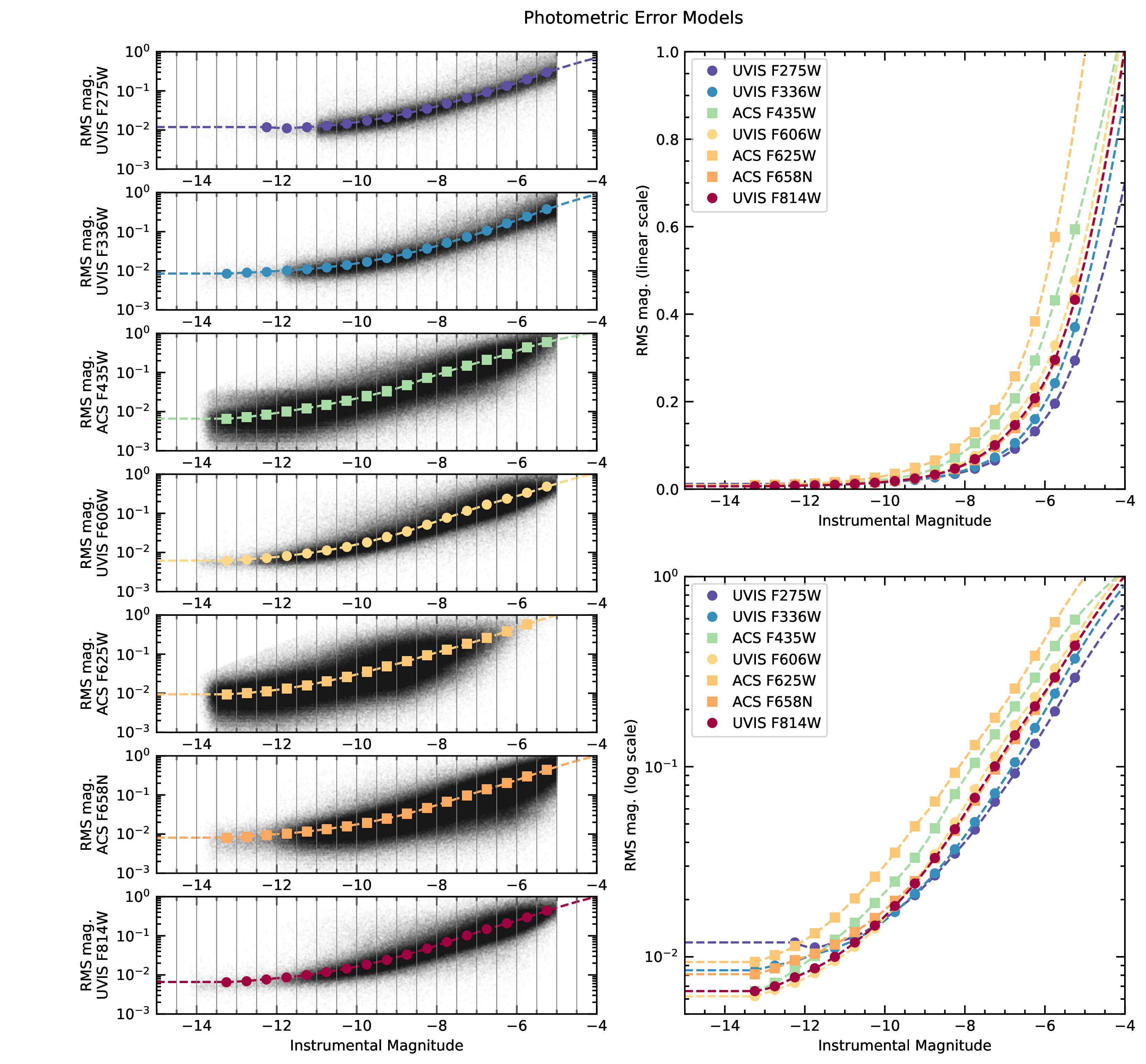}
  \caption{Empirical photometric error model for 7 different filters. The plots in the left panel show the RMS of the measured magnitudes of stars that have been detected in at least 3 exposures, plotted against their long-exposure instrumental magnitude (black dots) and the derived error model as colored line. In the two right panels, we compare the error models for the different filters in a linear and a logarithmic plot. All filters show similar curve of rising errors towards fainter magnitudes, but the errors of the two wide ACS filters (F435W, F625W) are higher than the rest. This is due to typically longer exposure times of these filters, which leads to higher crowding.}  
  \label{fig:error_model}
\end{figure*}

Before combining individual measurements, we want to find the dependence of the statistical photometric errors on the instrumental magnitude, to be able to properly weight the individual data points. Therefore, we create an empirical error model for each filter. For each filter, we choose one epoch and collect all stars that have been measured at least 3 times. Then, we determine the 67th percentile of the RMS of the instrumental magnitude of these stars in 0.5 magnitude wide bins. We quadratically interpolate between these values to obtain a smooth error model. The resulting error models are shown in Figure~\ref{fig:error_model} and we use these as 1-sigma errors on the individual measurements.

\subsection{Creation of a model for CTE effects}
As described in Section~\ref{subsec:firstpass}, we use CTE corrected input images of the \texttt{*flc.fits} type. However, the applied CTE correction underpredicts the evolution of the CTE loss for the most recent WFC3/UVIS observations. This leads to noticeable CTE effects in both the astrometry and photometry of observations taken after 2017, especially for images with a low background (i.e. those with a short exposure time and those with a very blue filter). Since the start of this work, \texttt{hst1pass} \citep{2022wfc..rept....5A} has been updated with a new and improved CTE correction. However, this correction is meant to be applied to astro-photometry measured on uncorrected \texttt{*flt.fits} type images. 

Since this improved CTE-correction routine is not currently included in \texttt{KS2}, to our \texttt{KS2}-based single-exposure catalogs we applied empirical corrections based on the comparison between the \texttt{hst1pass} runs on \texttt{*flt.fits} and \texttt{*flc.fits} images. To derive a model that can transfer the new corrections to our data, we first grouped the data in sets of the same filter and the same exposure time (ensuring the same background level in each such group). After this, we collected the residuals between the hst1pass-flt results and the KS2-flc results. We then modeled these residuals in an instrumental magnitude versus distance-to-amplifier space, to be able to calculate a correction for each measurement in our KS2 based catalogs.

The largest corrections are applied to the UV filters (F275W, F336W) of the GO-16777 program. For the faintest stars ($m_{\rm inst.}\sim -6$) at the largest distances from the amplifier the corrections can reach up to 1 mag, for brighter stars they are much lower.

\subsection{Combination of measurements}
\label{subsec:phot_comb}
In this step, we combine the photometry from different epochs and KS2 runs and also find the zero points to transform our instrumental magnitudes into the Vega magnitude system.

\begin{figure*}
  \centering
    \includegraphics[width=1.0\textwidth]{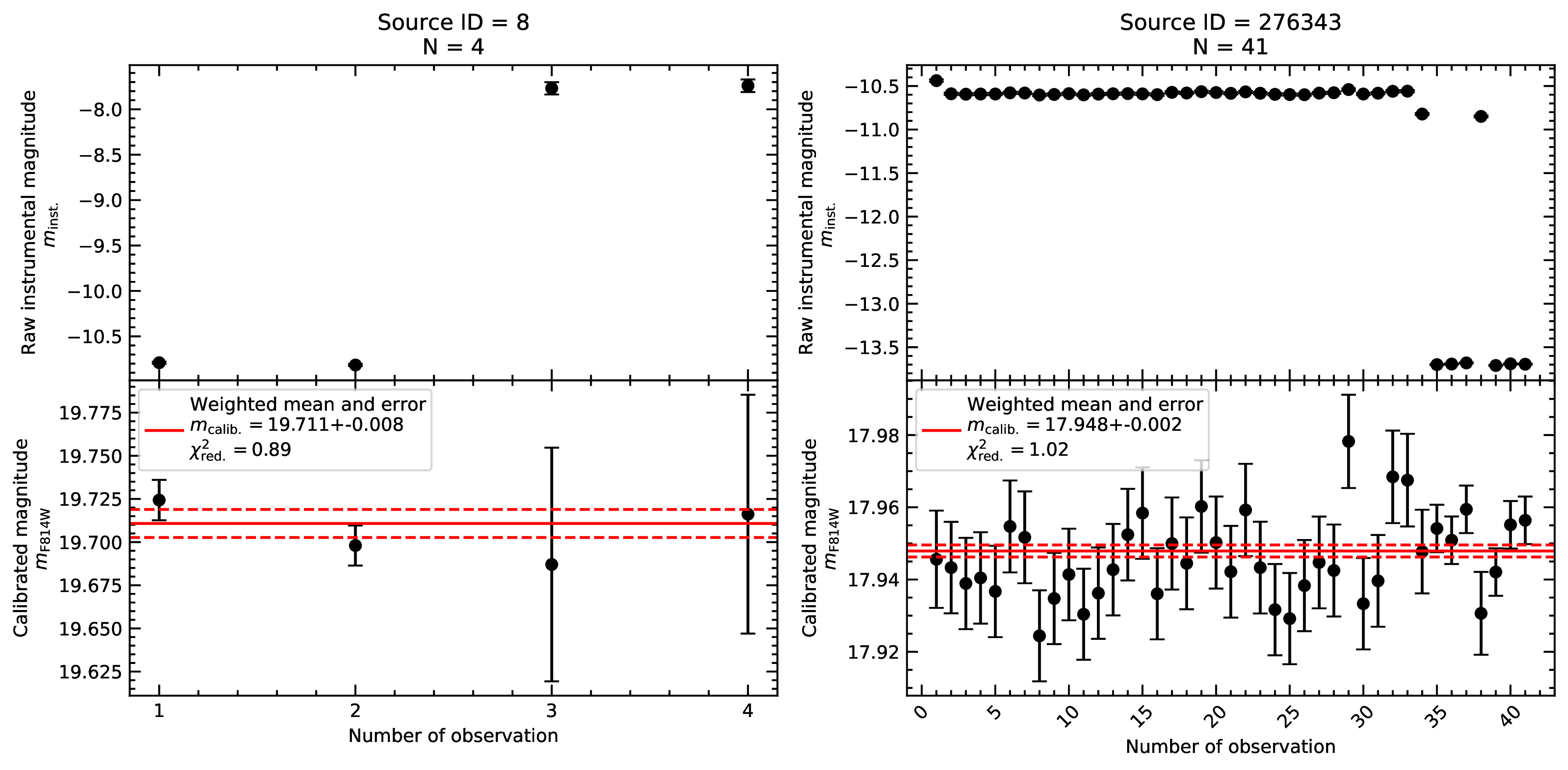}
  \caption{This figure shows the individual photometric measurements for two typical sources in the WFC3/UVIS F814W filter. The left panels are for a star measured in the outer regions, where we typically have 4 measurements (2 long and 2 short exposures from program GO-16777). The right panels are for a star in the central region where there is a much higher number of individual exposures and measurements. The upper panels show the raw (uncalibrated) instrumental magnitudes, the lower panels show the individual measurement after the zero point for each exposure is added, and the resulting error-weighted mean magnitude (red line).}
  \label{fig:photometry_example}
\end{figure*}

We follow an iterative approach in which we first crossmatch the individual single-image catalogs with the reference catalogs created in Section~\ref{subsec:photometric_ref} to determine the zero point for each exposure. 
We determine the zero point by calculating the difference between the instrumental magnitudes and the reference magnitudes. We calculate the 3.5\,$\sigma$ clipped mean of bright ($m_{\rm inst}<-9$), well-measured ($\mathtt{QFIT} > 0.95$) stars. Then, we combine measurements of all exposures using the error-weighted mean (with the empirical errors derived in Section~\ref{subsec:error_model}):
\begin{equation}
    m_{\rm combined} = \frac{\sum_{i=1}^{n} (\frac{m_{\rm inst,i}+ZP_i}{\sigma_{m,i}^2})}{\sum_{i=1}^{n}\frac{1}{\sigma_{m,i}^2}}
\end{equation}
The error of this weighted mean is:
\begin{equation}
    \Delta m_{\rm combined} = \sqrt{\frac{1}{\sum_{i=1}^{n}\frac{1}{\sigma_{m,i}^2}}}
\end{equation}
After this first iteration is done, we crossmatched the resulting calibrated average catalog again with the individual single-image catalogs and redetermined the zero points. When available, we use the same crossmatch as in the last proper motion iteration (as this crossmatch takes into account the motion of the stars). 
The new calibrated catalog allows us to also determine the zero point for those pointings for which there is not enough overlap (either spatially or in magnitude) with the initial reference catalog. We repeat this procedure 4 times. After 4 iterations no additional pointings can be added and all zero points have converged.
In addition to the weighted mean of the calibrated magnitudes, we also calculate several other statistical quantities based on the distribution of individual measurements (see Table~\ref{tab:cat_photometry}). An example of the combination of measurements in two typical photometric situations is shown in Figure~\ref{fig:photometry_example}.

\subsection{Empirical Photometric Corrections}
\label{subsec:phot_corr}
Even though \omc{} is located in a region of the sky with low total extinction ($E(B-V)=0.12$, \citealt{2010arXiv1012.3224H}) differential reddening of the order of up to $\pm$10\% has been reported in \cite{2017ApJ...842....7B} for the cluster's core. In addition to these physical effects, there are also small systematic zero point variations over the field caused by instrumental effects such as small variations of the PSF and the detector sensitivity, but also issues caused by the decreasing charge transfer efficiency of \textit{HST}'s ageing detectors.

All these effects broaden the observed CMD sequences of the different subpopulations and limit our ability to separate them. Therefore, we derive an empirical correction for spatially dependent photometry variations.

\subsubsection{Method}
\label{subsec:phot_corr_method}
The method we developed is adapted from the differential reddening correction described in detail in \cite{2017ApJ...842....7B}. In contrast to this work and due to the higher complexity of our dataset (two instruments, a larger time-span of the observations, and a much larger field with an irregular mosaic of observations) we could not use the assumption that all spatial photometric variations are caused by true physical extinction that follows a wavelength dependent reddening law. Instead, we determined a spatial photometric correction for each of the 6 filters by studying the behavior of a set of well-measured reference stars that all lie on a single sequence in the CMD. 

To define the set of reference stars we only used stars that had a photometric measurement in all of the 6 filters with full field coverage. After an initial quality selection, based on the QFIT and the scatter of the individual measurements, we manually selected a single subpopulation in the $m_{\rm F814W}~ vs.~ m_{\rm F275W}-m_{\rm F814W}$ and the $m_{\rm F814W}~ vs.~ m_{\rm F336W}-m_{\rm F435W}$ CMD, both with $ 15 < m_{F814W} < 19.0 $. This lower limit is enforced to limit the CTE effects which affect fainter magnitudes more strongly. We intentionally use these two CMDs which, together, allow the clean separation of a single sequence from the other subpopulations, ensuring the spread in the reference stars is only due to instrumental and DR effects. Our initial list of reference stars contains 70\,040 entries. It is updated once we perform the selection in differential reddening corrected CMDs. After 3 iterations it contains 59\,060 stars.

\begin{figure*}
  \centering
    \includegraphics[width=1.0\textwidth]{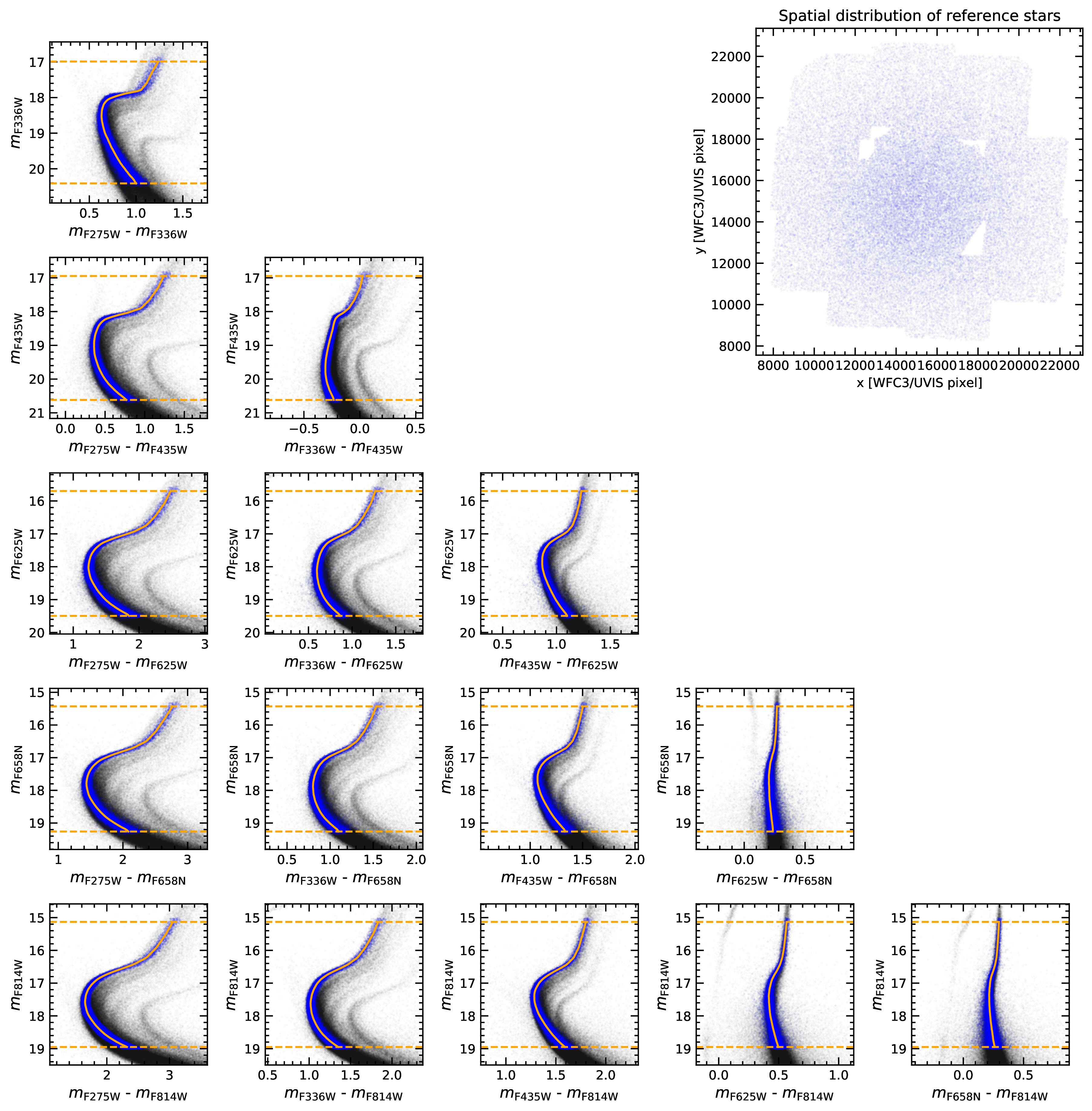}
  \caption{The plots in the lower left of this figure illustrate the multi-dimensional space in which we determined the photometric corrections. Each of the 15 smaller panels shows a color-magnitude diagram with a different filter combination. To determine the corrections, we determine the local photometric offsets with respect to fiducial lines (yellow solid lines) based on a single sequence of well-measured reference stars (blue). Yellow dashed lines denote the magnitude limits of the reference stars. The spatial distribution of the reference stars is shown in the upper right.}
  \label{fig:fiducial_lines}
\end{figure*}

Once the reference stars were identified, we determined the median color in bins of magnitude, for each of the 15 2-color CMDs that can be created with 6 filters (see Fig. \ref{fig:fiducial_lines}). These fiducial lines serve as the baseline with respect to which we compare the local distribution of magnitudes in the next step.

We determine the correction for each filter on an evenly spaced on-sky grid with pixel spacing of 100 WFC3/UVIS pixels (4\arcsec). For each point in our grid, we identify the 300 closest reference stars with a maximum search radius of 1000~WFC3/UVIS~pixel. For gridpoints with fewer than 150 neighbors we do not calculate a correction. Once the local set of reference stars was determined, we optimized the set of 6 photometric corrections by minimizing the squared sum of the deviations from the fiducial lines in each CMD. After the corrections have been determined for the six filters with full field coverage, we also calculated the optimal corrections value for the WFC3/UVIS F606W, which was only used in the central pointing. We interpolate the grid for each filter at every star location to obtain the photometric correction for each star. 

To quantify the statistical error of these correction values, we perform 20 bootstrap resamplings on each grid point correction value. This gives us an average error of $ \simeq 0.006$ mag for pixels with 300 reference star neighbors. Pixels with fewer than 300 neighbors have increased errors; those with fewer than 150 neighbors do not give us reliable correction estimations which is why we in this case we do not calculate a correction. However, this only affects a small area that the edge of the field.

We find a fairly narrow spread and uniform distribution for the error on our photometric corrections and therefore quote one value per filter for the error (see Table~\ref{tab:redd_coeff}). 

\subsubsection{Results}
We show the statistical properties of the correction in Table \ref{tab:redd_coeff} and detailed maps and histograms in Appendix~\ref{sec:appendixphotcorr} (Figure \ref{fig:corrections_acs} and \ref{fig:corrections_uvis}). The maps show various patterns, that can partially be attributed to physical differential reddening, but also transitions between different pointings. A detailed decomposition into those two components is out of the scope of this work and would not further improve the corrected photometry.

\begin{figure*}
  \centering
    \includegraphics[width=1.0\textwidth]{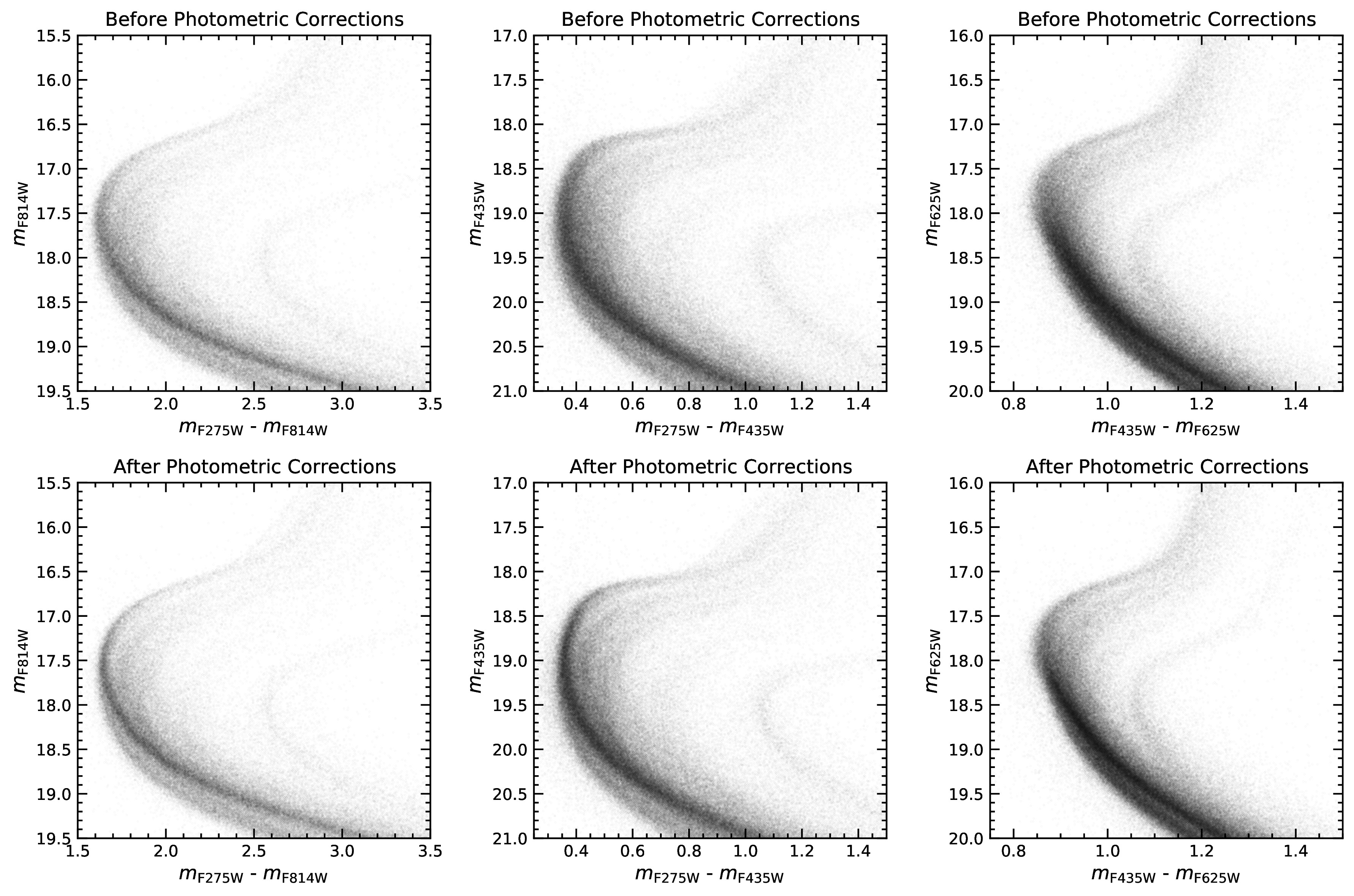}
  \caption{Color-magnitude diagrams (CMDs) of the turnoff region using various filter combinations before (top panels) and after (bottom panels) the photometric corrections have been applied. In all CMDs, the corrections lead to narrower sequences and a clearer separation of the different subpopulations.}
  \label{fig:corrections_before_after}
\end{figure*}

The effectiveness of the correction is demonstrated in various before-/after correction CMDs in Figure \ref{fig:corrections_before_after}. The corrections lead to narrower CMD sequences and a clearer separation of the different subpopulations in all CMDs.

\begin{table}[]
\caption{Statistical properties of the derived empirical photometric corrections for 7 filters}
\label{tab:redd_coeff}
\scriptsize
\centering
\begin{tabular}{lllllll}
\hline
Instrument & Filter & Min. & Med. & Max. & RMS & Error\footnote{Median error on correction determined using bootstrapping} \\ \hline
WFC3/UVIS & F275W &   -0.059 &    0.007 &    0.044 &    0.017 & 0.0050\\
WFC3/UVIS & F336W &   -0.033 &    0.003 &    0.035 &    0.010 & 0.0053\\
ACS/WFC & F435W &   -0.037 &    0.001 &    0.031 &    0.009 & 0.0053\\
WFC3/UVIS & F606W &   -0.027 &    0.002 &    0.036 &    0.010 & 0.0059\\
ACS/WFC & F625W &   -0.033 &    0.003 &    0.041 &    0.013 & 0.0056\\
ACS/WFC & F658N &   -0.033 &    0.004 &    0.041 &    0.013 & 0.0057\\
WFC3/UVIS & F814W &   -0.042 &   -0.001 &    0.041 &    0.015 & 0.0059\\ \hline
\end{tabular}
\end{table}

\subsection{Treatment of bright stars}
The individual photometric measurements discussed in the section above were all performed with the software \texttt{KS2}. However, this software is not able to measure saturated stars. This limits our completeness at the bright end, as the brightest red giant stars are saturated even in the shortest exposures of e.g. the F625W and F814W filters. Including precise photometry for these stars is still important, as they have the highest S/N in spectroscopic studies. We substituted the missing KS2 measurements with data from our single-image \texttt{hst1pass} catalogs, using the same zero-points as determined in the iterative procedure described in Section \ref{subsec:phot_comb}. 

\section{Comparison with literature catalogs and validation}
\label{sec:catalog_validation}

To validate our new astro-photometric measurements, we performed a search for residual color and magnitude trends (Appendix \ref{sec:appendixtrends}) as well as extensive comparisons with previously published \textit{HST} and \textit{Gaia} based catalogs. These are described in detail in Appendix \ref{sec:appendixlit}. In the following, we limit ourselves to a comparison of the general catalog properties and a summary of the direct astrometric/photometric comparisons.

\subsection{Comparison of general catalog properties and completeness}
We compare our astrometry and our photometry with the two other most recent high-precision catalogs for the central region of \omc{}: The \textit{HST}-based astrophotometric catalog published by \cite{2017ApJ...842....6B} and the \textit{Gaia} catalog (combining data from both \textit{DR3} and \textit{FPR}). The two comparison datasets are complementary for our catalog verification: While the \cite{2017ApJ...842....6B} catalog probes faint stars in the very center of \omc{} with a similar photometric methodology and a similar approach to measure relative proper motions (although with a significantly shorter temporal baseline), the \textit{Gaia} data is shallower but provides a larger field-of-view and gives us a fully independent comparison with absolute proper motions.

\begin{figure*}
  \centering
    \includegraphics[width=1.0\textwidth]{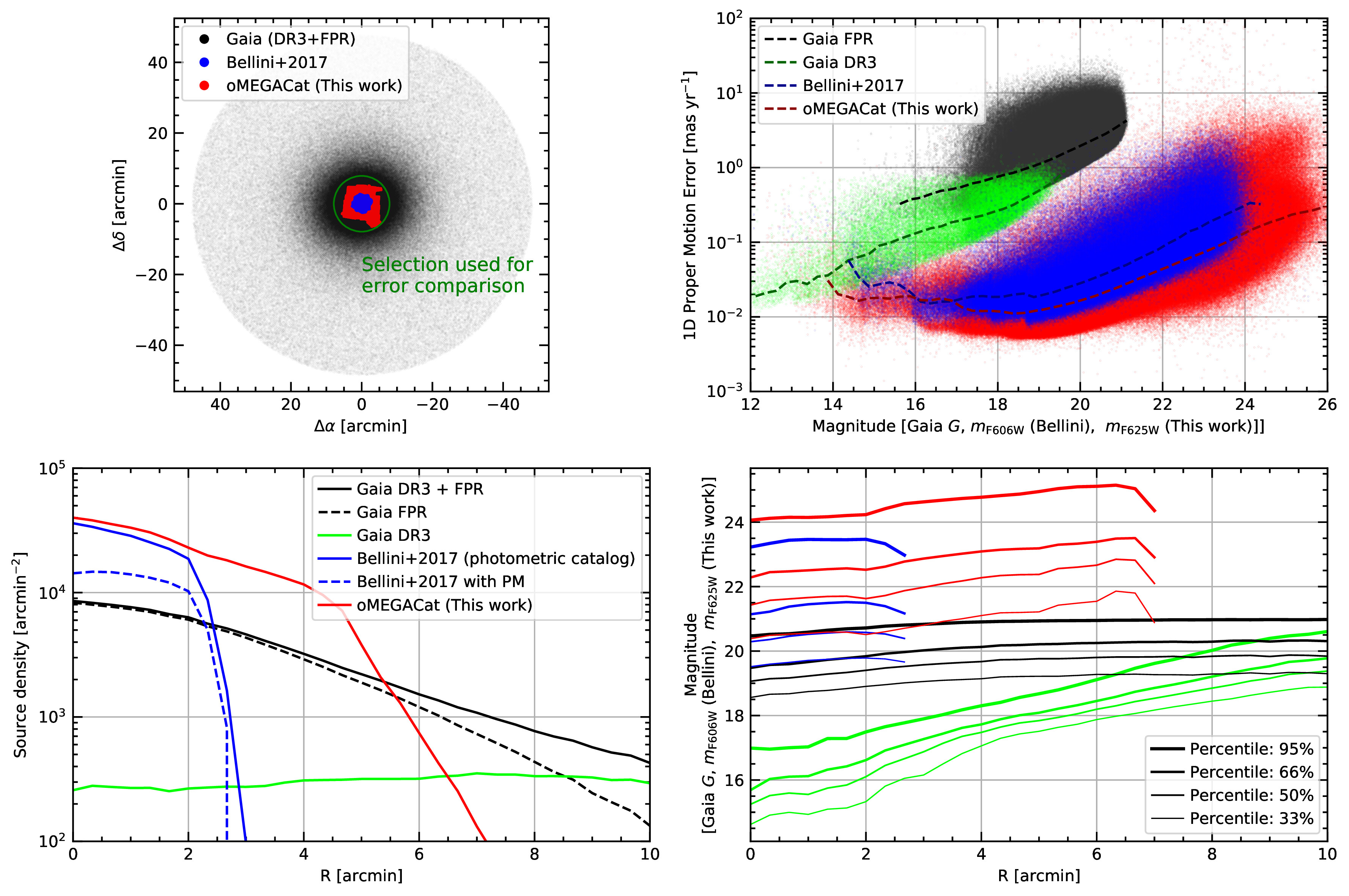}
  \caption{In this Figure we compare the general properties of three available high-precision datasets for the core of \omc{}: the new \textit{oMEGAcat} (this work), the astro-photometric catalog by \cite{2017ApJ...842....6B}, and the combined \textit{Gaia DR3 + FPR} catalog. \textit{Upper left:} Comparison of the on-sky footprint of the different catalogs. \textit{Upper right:} Comparison of proper-motion errors for different magnitudes. \textit{Lower left:} Comparison of the source density of the different catalogs at different radii. \textit{Lower right:} Comparison of different percentiles of the magnitude distribution for the different catalogs.}
  \label{fig:general_comparison_gaia_bellini}
\end{figure*}

In Figure~\ref{fig:general_comparison_gaia_bellini}, we compare various general properties of the three different datasets: the spatial coverage, the magnitude-dependent proper-motion errors, the source density as a function of radius, and the distribution of magnitudes. We can summarize our findings as follows:

The \cite{2017ApJ...842....6B} catalog is limited to a region with only half of the radius of the coverage of our catalog. In this inner region the source density of the photometric catalog is only slightly lower than in our new catalog which is expected, as the \cite{2017ApJ...842....6B} catalog is based on a similar dataset and the source detection was performed with the same software. However, if we restrict the comparison to stars where the \cite{2017ApJ...842....6B} catalog has proper motion measurements, the density in the literature drops by a factor of around two. This is also expected, as the astrometric part of the \cite{2017ApJ...842....6B} catalog is actually based on a previous data reduction \citep{2014ApJ...797..115B} with less sensitive photometry software and fewer available data. The proper motion errors of the two \textit{HST} catalogs show a similar dependence on the magnitude, however, the errors in our new catalog are typically lower by a factor of $\sim$2, due to the significantly longer temporal baseline of our catalog. In addition, our proper motion catalog reaches almost 2 magnitudes deeper.

For the comparison with the \textit{Gaia} catalog, we have to differentiate between the measurements published during the general (Early) Data Release 3 (\textit{DR3}), and the measurements published during the Focussed Product Release (\textit{FPR}) on \omc{}. Those two (disjunct) parts of the \textit{Gaia} catalog probe different regions on the sky and different magnitude regimes: While the \textit{Gaia DR3} data has all-sky coverage, the \textit{FPR} data is limited to a region of $r\leq\sim0.8^\circ$ around the cluster center. This is still significantly wider than the $r\sim7\arcmin$ region covered by our new catalog. The 1D proper motion errors of the \textit{Gaia DR3} reach a precision of $\sim20$\,µas\,yr$^{-1}$ for the brightest stars. At fainter magnitudes they are typically around one order of magnitude higher than the errors of our measurements. However, it is known that the nominal \textit{Gaia DR3} errors are underestimated in crowded fields \citep{2021MNRAS.505.5978V}. The two lower panels of Figure~\ref{fig:general_comparison_gaia_bellini} show that especially in the inner few arcminutes of \omc{} the completeness of \textit{Gaia DR3} is severely affected by crowding. Instead of an increase in the source density, the profile appears flat, and the magnitude distribution shifts towards brighter stars. This is expected: Due to the readout window strategy and its limited processing/downlink capabilities, the \textit{Gaia} satellite is only able to measure the brightest stars during its nominal operations.

The \textit{Gaia FPR} catalog on \omc{} is partially overcoming this crowding limitation, by using dedicated engineering images of the inner region of \omc{} that are processed on the ground. As it can be seen in the figure, the magnitude distribution and the source density profile of the combined catalog indicate a much better completeness towards the center. However, the new \textit{HST} catalog presented in this work still reaches around 3.5 magnitudes deeper, which leads to around 4 times more stars in the centermost region. At all magnitudes, the median proper-motion errors of the \textit{Gaia FPR} measurements are around a factor of $\sim$50 higher than the ones in our catalog.  

\subsection{Crossmatch and direct comparison of measurements}
We crossmatched our new catalog with both the \cite{2017ApJ...842....6B} and the \textit{Gaia} literature catalogs to directly compare the measurements. Details of this comparison are described in the Appendix~\ref{sec:appendixlit}. To summarize, both catalogs were almost fully included in our new catalogs. With a simple geometric match, we could recover more than 98\% of all literature sources, highlighting the completeness of our new catalog, but also the astrometric consistency with the previous works. There is good agreement between both the photometric and the astrometric parts of the \cite{2017ApJ...842....6B} catalog.

When comparing our newly measured proper motions with the \textit{Gaia} catalog, more than 99\% of the \textit{Gaia} sources within our field of view can be recovered in the \textit{HST} catalog. The differences between the measured positions in both catalogs can be explained with the absolute motion of \omc{}, but also the internal dispersion (see Figure \ref{fig:gaia_footprint}). For the further tests described in Appendix~\ref{sec:appendixlit}, we restricted ourselves to a subsample of sources well-measured in both catalogs. This restricts us to relatively bright stars (Figure \ref{fig:gaia_footprint}, middle) and also highlights the crowding limitations of the \textit{Gaia} catalog in the crowded cluster center (Figure \ref{fig:gaia_footprint}, left). When comparing the proper motions, one immediately notices the fundamental difference between our \textit{locally measured, relative} proper motions and the \textit{absolute} \textit{Gaia} measurements. The residuals of the comparison can be explained by bulk motion of \omc{} but also its rotation in the plane of the sky. This is used in the following section to measure the cluster's rotation.
 
\begin{figure*}
  \centering
    \includegraphics[width=1.0\textwidth]{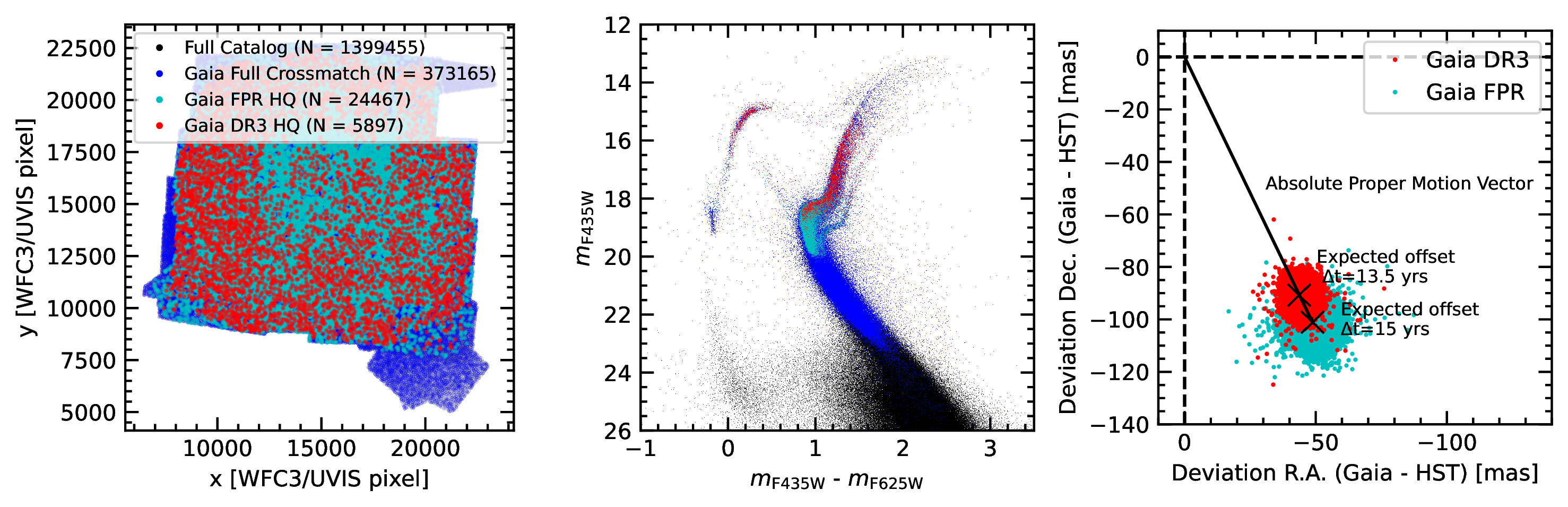}
  \caption{\textbf{Left panel:} Footprint of our proper-motion catalog and all \textit{Gaia} stars that could be crossmatched (blue). High-quality stars from \textit{Gaia FPR} are marked in cyan, and high-quality stars from \textit{Gaia DR3} are marked in red. \textbf{Middle panel:} The same sample of stars, but plotted as a color-magnitude diagram. One can see that all HQ \textit{Gaia DR3} measurements are limited to stars brighter than the main-sequence turn-off, while the \textit{Gaia FPR} sample reaches slightly deeper. \textbf{Right panel:} Absolute deviation between the positions from our \textit{HST} catalog and the two HQ \textit{Gaia} subsamples. The shift of the centroid can be explained by the absolute proper motion of \omc{}, while the spread of the distribution is caused by the displacement due to the velocity dispersion.}
  \label{fig:gaia_footprint}
\end{figure*}

\section{Measuring \omc{}'s Rotation curve and Inclination}
\label{sec:rotation}
\subsection{Measuring the rotation curve}
\label{subsec:rotation_sky}
Measurements of the plane-of-sky rotation of globular clusters face a fundamental challenge: The available \textit{HST}-based proper-motion catalogs (including this work) were all created with relative proper motions that were measured with a local approach that determines the stellar motions with respect to other neighboring cluster stars. This erases both the bulk motion and the rotation signature of the cluster from the proper motions. In principle, these quantities can be recovered by searching for extragalactic background sources that show the local bulk motion with a flipped sign (see \citealt{2003AJ....126..772A,2003AJ....126..247B,2017ApJ...844..167B} for the introduction of this method and \citealt{2018ApJ...854...45L} for its application to \omc). Also in the vector point diagram created with our new catalog, an overdensity of background sources at the inverse absolute proper motion value is visible. However, due to the low number of these background-objects and their typically faint magnitudes, it is difficult to study a varying velocity field such as that of rotation.

Absolute proper-motion catalogs such as the \textit{Gaia} catalogs do not suffer from these limitations and have been used to measure the rotations of many globular clusters (e.g. \citealt{2018MNRAS.481.2125B,2019MNRAS.485.1460S} for \textit{Gaia DR2}, \citealt{2021MNRAS.505.5978V} for \textit{Gaia EDR3}). However, in the crowded cluster centers only a few stars are measured, and additionally, the rotation signal is hidden by the velocity dispersion of individual stars.

By combining our new relative proper-motion catalog with the absolute measurements from \textit{Gaia DR3 \& FPR}, we can overcome the limitations of these past works: By calculating the difference between the absolute \textit{Gaia} proper motion and our relative proper motions, we obtain a direct measurement of the bulk-motion and any local proper motion trends. This works, as the \textit{Gaia} proper motions can be seen as a superposition of the bulk motion of the cluster, locally varying systematic motion such as rotation, and the random motion of individual stars, while our relative proper motions only contain the random motions relative to the bulk motion.

\begin{figure*}
  \centering
    \includegraphics[width=1.0\textwidth]{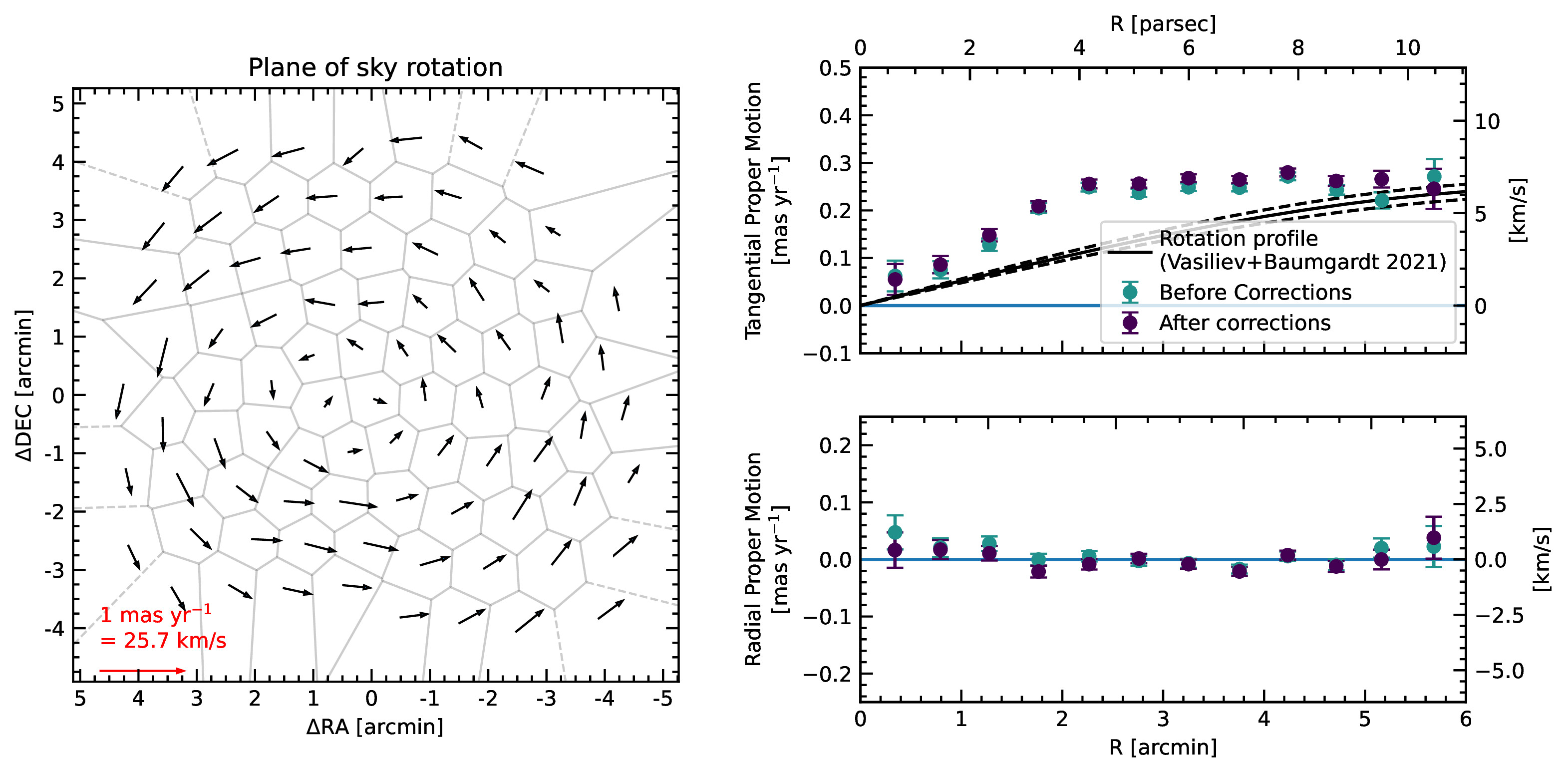}
  \caption{Plane of sky rotation determined in 2-dimensional Voronoi bins (\textit{left}) and as radial profiles (\textit{right}). The numerical values of the rotation profile are also listed in Appendix~\ref{sec:appendixrotationtable}; Table~\ref{tab:rotationprofile}. In comparison, the rotation profile from \cite{2021MNRAS.505.5978V} shows a significantly shallower increase in the rotation curve. However, it is based on an interpolation of the rotation data towards the center and there were no measurements available to constrain the rotation at smaller radii.}
  \label{fig:rotation}
  \end{figure*}

We study these differences for a subset of 30\,364 well-measured stars (see Appendix~\ref{sec:appendixlit}) both in 2-dimensional maps and radially. To this aim, we used the Voronoi binning code of \citet{2003MNRAS.342..345C} to create two-dimensional bins containing $\sim$250 sources each, and manually created radial bins of $0.5\arcmin$ width. The results, displayed in Figure~\ref{fig:rotation}, show a clear rotation pattern with a gradual increase of the rotation in the inner 2 arcminutes flattening out at an amplitude of $\sim$0.3~mas yr$^{-1}$. At an assumed cluster distance of 5.43~kpc \citep{2021MNRAS.505.5957B} this corresponds to a rotation of around 7~km s$^{-1}$, similar to what has been observed using MUSE line-of-sight velocity data \citep{2018MNRAS.473.5591K}. We provide the numerical values of the rotation profile in Appendix~\ref{sec:appendixrotationtable}; Table~\ref{tab:rotationprofile} and in a machine readable format. Other proper motion studies measure a similar rotation amplitude, but were typically limited to regions at larger distances from the center \citep{2000AA...360..472V,2002ASPC..265...41V,2006A&A...445..513V,2018MNRAS.481.2125B,2021MNRAS.505.5978V}. This also explains the difference between our measurements and the rotation profile derived in \cite{2021MNRAS.505.5978V}, see Figure \ref{fig:rotation}. In the innermost region ($r<3\arcmin$), we see a much steeper increase of the rotation profile than described by \cite{2021MNRAS.505.5978V}, however, this region was previously unconstrained due to the lack of data at small radii.

\subsection{Measurements of \omc{}'s inclination}

As demonstrated by \citet{2006A&A...445..513V}, the availability of both proper motions and line-of-sight velocities enables a direct and nearly model-independent way to measure the inclination $i$ of a stellar system, solely based on the assumption of axisymmetry. This is due to the following relation \citep[eq.~8 in][]{2006A&A...445..513V} between the mean proper motion along the system's projected semi-minor axis, $\langle\mu_{y'}\rangle$, and the mean line-of-sight velocity, $\langle v_{z'}\rangle$.
\begin{equation}
    \langle v_{z'}\rangle(x',\,y') = 4.74\,D\,\tan i\,\langle\mu_{y'}\rangle(x',\,y') 
\label{eq:inclination}
\end{equation}
Note that we follow \citet{2006A&A...445..513V} in that $x'$ and $y'$ denote the cluster-centric coordinates along the projected semi-major and semi-minor axes of $\omega$~Cen.

To investigate the three-dimensional rotation of $\omega$~Cen, we combined the proper motion sample described in Sec.~\ref{subsec:rotation_sky} with the MUSE catalog presented in \citet{2023ApJ...958....8N} and kept all stars that appear in all three data sets (i.e., HST, Gaia, and MUSE). We used the same Voronoi bins as shown in Fig.~\ref{fig:rotation} to measure mean proper motions and line-of-sight velocities across the face of the cluster.

\begin{figure*}
  \centering
    \includegraphics[width=1.0\textwidth]{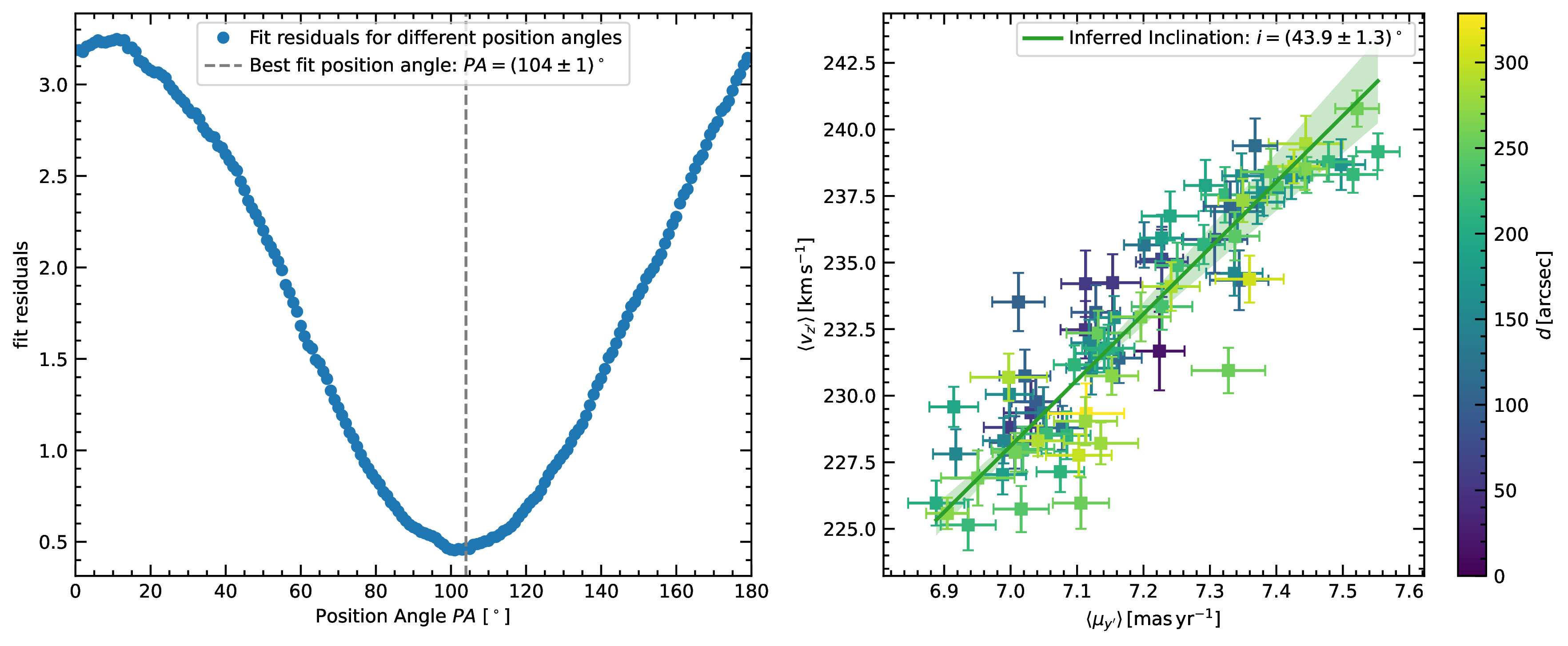}
  \caption{Determination of position angle and inclination. The left panel shows the scatter in the relation between mean 
 line-of-sight velocity and mean semi-minor-axis proper motion defined by eq.~\ref{eq:inclination}, for the different Voronoi bins and for a wide range of assumed position angles of $\omega$~Cen. For the position angle yielding the least amount of scatter, the right panel shows the aforementioned relation. The results obtained in the various Voronoi bins are color-coded by the distance of each bin to the cluster center.}
  \label{fig:pa_and_incl}
\end{figure*}

Inferring the inclination of $\omega$~Cen via eq.~\ref{eq:inclination} requires an assumption about the orientation of the cluster in the plane of the sky. \citet{2006A&A...445..513V} determined a position angle of the semi-major axis of $PA=100^\circ$ (measured north to east) by fitting elliptical isophotes to a DSS image of $\omega$~Cen. Here, we follow a different approach in that we determine $\langle v_{z'}\rangle$ and $\langle\mu_{y'}\rangle$ in every Voronoi bin for different assumed position angles and fitting a straight line to the relation between the two. Applying eq.~\ref{eq:inclination}, we adopt the position angle that minimizes the fit residuals between the two. The result of this exercise is shown in the left panel of
Fig.~\ref{fig:pa_and_incl}. The fit residuals show a well-defined minimum at a position angle close to the value of $PA=100^\circ$  obtained by \citet{2006A&A...445..513V}. By fitting a quadratic function to the fit residuals within $10^\circ$ of the minimum, we obtain $PA=104\pm1^\circ$. Note that we adopt our stepsize in position angle as the uncertainty, as the nominal uncertainty of the minimum of the quadratic fit is smaller.

Adopting $PA=104^{\circ}$, we show the relation between $\langle v_{z,}\rangle$ and $\langle\mu_{y'}\rangle$ in the right panel of Fig.~\ref{fig:pa_and_incl}. The values and uncertainties for each data point were determined via a maximum likelihood analysis, where each component of the velocity distribution per Voronoi bin was matched to a two-parameter Gaussian model (mean velocity and velocity dispersion) using the Markov-Chain Monte Carlo code \textsc{emcee} \citep{2013PASP..125..306F}. We observe a strong correlation between the two quantities, as expected based on eq.~\ref{eq:inclination}. Nevertheless, it is interesting to note that the individual data points show a larger scatter around the best-fitting linear relation than expected based on their uncertainties. To investigate if this trend is indicative of deviations from axisymmetry, we colour-code the data points by the distances of the corresponding Voronoi bins to the cluster center. However, there is no obvious trend that bins at specific distances show larger deviations.

The linear fit included in the right panel of Fig.~\ref{fig:pa_and_incl} corresponds to $D\tan i = 5.23\pm0.23~{\rm kpc}$. If we adopt again a distance of $D=(5.43\pm0.05)~{\rm kpc}$ \citep{2021MNRAS.505.5957B}, we obtain an inclination of $i=(43.9\pm1.3)^\circ$. This value is in good agreement with previous estimates of the inclination of $\omega$~Cen. \citet{2006A&A...445..513V} derived a value of $i=48^\circ\,(+9\, -7)^\circ$, while \citet{2019MNRAS.485.1460S} found $i=(39.2\pm4.4)^\circ$.

\section{Released data products and recommended use}
\label{sec:data_products}
We make our catalog public in the form of fits tables and Machine Readable ASCII files in a repository hosted by Zenodo ( \href{https://zenodo.org/doi/10.5281/zenodo.11104046}{doi:10.5281/zenodo.11104046}). In the following section, we describe the content of the different published files.
\subsection{Astrometric Catalog}
\label{subsec:data_products_astro}
We publish a single table that contains the astrometric data including precise positions, proper motions, and several diagnostic parameters. For a description of the different columns see Appendix~\ref{sec:appendixtablecolumns}; Table \ref{tab:cat_astrometry}. Our catalogs contains all sources which were recovered in at least two epochs (see Section \ref{subsec:crossmatch}). For a small fraction of these sources, no proper motion measurement was possible (either due to saturation or because the proper motion fit did not converge). For these sources we only report the measured position.
\subsubsection{Exemplary selection of a high-quality subset}
Depending on the specific science case one has to restrict the dataset to obtain a subset with the necessary precision and reliability. 
In the following we discuss how a quality selection combining several diagnostics can be assembled. This exemplary selection is also demonstrated in our example use case notebook and the resulting flag is published with the catalog. The vector-point diagram shown in Figure~\ref{fig:vpd} is also using these selctions, demonstrating their effectiveness.

Typically one would start with requiring the formal proper motion error on both components to be below a desired limit (e.g. $\sigma\mu_\delta<0.2$\,mas\,yr$^{-1}$). In addition, it is recommended to use the reduced chi square of the linear fit of both proper motions to limit the sample to stars with a well-behaved measurement, here we adopt a limit of $\chi^2_{red}<5$.

For our high-quality selection we additionally require a baseline of at least 10~years to obtain a more uniform dataset. The baseline cut is met throughout most of the field of view, however some of the outer regions with partial overlap are lost (see also Fig. \ref{fig:pm_baseline}). Finally, to reject stars where a lot of measurements were clipped, we require a fraction of used measurements $N_{used}/N_{found}>0.8$.
The combination of these criteria is met for 1\,024\,768 of 1\,395\,781 stars with a proper motion measurement.
\subsubsection{Recommended use of local corrections}
The local astrometric corrections (see Section \ref{subsec:corrections}) use the proper motions of neighbouring stars with similar magnitudes to correct for residual systematic effects. The columns \texttt{pmra\_corrected} and \texttt{pmdec\_corrected} contain the proper motions resulting after the local corrections are applied (with the corresponding errors in \texttt{pmra\_corrected\_err} and \texttt{pmdec\_corrected\_err}). We recommend using these corrections for studies including faint stars in the outer fields (where systematic trends caused by CTE are strongest, see Figure~\ref{fig:pm_corrections}). In the centermost region, where systematic spatial trends are less of an issue due to the higher number of pointings and rotation angles, it is a trade-off between larger statistical errors, due to the error on the correction, and residual spatial trends.
\subsubsection{Crossmatches with other catalogs}
To facilitate future investigations, we include the results of a crossmatch with two literature catalogs in our astrometric table:

The column \texttt{Nitschai\_ID} contains the ID (Column \texttt{MUSE} in the MUSE catalog) of stars that have been matched with \textit{oMEGACat} I MUSE catalog \citep{2023ApJ...958....8N}. For the MUSE cross-match we used a matching radius of 40~mas and also required that sources were measured in both the F435W and F625W filter. This allowed us to apply an additional photometric criterion, as the MUSE catalog also contains the photometry from the \cite{2010ApJ...710.1032A} catalog. We required that there were no significant deviations in the photometry between the two catalogs. Finally, if multiple sources from the \textit{HST} catalog lay within matching radius of a MUSE source, we only kept the closest crossmatch. This leads to successful crossmatch for 307\,030 of the 342\,797 stars in the MUSE catalog.

The column \texttt{gaia\_id} contains the gaia source ID of stars from a crossmatch with both \textit{Gaia FPR \& DR3} (the column \texttt{gaia\_origin} specifies the data release). Due to the larger astrometric errors in the \textit{Gaia} catalogs, we used a matching radius of 160~mas. We did not apply a photometric cut, however this can be used to further refine the selection. Again we only kept the closest cross-match in the case of multiple sources within the matching radius. In total 373\,291 stars match these criteria. To facilitate further comparisons between our dataset and the Gaia measurements we also include several key quantities from the crossmatched \textit{Gaia} catalog in our release data products, including the source positions measured with \textit{Gaia}, the absolute \textit{Gaia} proper motion and the \textit{Gaia G} photometry. The \textit{Gaia} proper motions can be also used to substitute the missing proper motions for stars too bright for \textit{HST} measurements. In that case, the different definitions of the proper motions (oMEGACat: relative; \textit{Gaia}: absolute) have to be taken into account.
In addition, we add the flag \texttt{gaia\_hq\_subset} for the high quality subset that has been used for the rotation curve determination. 

\subsection{Photometric Catalog}
\label{subsec:data_products_phot}
\subsubsection{Recommended usage of corrections and errors}
We publish a table with the photometric information for each of the 3 ACS/WFC filters (F435W, F625W, F658N) and the 4 WFC3/UVIS filters (F275W, F336W, F606W, F814W). For a description of the different columns see Appendix~\ref{sec:appendixtablecolumns}; Table~\ref{tab:cat_photometry}. In general, we recommend the use of our empirical photometric corrections (Section \ref{subsec:phot_corr_method}), although this slighlty reduces the coverage. 

Just like for the astrometric catalog, in the following section we explain how to select a sample of well-measured stars, with the caveat that each science case might have different requirements for these selections.

For most photometric use-cases we recommend to use the weighted mean of method 1 (see Section \ref{subsec:ks2}) photometry (\texttt{m1\_weighted\_mean}). The corresponding weighted mean error is saved as column \texttt{m1\_weighted\_mean\_error}. This error may be underestimated in cases of crowding, therefore, we recommend scaling it with the square-root of the \texttt{chi2\_red} whenever $\chi^{2}_{red.} > 1.0$.
When applying the empirical photometric corrections determined in Section \ref{subsec:phot_corr_method} (which we also recommend for most cases), one additionally has to add the error on the correction (Table~\ref{tab:redd_coeff}) in quadrature. For the convenience of the user, we provide the corrected photometry and the combined error in the first two columns \texttt{corrected\_mag} and \texttt{corrected\_mag\_error} in the published data products.  We also note here that the absolute zeropoints have reported uncertainties of $\sim$1\% (ACS, \citealt{2016AJ....152...60B}) and 2-3\% (UVIS, \citealt{2022AJ....164...32C}), which corresponds to absolute uncertainties at the 0.02-0.03 mag level. This absolute uncertainty does not affect the internal consistency of our catalog (which is ensured by the corrections), but has to be taken into account when comparing the data e.g. with isochrone models. 
\subsubsection{Caveats about different magnitude regimes}
We remind the user, that bright saturated stars were not measured with \texttt{KS2}, instead we substituted their \texttt{hst1pass} measurement. We mark all stars for which this was the case with the \texttt{brightlist\_flag}.
We also caution the user that brighter stars often only have one short exposure measurement in some of our filters (ACS/WFC F435W, F625W), therefore all photometric selections that require more than one measurement can reduce the coverage and completeness for these otherwise well-measured bright stars.

For faint main-sequence stars ($m_{\rm F275W}>22$, $m_{\rm F336W}>21$), uncorrected charge-transfer efficiency effects introduce systematic spatial variations in the Filters WFC3/UVIS F275W and F336W. This mostly affects the outer pointings which were taken recently with the aging detectors, while the center is less affected.

Finally we remind the user, that the photometric corrections were derived using reference stars in the magnitude region ($15.0 < m_{F814W} < 19.0$), see also Section \ref{subsec:corrections}. Magnitude independent effects are corrected nevertheless, but the corrections are most effective in this region.

\subsubsection{Quality criteria }
Several quality criteria measure how well the PSF describes the measured flux of the source. This includes the QFIT parameter (the linear correlation coefficient between the PSF and the measured source flux), the RADXS parameter (a measure whether a source is more extended or sharper than expected from the PSF \citealt{2008ApJ...678.1279B}); and the o value (the ratio between the flux of source and of neighboring stars). These parameters are determined in each individual exposure, when combining measurements with the magnitude-error weighted mean (see Section \ref{sec:photometry}), we also calculate a mean of these quality parameter, which is the value we report in the catalogs.

For stars with fainter magnitude, the \texttt{QFIT} parameter worsens due to their lower signal-to-noise. Therefore, it is recommended to use magnitude dependent thresholds.

\subsubsection{Exemplary photometric correction} Our exemplary selection rejects stars below the 10th\texttt{ QFIT} percentile 0.5 mag wide bins. Stars with \texttt{QFIT}$>0.98$ are always kept, while stars with \texttt{QFIT}$<0.4$ are always rejected.
The only other criterion we apply is \texttt{o}$<0.5$ i.e. the stars flux within the fit aperture is at least twice as high as the flux of neighboring sources. The resulting selections is published in column \texttt{phot\_hq\_flag} in the photometric tables.

Table~\ref{tab:phot_counts} lists the number of stars that match the combined photometric criteria in the different filters.

\begin{centering}
\begin{table}[]
\caption{Number of stars in each filter different that are saturated ($N_{\rm Sat.}$) or are matching our exemplary quality criterion ($N_{\rm HQ}$) compared to the total number of measurements available in that filter $N_{\rm Total}$  }
\label{tab:phot_counts}
\scriptsize
\begin{tabular}{lllll}
\hline
Instrument & Filter & $N_{\rm Sat.}$ & $N_{\rm HQ}$ & $N_{\rm Total}$\\ \hline
WFC3/UVIS & F275W &     697 &  599477 &  825061 \\
WFC3/UVIS & F336W &     691 &  761759 & 1105255 \\
ACS/WFC & F435W &    1383 &  883083 & 1355786 \\
WFC3/UVIS & F606W &    2088 &  435618 &  622052 \\
ACS/WFC & F625W &    3944 &  990139 & 1395979 \\
ACS/WFC & F658N &    4418 &  953034 & 1387347 \\
WFC3/UVIS & F814W &    2090 & 1045201 & 1335929 \\ \hline
\end{tabular}
\end{table}
\end{centering}

\subsection{Stacked images}
Along our astro-photometric catalog, we publish stacked images for the 7 filters, for which we release photometric information.
The stacked images are normalized to the typical exposure time for the respective filter (see Table \ref{tab:exp_times}) and combine images from all epochs. 
Note that the exact flux distribution of sources in the individual images is not preserved in the stacked images, and therefore their main use should be as a high-quality representation of the scene rather than for PSF fitting. The images contain precise WCS information in their header and are also compatible with the pixel coordinates in our astrometric catalog (apart from a shift of $[5000, 5000]$ pixels to allow for a smaller image size). An RGB image based on the filters WFC3/UVIS F275W, F336W and F814W can be found in Figure~\ref{fig:rgb_stack}.

\begin{table}[]
\caption{Exposure times to which the published stacked images are normalized.}
\label{tab:exp_times}
\scriptsize
\begin{tabular}{lll}
\hline
Instrument & Filter & Exposure time \\ \hline
ACS/WFC    & F435W  & 340\,s        \\
ACS/WFC    & F625W  & 340\,s        \\
ACS/WFC    & F658N  & 440\,s        \\ \hline
WFC3/UVIS  & F275W  & 773\,s        \\
WFC3/UVIS  & F336W  & 475\,s        \\
WFC3/UVIS  & F606W  & 40\,s         \\
WFC3/UVIS  & F814W  & 250\,s        \\ \hline
\end{tabular}
\end{table}

\subsection{Public examples on catalog usage}
Together with the data products we publish an IPython notebook \footnote{Again this notebook is made publicly accessible in a Zenodo repository (\href{https://zenodo.org/doi/10.5281/zenodo.11104046}{doi:10.5281/zenodo.11104046})} that can be used as starting point for the usage of our catalog. The notebook includes:
\begin{itemize}
    \item Selection of high-quality astrometric measurements and plot of a vector-point diagram
    \item Comparison of \textit{Gaia} and \textit{HST} proper motions
    \item Selection of high-quality photometric measurements and plot of several CMDs
    \item An exemplary calculation on how to propagate the stellar motions from the new catalog to any given epoch while properly accounting for the absolute motion of the cluster and the relative motion of the individual stars
    \item Plots of the stacked images overlaid with data from the catalog 
\end{itemize}
\section{Conclusions}
\label{sec:conclusions}
In this second paper of the \textit{oMEGACat} series we describe the creation of a deep, \textit{HST} based astrometric and photometric catalog covering the cluster \omc{} out to its half-light radius. The full catalog is made public along with this publication.

The catalog contains high-precision proper-motion measurements for around 1.4 million stars, more than any other space- or ground-based catalog of \omc.
For bright stars ($m_{\rm F625W}\approx 18$) we reach a median 1D proper-motion error of 0.011~mas yr$^{-1}$. In the well-covered inner region, this median error decreases down to 0.007~mas yr$^{-1}$, corresponding to a velocity of only 0.15\,km\,s$^{-1}$ at the distance of \omc{}.
We corrected our proper motions from residual systematic effects using an approach that measures the net-motion of neighboring cluster stars.

Our catalog also contains photometry in 6 filter bands (WFC3/UVIS: F275W, F336W, F625W; ACS/WFC: F435W, F625W, F658N) for the full field and an additional filter (F606W) with especially good coverage in the centermost region. This filter set allows the separation of the various, complex stellar subpopulations hosted by \omc{}. 

We compare our catalog with the available literature catalogs (\citealp{2017ApJ...842....6B}; \textit{Gaia DR3} \citealp{2021AA...649A...1G,2021A&A...649A...2L}; \textit{Gaia FPR:} \citealp{2023AA...680A..35G}) and can confirm a generally good agreement, with our catalog having a significantly higher proper motion precision and reaching fainter magnitudes than all the previous works.

Our catalog is complementary to the recently published, large spectroscopic catalog \citep{2023ApJ...958....8N}, covering the same region on the sky and containing line-of-sight velocity and metallicity measurements for more than 300\,000 stars.

As a first science result, we determined the plane-of-sky rotation curve of \omc{} with unprecedented resolution using a combination of our relative proper motions and the absolute proper motions from \textit{Gaia}. In addition, we obtain a precise measurement of \omc{}'s inclination of $i=(43.9\pm1.3)^\circ$.

The combined \textit{oMEGACat} catalogs are already enabling a broad range of interesting science. Ongoing projects are the study of the Age-Metallicity relation of \omc{} (Clontz et al. in prep.), the automated separation of subpopulations based on photometry and metallicity (Clontz et al. in prep), the discovery of fast-moving stars indicative of an intermediate-mass black hole \citep{2024Natur.631..285H}, the search for spatial differences in the metallicity distribution \citep{2024ApJ...970..152N} and the extraction of individual abundances using stacked spectra (Di Stefano et al. in prep.). The MUSE data of the centermost region is well matched to the depth of our proper motion catalog and has revealed a counter-rotating core in the centermost \citep{2024MNRAS.528.4941P} region. We plan to use the combined dataset to create a dynamical model of this region (Pechetti et al.) and eventually the whole cluster.

\begin{acknowledgments}
A.B. acknowledges support from STScI grant GO-15857.
A.B. C.C.,M.A.C., and A.C.S. acknowledge support from STScI grant GO-16777. Based on archival and new observations with the NASA/ESA Hubble Space Telescope, obtained at the Space Telescope Science Institute, which is operated by AURA, Inc., under NASA contract NAS 5-26555. 
SK acknowledges funding from UKRI in the form of a Future Leaders Fellowship (grant no. MR/T022868/1). 
A.F.K. acknowledges funding from the Austrian Science Fund (FWF) [grant DOI 10.55776/ESP542]. 
This work has made use of data from the European Space Agency (ESA) mission
{\it Gaia} (\url{https://www.cosmos.esa.int/gaia}), processed by the {\it Gaia}
Data Processing and Analysis Consortium (DPAC,
\url{https://www.cosmos.esa.int/web/gaia/dpac/consortium}). Funding for the DPAC
has been provided by national institutions, in particular, the institutions
participating in the {\it Gaia} Multilateral Agreement.
\end{acknowledgments}

%

\vspace{5mm}
\facilities{}
Gaia, HST

\software{}
astropy \citep{2022ApJ...935..167A}, matplotlib \citep{2007CSE.....9...90H}, numpy \citep{2020Natur.585..357H}, scipy \citep{2020NatMe..17..261V}, IPhyon \citep{2007CSE.....9c..21P}, hst1pass \citep{2022wfc..rept....5A}



\clearpage
\appendix

\section{Dataset}
\label{sec:appendixdataset}
Tables \ref{tab:uvisdata} and \ref{tab:acsdata} show show information on all individual exposures used for the creation of our catalog. A compilation of the full
data set is also archived at doi:\dataset[10.17909/26qj-g090]{http://dx.doi.org/10.17909/26qj-g090}.
\begin{table*}[h]
\caption{List of all \textit{HST} WFC3/UVIS observations used for our astrophotometric measurements}
\label{tab:uvisdata}
\tiny
\begin{tabular}{llllll}
\hline
GO    & PI        & Filter & N x Exp. Time       & Min. - Max. Epoch  & Field                                   \\ 
   &        &  &      &  (year$ - 2000$) &                                    \\ \hline
11452 & J. Kim Quijano & F275W &  1 $\times$   35;  9 $\times$  350;  & 9.53740 - 9.53832 & Center \\
 &  & F336W &  1 $\times$   35;  9 $\times$  350;  &  &  \\
 &  & F438W &  1 $\times$   35;  &  &  \\
 &  & F606W &  1 $\times$   35;  &  &  \\
 &  & F814W &  1 $\times$   35;  &  &  \\ \hline
11911 & E. Sabbi & F275W & 22 $\times$  800;  & 10.03462 - 10.50779 & Center \\
 &  & F336W & 19 $\times$  350;  &  &  \\
 &  & F390W & 15 $\times$  350;  &  &  \\
 &  & F438W & 25 $\times$  350;  &  &  \\
 &  & F555W & 18 $\times$   40;  &  &  \\
 &  & F606W & 27 $\times$   40;  &  &  \\
 &  & F775W & 16 $\times$  350;  &  &  \\
 &  & F814W & 27 $\times$   40;  &  &  \\ \hline
12094 & L. Petro & F606W &  9 $\times$   40;  & 10.31645 - 10.31699 & Center \\ \hline
12339 & E. Sabbi & F275W &  9 $\times$  800;  & 11.12532 - 11.22895 & Center \\
 &  & F336W &  9 $\times$  350;  &  &  \\
 &  & F438W &  9 $\times$  350;  &  &  \\
 &  & F555W &  9 $\times$   40;  &  &  \\
 &  & F606W &  9 $\times$   40;  &  &  \\
 &  & F814W &  9 $\times$   40;  &  &  \\ \hline
12353 & V. Kozhurina-Platais & F606W & 13 $\times$   40;  & 10.95038 - 11.56642 & Center \\ \hline
12580 & A. Renzini & F275W &  2 $\times$  909;  2 $\times$  914;  2 $\times$ 1028;  2 $\times$ 1030;  2 $\times$ 1267;  & 12.18821 - 12.32857 & Southwest \\
 &  & F336W &  2 $\times$  562;  2 $\times$  565;  1 $\times$  945;  1 $\times$  953;  &  &  \\
 &  & F438W &  4 $\times$  200;  2 $\times$  210;  &  &  \\ \hline
12694 & K. Long & F606W &  2 $\times$  350;  & 12.15906 - 12.32275 & Center\\ \hline
12700 & A. Riess & F775W &  2 $\times$  450;  & 12.48594 - 12.48620 & Center \\ \hline
12714 & V. Kozhurina-Platais & F606W &  4 $\times$   40;  & 12.18563 - 12.18574 & Center \\ \hline
12802 & J. MacKenty & F336W & 29 $\times$   10;  8 $\times$  700;  & 12.56737 - 12.56800 & Center \\ \hline
13100 & V. Kozhurina-Platais & F606W &  3 $\times$   40;  9 $\times$   48;  & 12.95411 - 13.22892 & Center \\ \hline
13570 & V. Kozhurina-Platais & F606W &  9 $\times$   40;  & 13.95233 - 14.68451 & Center \\ \hline
14031 & V. Kozhurina-Platais & F606W & 19 $\times$   40;  5 $\times$   60;  1 $\times$  120;  & 15.02381 - 15.46965 & Center \\ \hline
14393 & V. Kozhurina-Platais & F606W & 19 $\times$   40;  3 $\times$   60;  & 15.94769 - 16.48538 & Center \\ \hline
14550 & V. Kozhurina-Platais & F606W &  9 $\times$   60;  & 17.08296 - 17.46694 & Center \\ \hline
14759 & T. Brown & F275W &  3 $\times$  765;  3 $\times$  850;  & 16.94372 - 17.28252 & Southeast \& Southwest \\
 &  & F336W &  3 $\times$  630;  3 $\times$  765;  &  &  \\
 &  & F438W &  3 $\times$  630;  3 $\times$ 1025;  &  &  \\ \hline
15000 & V. Kozhurina-Platais & F606W &  9 $\times$   60;  & 18.00372 - 18.51208 & Center \\ \hline
15593 & V. Kozhurina-Platais & F606W &  9 $\times$   60;  & 19.08460 - 19.54426 & Center \\ \hline
15594 & V. Kozhurina-Platais & F438W &  2 $\times$   50;  6 $\times$  697;  & 19.15842 - 19.65534 & Center \\
 &  & F606W &  2 $\times$   50;  6 $\times$  697;  &  &  \\
 &  & F814W &  3 $\times$   50;  9 $\times$  697;  &  &  \\ \hline
15733 & V. Kozhurina-Platais & F606W &  6 $\times$   60;  & 20.08423 - 20.16737 & Center \\ \hline
15857 & A. Bellini & F275W &  1 $\times$  710;  1 $\times$  730;  & 21.15030 - 21.15038 & Southwest \\
 &  & F336W &  1 $\times$  497;  1 $\times$  520;  &  &  \\ \hline
16117 & M. Reinhart & F606W &  4 $\times$   15;  4 $\times$  400;  & 20.45939 - 20.45956 & Center \\ \hline
16413 & V. Kozhurina-Platais & F606W & 12 $\times$   60;  & 21.14014 - 21.58543 & Center \\ \hline
16441 & J. Anderson & F606W &  8 $\times$    4;  4 $\times$  800;  & 21.00757 - 21.00781 & Center \\ \hline
16588 & V. Kozhurina-Platais & F606W &  9 $\times$   60;  & 22.03211 - 22.48428 & Center \\ \hline
16777 & A. Seth & F275W & 10 $\times$  700; 20 $\times$  773;  & 22.62801 - 23.09828 & Ring reaching r$_{\rm HL}$\\
 &  & F336W & 20 $\times$   40; 30 $\times$  475;  &  &  Excluding Center\\
 &  & F814W & 20 $\times$   15; 20 $\times$  250;  &  &  \\ \hline
17023 & C. Martlin & F606W &  3 $\times$   60;  & 23.04349 - 23.04357 & Center \\ \hline

\end{tabular}
\end{table*}

\begin{table*}[h]
\caption{List of all \textit{HST} ACS/WFC observations used for our astrophotometric measurements}
\label{tab:acsdata}
\tiny
\begin{tabular}{llllll}
\hline
GO    & PI        & Filter & N x Exp. Time       & Min. - Max. Epoch  & Field                                   \\ 
   &        &  &      &  (year$ - 2000$) &                                   \\ \hline
9442 & A. Cool & F435W &  9 $\times$   12; 27 $\times$  340;  & 2002.48916 - 2002.49745 & 3x3 grid \\
 &  & F625W &  9 $\times$    8; 27 $\times$  340;  &  &   covering 10$\arcmin$x10$\arcmin$\\
 &  & F658N & 36 $\times$  440;  &  &  \\ \hline
10252 & J. Anderson & F606W &  1 $\times$   15;  5 $\times$  340;  & 2004.94612 - 2004.94636 & South East \\
 &  & F814W &  1 $\times$   15;  5 $\times$  340;  &  &  \\ \hline
10775 & A. Sarajedini & F606W &  2 $\times$    4;  8 $\times$   80;  & 2006.16055 - 2006.55718 & Center \\
 &  & F814W &  2 $\times$    4;  8 $\times$   90;  &  &  \\ \hline
12193 & J. Lee & F606W &  1 $\times$  200;  1 $\times$  500;  & 2011.53248 - 2011.53254 & North West \\
 &  & F814W &  1 $\times$  400;  &  &  \\ \hline
13066 & L. Smith & F435W &  9 $\times$    6;  9 $\times$  339;  & 2012.63115 - 2012.63150 & Center \\
 &  & F606W &  1 $\times$  339;  &  &  \\ \hline
13606 & J. Anderson & F435W &  4 $\times$  339;  & 2013.95179 - 2013.95223 & Center \\
 &  & F606W &  4 $\times$   80;  &  &  \\
 &  & F814W &  4 $\times$   90;  &  &  \\ \hline
15594 & V. Kozhurina-Platais & F435W &  2 $\times$   42;  6 $\times$  647;  & 2019.15842 - 2019.65534 & Center \\
 &  & F606W &  2 $\times$   42;  6 $\times$  656;  &  &  \\
 &  & F814W &  2 $\times$   42;  6 $\times$  656;  &  &  \\ \hline
15764 & N. Hathi & F435W &  2 $\times$  339;  & 2020.10328 - 2020.54298 & Center \\
 &  & F475W &  2 $\times$  339;  &  &  \\
 &  & F555W &  1 $\times$  339;  &  &  \\
 &  & F606W &  2 $\times$  339;  &  &  \\
 &  & F625W &  1 $\times$  339;  &  &  \\
 &  & F658N &  1 $\times$  339;  &  &  \\
 &  & F775W &  2 $\times$  339;  &  &  \\
 &  & F814W &  2 $\times$  339;  &  &  \\ \hline
15857 & A. Bellini & F606W &  2 $\times$  417;  2 $\times$  668;  1 $\times$  671;  2 $\times$  700;  3 $\times$  757;  & 2020.70743 - 2021.15038 & South West \\
 &  & F814W &  3 $\times$  337;  3 $\times$  379;  &  &  \\ \hline
16380 & M. Chiaberge & F606W &  6 $\times$   40;  6 $\times$  150;  6 $\times$  400;  & 2021.54871 - 2021.55220 & Center \\ \hline
16384 & Y. Cohen & F435W &  2 $\times$  337;  & 2021.14867 - 2021.65577 & Center \\
 &  & F475W &  2 $\times$  337;  &  &  \\
 &  & F555W &  2 $\times$  337;  &  &  \\
 &  & F606W &  3 $\times$  337;  &  &  \\
 &  & F625W &  2 $\times$  337;  &  &  \\
 &  & F658N &  1 $\times$  350;  &  &  \\
 &  & F775W &  3 $\times$  337;  &  &  \\
 &  & F814W &  2 $\times$  337;  &  &  \\ \hline
16520 & N. Hathi & F435W &  2 $\times$  337;  & 2022.19657 - 2022.61983 & Center \\
 &  & F475W &  2 $\times$  337;  &  &  \\
 &  & F555W &  1 $\times$  337;  &  &  \\
 &  & F606W &  2 $\times$  337;  &  &  \\
 &  & F625W &  1 $\times$  337;  &  &  \\
 &  & F658N &  1 $\times$  350;  &  &  \\
 &  & F775W &  2 $\times$  337;  &  &  \\
 &  & F814W &  2 $\times$  337;  &  &  \\ \hline
16968 & N. Hathi & F435W &  1 $\times$  337;  & 2023.08588 - 2023.08596 & Center \\
 &  & F475W &  1 $\times$  337;  &  &  \\
 &  & F606W &  1 $\times$  337;  &  &  \\
 &  & F775W &  1 $\times$  337;  &  &  \\
 &  & F814W &  1 $\times$  337;  &  &  \\ \hline
\end{tabular}
\end{table*}

\clearpage
\section{Photometric Corrections}
\label{sec:appendixphotcorr}
\begin{figure*}[h]
  \centering
    \includegraphics[width=1.0\textwidth]{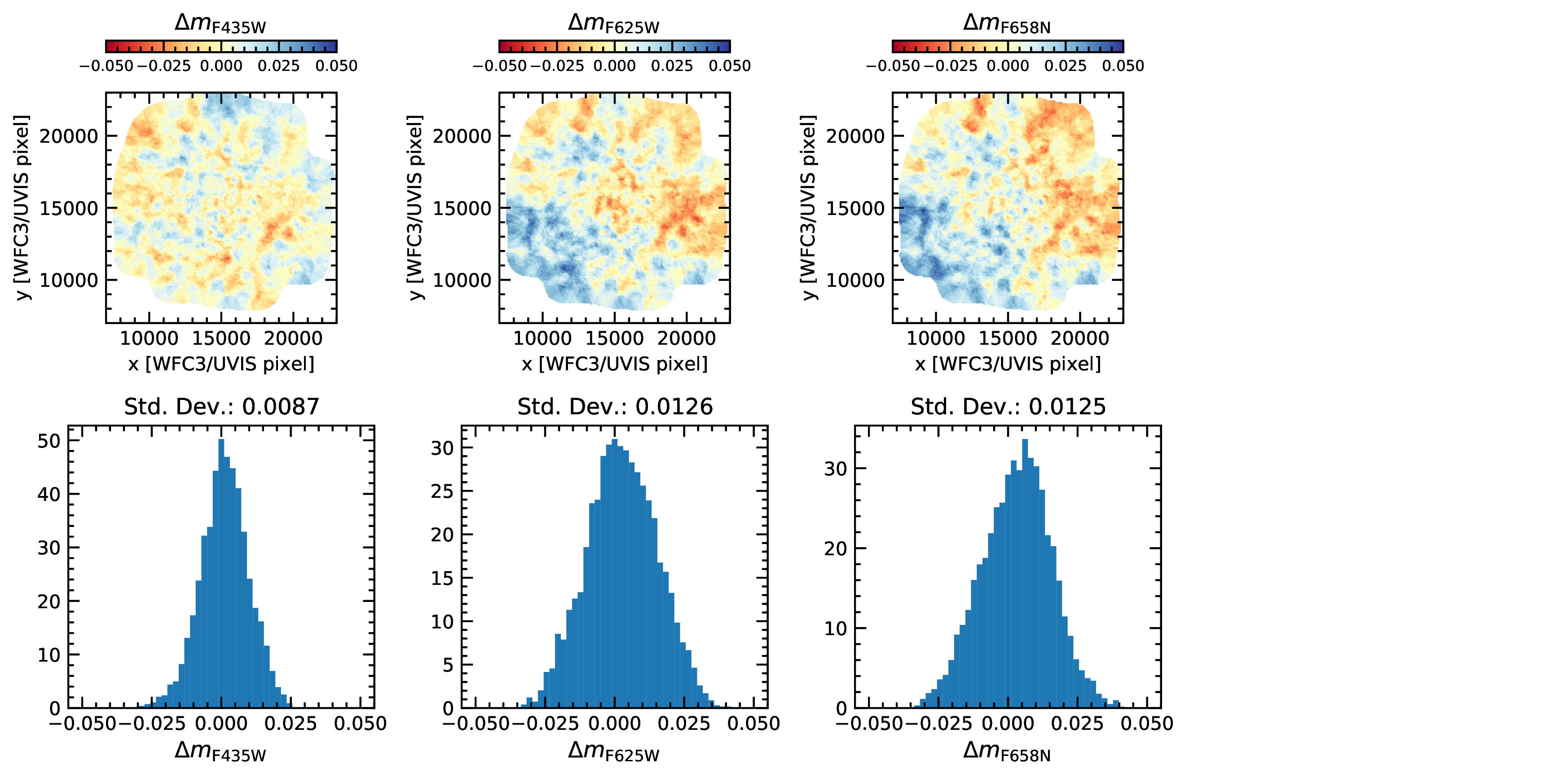}
  \caption{The upper row in this figure shows maps of the empirical photometric corrections (see Section~\ref{subsec:corrections}) for each of the 3 ACS/WFC filters in our dataset. They are a combination of physical differential reddening and instrumental/calibration effects. The lower panel shows histograms of the distribution of correction values. }
  \label{fig:corrections_acs}
\end{figure*}

\begin{figure*}[h]
  \centering
    \includegraphics[width=1.0\textwidth]{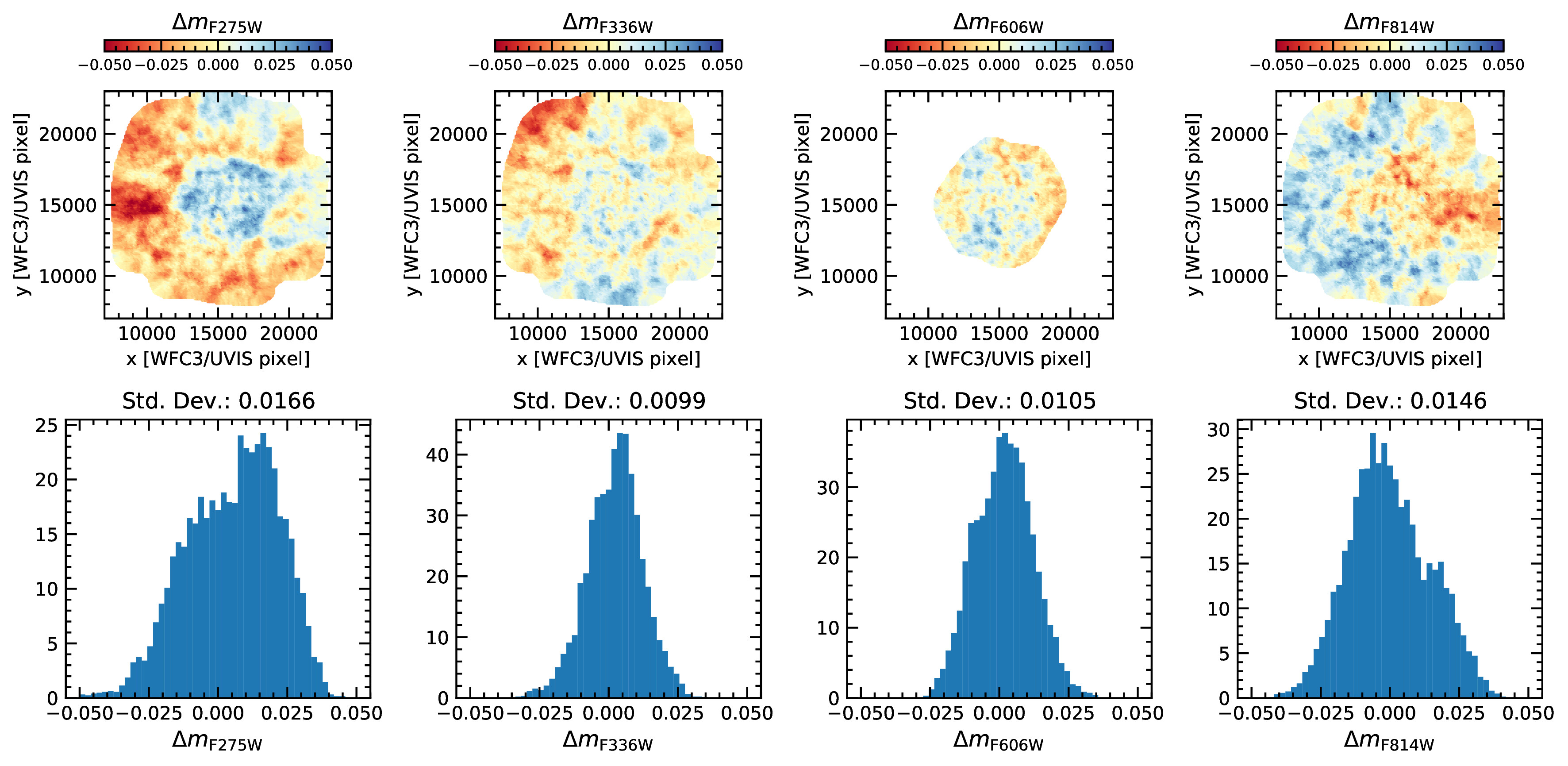}
  \caption{The upper row in this figure shows maps of the empirical photometric corrections (see Section~\ref{subsec:corrections}) for each of the 4 WFC3/UVIS filters in our dataset. They are a combination of physical differential reddening and instrumental/calibration effects. The lower panel shows histograms of the distribution of correction values. }
  \label{fig:corrections_uvis}
\end{figure*}

Figures \ref{fig:corrections_acs} and \ref{fig:corrections_uvis} show maps of the spatially variable photometric corrections derived in Section \ref{subsec:phot_corr}. These corrections are a superposition of differential reddening, which has a physical origin, and instrumental effects and zero-point variations.

\clearpage
\section{Catalog validation: Search for systematic effects in magnitude and color}
\label{sec:appendixtrends}
By construction, the a-posteriori corrections described in Section~\ref{subsec:corrections} identify any systematic spatial trends within certain magnitude ranges and remove them. Figure~\ref{fig:pm_corrections} shows that these trends were strongest for faint magnitudes and that the a-posteriori corrections could remove them efficiently.

\begin{figure*}[h]
  \centering
    \includegraphics[width=1.0\textwidth]{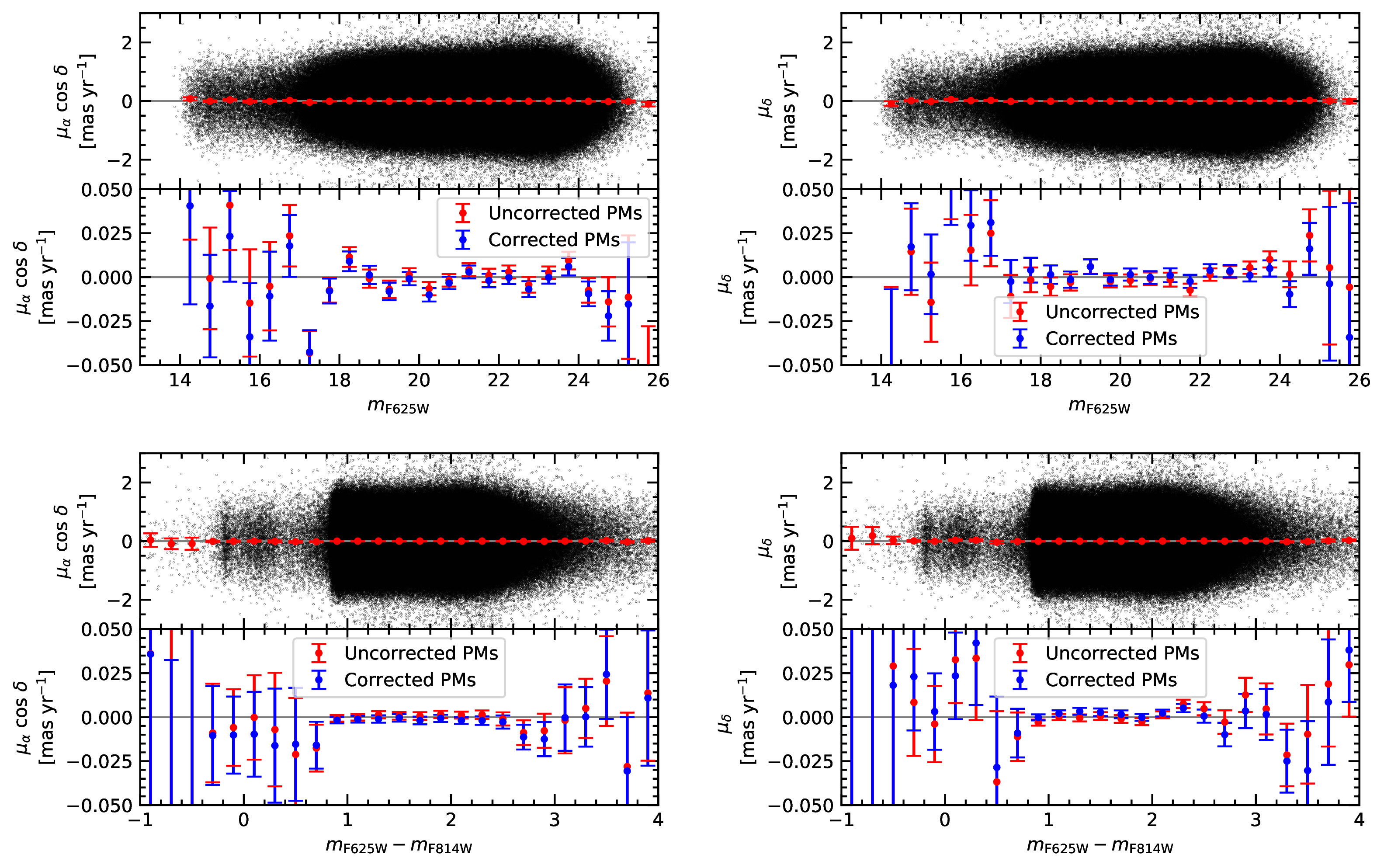}
  \caption{Illustration of our search for global systematic effects as a function of magnitude and color. The first row shows the proper-motion measurements plotted against the magnitude for the right ascension (left) and declination (right) components. The individual measurements are shown as black dots, the median proper motion in 0.5 mag wide bins with red error bars. As the median proper motion is very close to zero, we show a more detailed view in the second row. Neither a-posteriori corrected nor uncorrected proper motions show any significant systematic trend. Similarly in rows 3 and 4 we show the two proper-motion components plotted against their $m_{\rm F625W}-m_{\rm F814W}$ color index. Also here, there are no significant trends visible.}
  \label{fig:pm_systematics}
\end{figure*}

In addition to the local trends in magnitude, we also searched for global trends in magnitude and color, by dividing the proper motions in $m_{\rm F625W}$-magnitude, or $m_{\rm F625W}-m_{\rm F814W}$ color bins (see Figure~\ref{fig:pm_systematics}). We calculated the median of the two proper-motion components in each bin and did not see any significant deviation from zero, neither for the raw nor for the a-posteriori corrected proper motions (see Figure \ref{fig:pm_systematics}).

\clearpage
\section{Comparisons with literature catalogs}
\label{sec:appendixlit}
\subsection{Crossmatch and direct comparison with \cite{2017ApJ...842....6B}}
The deepest and most precise photometric and astrometric catalog of the central region of \omc{} beside this work was published by \cite{2017ApJ...842....6B}. The authors published \texttt{KS2} (see Section~\ref{subsec:ks2})  photometry for 26 filters of WFC3/UVIS and WFC3/IR. In addition, they crossmatched and published the proper-motion catalog from \cite{2014ApJ...797..115B} along with the photometric catalogs. In our catalog we include 9 years of additional data and use improved analysis tools, (\citealp{2014ApJ...797..115B} do not use second-pass photometry for the proper-motion measurements, no focus variable PSF models were available at that time) and therefore, we expect significantly smaller astrometric errors in our new catalog. We crossmatched both catalogs to see whether the photometry and astrometry results are consistent, at least in the core region included in both catalogs.

\begin{figure*}[h]
  \centering
    \includegraphics[width=1.0\textwidth]{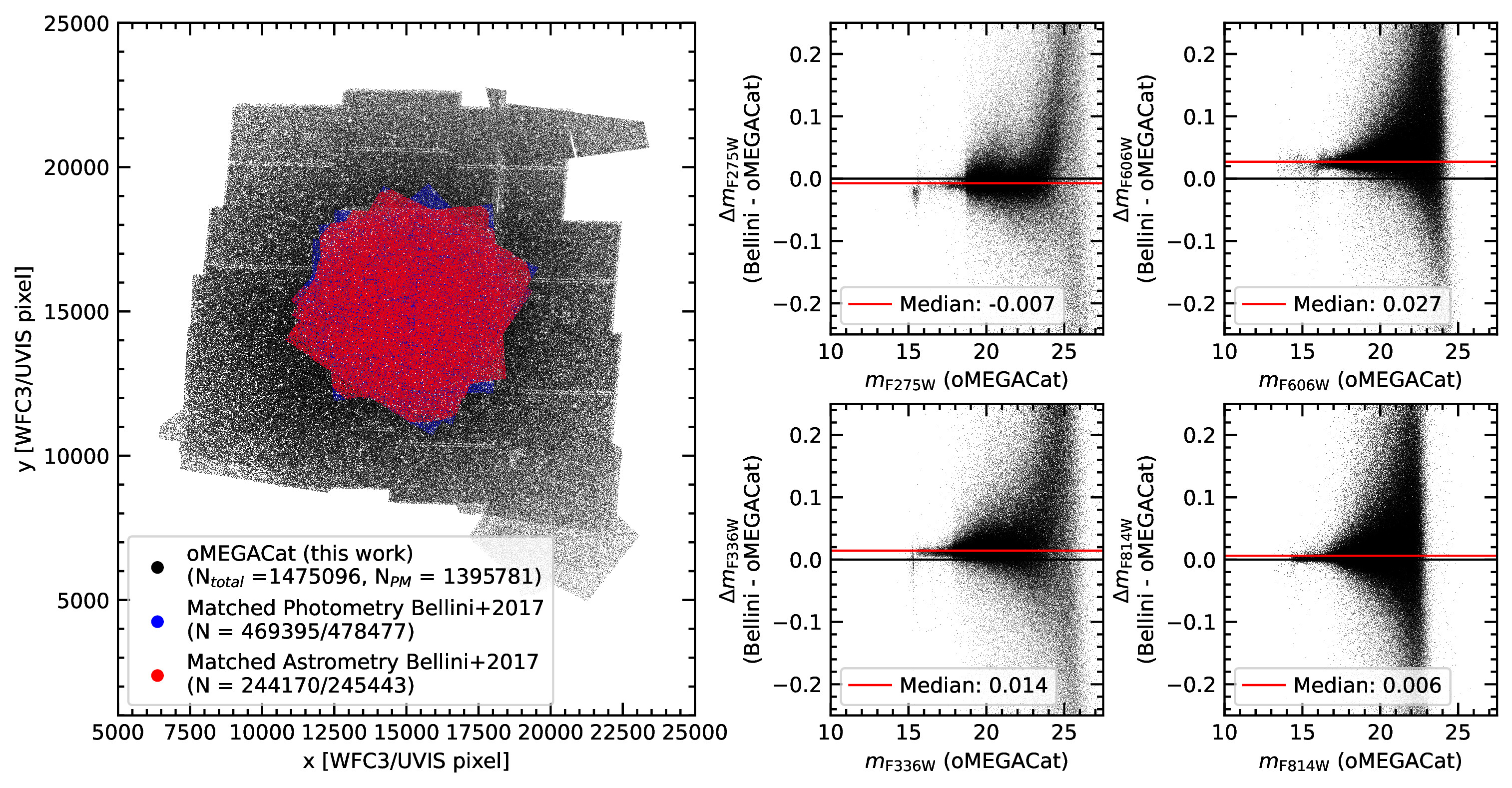}
  \caption{\textit{Left:} Footprint of our proper-motion catalog (black) and the photometric (blue) and astrometric (red) catalogs published in \cite{2017ApJ...842....6B}. \textit{Right:} Comparison of calibrated photometry between this work and the \cite{2017ApJ...842....6B} photometric catalog for 4 WFC3/UVIS filters (F275W, F336W, F606W, F814W)}
  \label{fig:bellini_photometry}
\end{figure*}

We transformed the pixel-based coordinate system of \cite{2017ApJ...842....6B} to our own reference system using 6 parameter linear transformations. After that, we used a matching radius of 1~WFC/UVIS pixel (40~mas) to crossmatch stars. This radius is large enough, as the reference epochs of the two catalogs are similar (Bellini et al.: 2007.0; This work: 2012.0). In 5 years the stars will have an RMS displacement of just 0.08~pixel. As it can be seen in Figure~\ref{fig:bellini_photometry}, the Bellini et al. catalog is fully contained within our larger field and most stars can be crossmatched (465\,362/478\,477 for the photometric and 242934/245443 for the astrometric catalog).

While the Bellini et al. catalog contains photometry for 26 filters, the focus of our study was on the 6 (3 ACS/WFC, 3 WFC3/UVIS) filters for which we have full coverage out to the half-light radius and the WFC3/UVIS F606W filter for which we have the largest number of photometric measurements in the center. Therefore, there are only 4 filters that we can compare. We show the difference of the method~1 (see Section~\ref{subsec:ks2}) photometry for these 4 filters in Figure~\ref{fig:bellini_photometry}. The overall agreement is good and only very small systematic shifts of the zero point can be observed (F275W: -0.007; F336W: 0.014; F606W: 0.027; F814W: 0.006). We attribute those differences to the updated instrumental zero point values (see Section~\ref{subsec:photometric_ref}) and the slightly different radii used to create the aperture-photometry-based reference systems. They are of the same order of magnitude as the reported uncertainties on the absolute flux calibration \citep{2022AJ....164...32C}.

\begin{figure*}[h]
  \centering
    \includegraphics[width=1.0\textwidth]{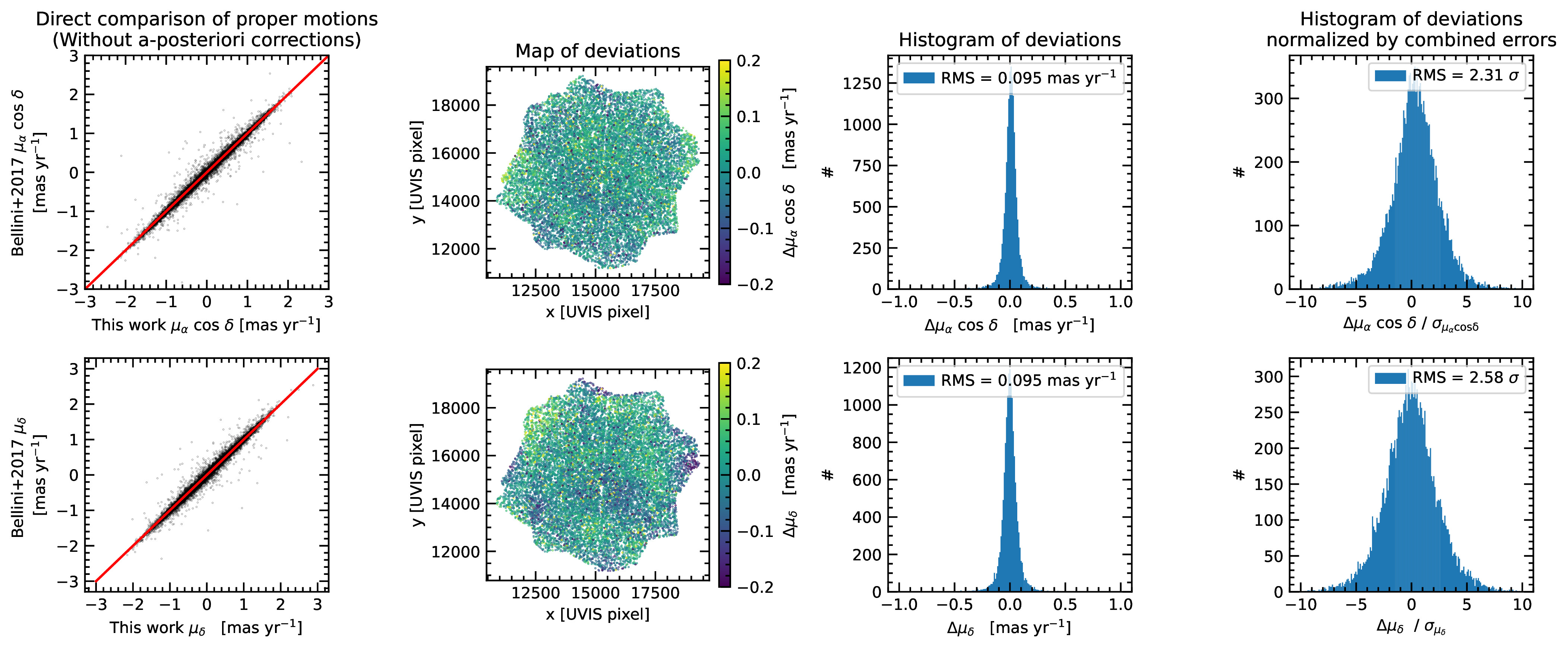}
  \caption{Comparison of uncorrected proper motions with \cite{2017ApJ...842....6B} for the right-ascension (upper row) and the declination proper-motion component (lower row). The panels first-from-the-left show a direct comparison of the proper-motion components. The red line is the plane-bisector and not a fit to the data. The second-from-the-left panels show how the proper motion difference between the two datasets varies over the field. Some mild systematic trends are visible, as expected for the uncorrected, amplifier-based, proper motions. The third-from-the-left panels show a histogram of the differences between the measurements, the fourth-from-the-left panels show the same differences but divided by the combined proper-motion error.}
  \label{fig:bellini_uncorrected}
\end{figure*}

To directly compare the proper motions, we restrict our analysis to stars brighter than $m_{\rm F606W} = 18$, as their statistical errors are lower and potential systematic effects are easier to detect. Figure~\ref{fig:bellini_uncorrected} shows the comparison of the raw proper motions. While the overall agreement is good (RMS of difference $\sim$0.09~mas yr$^{-1}$ in both components), one can see some low spatial frequency effects with amplitudes of up to $\sim$0.1~mas yr$^{-1}$. These systematic deviations cannot be explained using the individual proper-motion errors alone (the error distributions are 2.31/2.56 times wider than what would be expected from the proper motions errors alone). This is not unexpected for the raw proper motions and can be attributed to CTE effects and residual distortion and is also why we employed the local a-posteriori corrections (see Section~\ref{subsec:corrections}).

\begin{figure*}[h]
  \centering
    \includegraphics[width=1.0\textwidth]{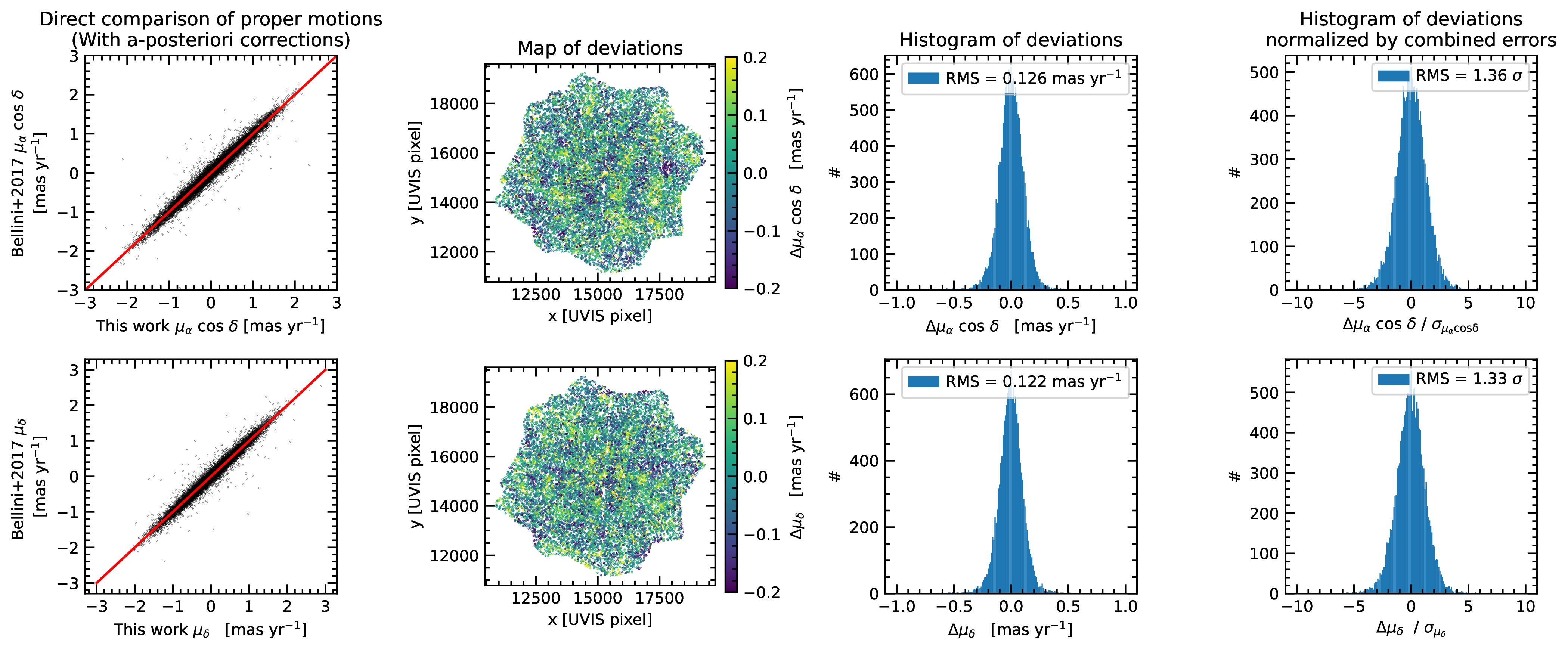}
  \caption{Similar to Figure~\ref{fig:bellini_uncorrected}, but instead of the amplifier-based proper motions, we apply the local a-posteriori corrections in both catalogs before the comparison. Note how the spatial variation of the deviations (second-from-left panels) and the normalized distribution of deviations (rightmost panels) change with respect to Figure~\ref{fig:bellini_uncorrected}.}
  \label{fig:bellini_corrected}
\end{figure*}

\begin{figure*}
  \centering
    \includegraphics[width=1.0\textwidth]{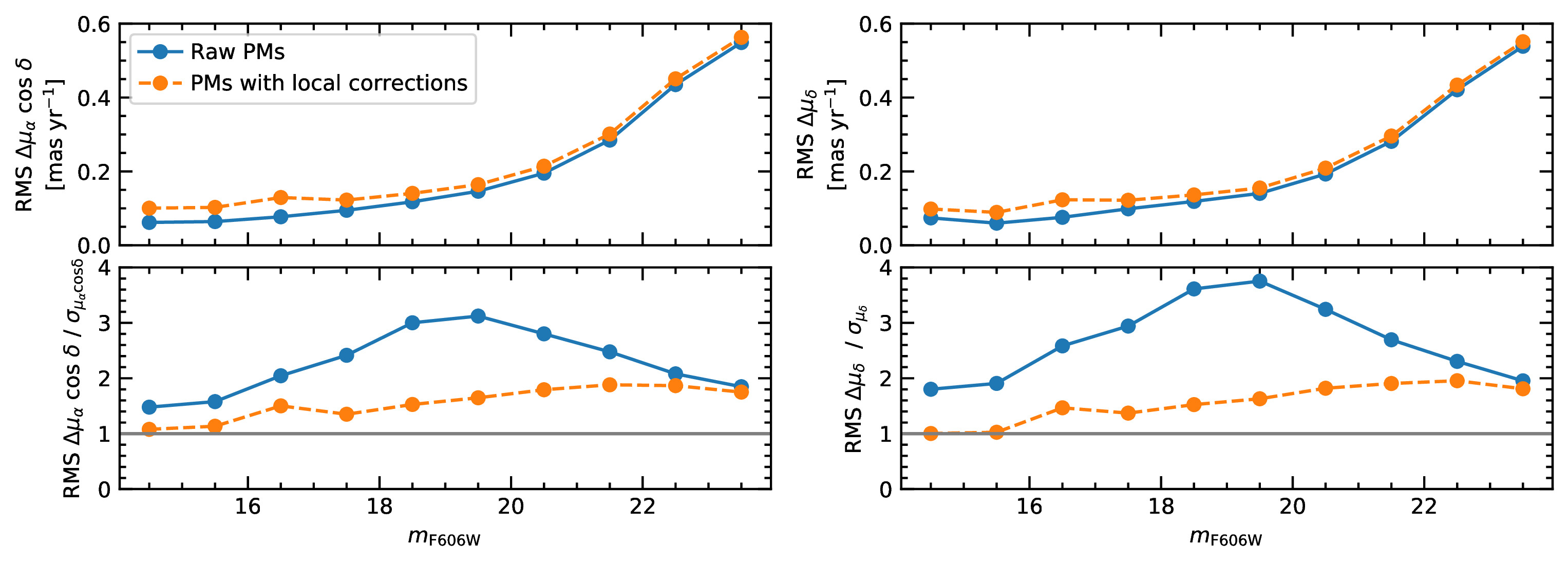}
  \caption{Analysis of how the dependence of the RMS of the deviation between this work and the \cite{2017ApJ...842....6B} catalogs changes with magnitude. The upper panels show the RMS of the absolute deviation of the two proper-motion components, and the lower panels show the RMS of the relative (i.e. scaled by the combined error) proper-motion components.}
  \label{fig:bellini_rms}
\end{figure*}

In Figure~\ref{fig:bellini_corrected}, we compare the proper motions after the local a-posteriori corrections have been applied in both catalogs. As expected, the low spatial frequency pattern has disappeared. Instead, we now can see some granularity which is most likely caused by the spatial scale of the local corrections and the limited number of available reference stars. The errors of the corrected proper motions do account for this additional uncertainty and, therefore, the distribution of the residuals is now much more compatible with the errors in the proper-motion catalogs (1.36 / 1.33 times wider than what would be expected from the proper motions errors). The magnitude dependence of the deviations between the proper-motions from the two catalogs is compared in Figure~\ref{fig:bellini_rms}, as expected the deviations increase for fainter stars.


\clearpage
\subsection{Crossmatch and direct comparison with \textit{Gaia} DR3 \& FPR}
\begin{figure*}[h]
  \centering
\includegraphics[width=1.0\textwidth]{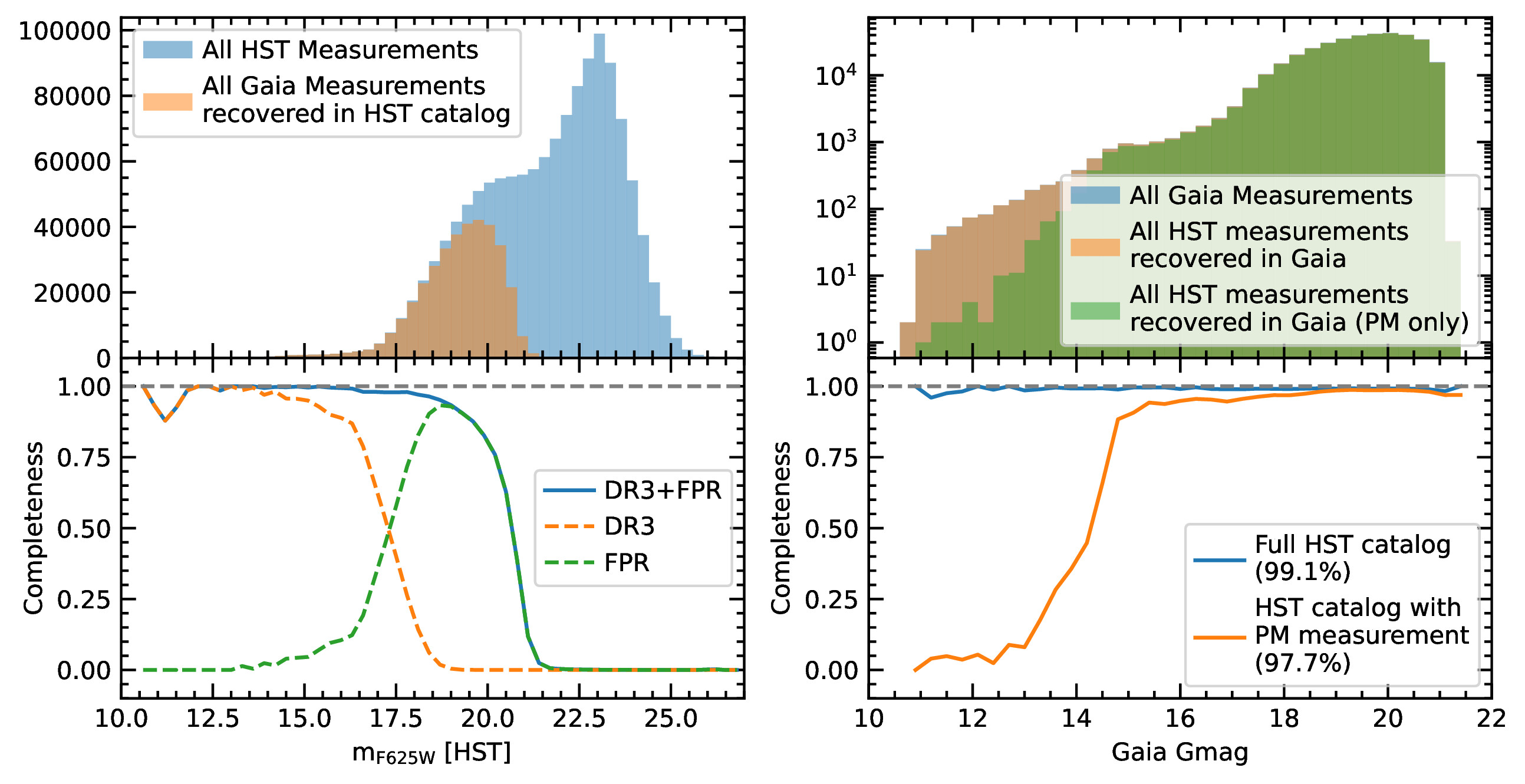}
  \caption{The upper panels show histograms of the magnitude distribution of both our new \textit{oMEGACat HST} catalog and the combined \textit{Gaia DR3+FPR} catalog, and the respective cross-matches between them. The ratio of recovered stars over all stars gives us the relative completeness between the catalogs, which is shown in the lower panels. The combined \textit{Gaia} catalog shows high completeness ($>90\%$) with respect to \textit{HST} until it sharply drops at faint magnitudes. A 50\% level of completeness is reached around $m_{\rm F625W}\sim21$. The \textit{HST} catalog is complete ($>99\%$) with respect to \textit{Gaia} over the full \textit{Gaia} magnitude range, however at bright magnitudes (\textit{Gaia}\,Gmag $< 14.5$) typically no \textit{HST} proper motions are available due to saturation in the \textit{HST} images.}
  \label{fig:gaia_completeness}
\end{figure*}

As already described above, the \textit{Gaia DR3} is affected by crowding and has been extended by the \textit{Gaia FPR}. Here, we crossmatch the combined \textit{DR3 + FPR Gaia} catalog with our new \textit{HST} catalog. Using a simple geometric cutoff of 160~mas, we find 373\,291 stars in common with our catalog (27\,123 from \textit{DR3} and 346\,168 from the \textit{FPR}). When comparing the completeness of the two catalogs with respect to each other, see Figure \ref{fig:gaia_completeness}, one can see very good agreement. When using our new \textit{HST} catalog as reference, the combined \textit{Gaia DR3 + FPR} catalog shows good completeness for magnitudes brighter than $m_{\rm F625W}\sim19$, for fainter stars the \textit{Gaia} completeness drops sharply. On the other hand, if one uses the \textit{Gaia} catalog as a reference and restricts it to the region covered by our \textit{HST} catalog, almost all ($>99\%$) \textit{Gaia} sources can be recovered in the \textit{HST} catalog over the full magnitude range. However, the sources brighter than \textit{Gaia G}$ ~14.5$ typically have no \textit{HST} proper motion measurements, due to saturation in the \textit{HST} data.

As many of the crossmatched \textit{Gaia} sources have limited accuracy and are affected by crowding, for the following analysis we limited ourselves to a subset of well-measured stars. We used a simple proper motion error cut of 0.6\,mas\,yr$^{-1}$ in both components and both datasets. In addition, we restrict the matching radius to 40~mas (1 WFC3/UVIS pixel) for the high-quality subset and only use stars with a minimum proper-motion baseline of 10 years.
This high-quality (HQ) subset of crossmatched stars contains 5\,897 entries from \textit{DR3} and 24\,467 from the \textit{FPR}. In Figure~\ref{fig:gaia_footprint} (left), it can be seen that, while the actual stellar density increases towards the center, the number of well-measured Gaia stars decreases. Also, it can be seen in the color-magnitude diagram in Figure~\ref{fig:gaia_footprint} (middle panel) that the overlap between the catalogs is limited to a relatively small magnitude range of mostly evolved stars. In the right panel of Figure~\ref{fig:gaia_footprint}, we study the positional residuals between our catalog and well-measured stars from the \textit{Gaia} catalogs. Given the different reference epochs of the catalogs (our catalog: 2012.0, \textit{Gaia DR3}: 2016.0; \textit{Gaia FPR:} 2017.5), and the fact that we anchored our (co-moving) astrometric reference system on observations from 2002.5, we expect both a systematic shift (based on the absolute cluster motion with respect to 2002.5), and random deviations (caused by the individual stars random motion). As it can be seen in the figure, the absolute deviation is around 101~mas for \textit{DR3}, and 112~mas for \textit{FPR}, which is compatible with what we expect from the known absolute cluster motion from our initial 2002.5 epoch towards the respective \textit{Gaia} reference epochs. The random position deviation has an RMS of $\sim$\,3~mas (\textit{DR3}) and 4~mas (\textit{FPR}), also compatible with the displacement expected from the velocity dispersion of the cluster.

\begin{figure*}[h]
  \centering
\includegraphics[width=1.0\textwidth]{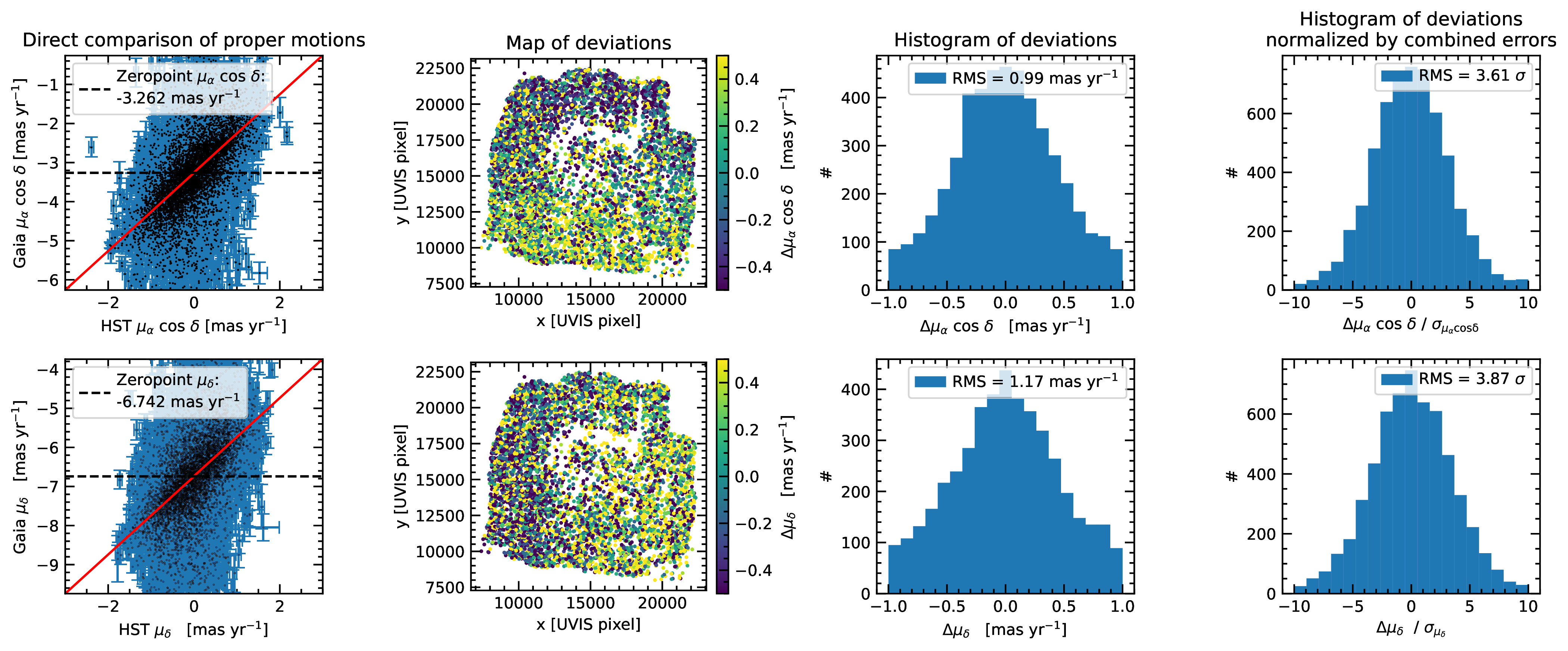}
  \caption{Similar to Figures \ref{fig:bellini_uncorrected} and \ref{fig:bellini_corrected}, but in this figure we compare the corrected relative proper motions from this work with the absolute proper motions from the \textit{Gaia DR3}. In the second-from-left panel one can see the dearth of \textit{Gaia DR3} stars in the center of \omc, but also some completeness issues of our \textit{oMEGACat} at very bright magnitudes. The visible systematic trends are the clear imprint of the cluster's rotation, which is further studied in Section~\ref{sec:rotation} and Figure~\ref{fig:rotation}.}
  \label{fig:gaia_match_dr3}
\end{figure*}

\begin{figure*}[h]
  \centering
  \includegraphics[width=1.0\textwidth]{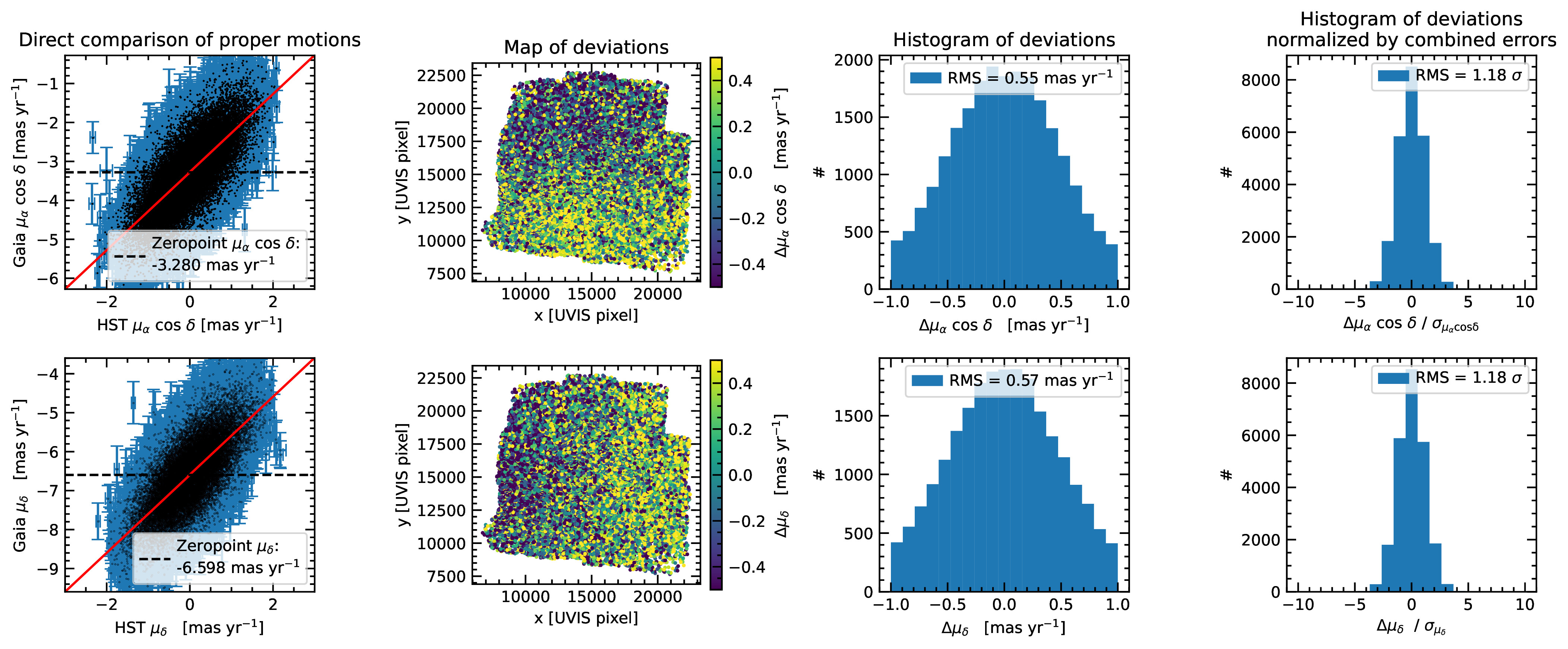}
  \caption{Similar to Figure  \ref{fig:gaia_match_dr3}, but in this figure we compare the corrected relative proper motions from this work with the absolute proper motions from the \textit{Gaia FPR}. In comparison with \textit{Gaia DR3} the completeness in the center is much better and astrometric residuals are lower. Just as in Figure~\ref{fig:gaia_match_dr3} one can see the clear imprint of the cluster's rotation, which is further studied in Section~\ref{sec:rotation} and Figure~\ref{fig:rotation}.}
  \label{fig:gaia_match_fpr}
\end{figure*}

Figure~\ref{fig:gaia_match_dr3} (\textit{DR3}) and \ref{fig:gaia_match_fpr} (\textit{FPR}) show the direct comparison and the differences between the proper motion components between the \textit{Gaia} datasets and our new catalog. While there is an overall good agreement in terms of pure proper-motion values after the absolute proper motion of \omc{} is accounted for, there is one large-scale systematic trend that leads to differences of up to $\sim$0.3~mas yr$^{-1}$. The reason for these systematic differences lies in the different approaches used to measure proper motions: While \textit{Gaia} measures absolute proper motions anchored to a fixed reference frame, we measure our \textit{HST} proper motions relative to the bulk motion of cluster stars. We further discuss these systematic differences between the two catalogs in the following Section \ref{sec:rotation}, where we use them to obtain a new, accurate measurement of \omc's rotation curve.

\clearpage
\section{Description of Columns in the Data Product Tables}
\label{sec:appendixtablecolumns}
Table \ref{tab:cat_astrometry} shows the content of the astrometric catalog with explanations for each individual column (see also Section \ref{subsec:data_products_astro}). Table \ref{tab:cat_photometry} shows the content of each of our seven photometric catalogs with explanations for each column (see also Section \ref{subsec:data_products_phot}). The tables can be downloaded from Zenodo: \href{https://zenodo.org/doi/10.5281/zenodo.11104046}{doi:10.5281/zenodo.11104046}.

\begin{table*}[h]
\caption{Content of the astrometric catalog.}
\label{tab:cat_astrometry}
\scriptsize
\begin{tabular}{p{3cm}p{8cm}l}
\hline
Column   & Description        & Unit \\ \hline
\texttt{ID}                     & oMEGACat II Identifier                                            & -                \\
\texttt{RA}                     & Right Ascension $\alpha$                                           & degree                 \\
\texttt{DEC}                    & Declination $\delta$                                           & degree                 \\
\texttt{x}                      & x coordinate in pixel based coordinate system          & 40~mas ($\sim$ 1 WFC3/UVIS pixel)                 \\
\texttt{y}                      & y coordinate in pixel based coordinate system          & 40~mas ($\sim$ 1 WFC3/UVIS pixel)                 \\
\texttt{pmra}                   & Proper motion in R.A. direction $\mu_\alpha \cos \delta$ & mas~yr$^{-1}$             \\
\texttt{pmdec}                  & Proper motion in Dec. direction $\mu_\delta$ & mas~yr$    ^{-1}$             \\
\texttt{pmra\_err}               & Proper motion error in R.A. direction $\sigma\mu_\alpha \cos \delta$ & mas~yr$^{-1}$             \\
\texttt{pmdec\_err}              & Proper motion error in Dec. direction $\sigma\mu_\delta \delta$ & mas~yr$^{-1}$             \\
\texttt{ra\_err\_measured}        & Error on R.A. position measurement          & mas                 \\
\texttt{dec\_err\_measured}       & Error on Dec. position measurement   & mas                 \\
\texttt{chi2x}                  &  Reduced $\chi^2$ for PM fit in R.A. direction & -                 \\
\texttt{chi2y}                  &  Reduced $\chi^2$ for PM fit in Dec. direction & -                 \\
\texttt{uuu}                    &  Flag indicating whether a star was used as reference star   & -                 \\
\texttt{nfound}                 &  Number of astrometric measurements available for PM Fit           & -                 \\
\texttt{nused}                  & Number of astrometric measurements actually used for PM Fit     & -                 \\
\texttt{baseline}               & Temporal baseline of the PM Fit              & years                 \\
\texttt{pmra\_corrected}         & Locally corrected proper motion in R.A. direction   & mas~yr$^{-1}$                  \\
\texttt{pmdec\_corrected}        & Locally corrected proper motion in Dec. direction   & mas~yr$^{-1}$   \\
\texttt{pmra\_corrected\_err}     & Error on locally corrected proper motion in R.A. direction   & mas~yr$^{-1}$                      \\
\texttt{pmdec\_corrected\_err}    & Error on locally corrected proper motion in Dec. direction   & mas~yr$^{-1}$                     \\
\texttt{n\_correction\_stars}     & Number of stars used for local a posteriori correction          & -                 \\
\texttt{rmax\_correction\_stars}  & Maximum distance to reference stars used for local a posteriori correction                                            & 40~mas ($\sim$ 1 WFC3/UVIS pixel)                 \\
\texttt{nitschai\_id}            & ID in \textit{oMEGACat I} MUSE spectroscopic catalog \citep{2023ApJ...958....8N}   & -                 \\

\texttt{gaia\_id}                & Gaia Source Identifier                                                                                                      & -                 \\
\texttt{gaia\_origin}            & Gaia Data Release in which crossmatched sources were published                                                              & -                 \\
\texttt{gaia\_ref\_epoch}         & Gaia reference epoch (2016.0 for DR3, 2017.5 for FPR)                                                                       & years                 \\
\texttt{gaia\_ra}                & Gaia Right Ascension                                                                                                        &  degree                \\
\texttt{gaia\_ra\_err}            & Gaia Right Ascension Error                                                                                                  & mas                 \\
\texttt{gaia\_dec}               & Gaia Declination                                                                                                            & degree\\
\texttt{gaia\_dec\_err}           & Gaia Declination Error                                                                                                      & mas                 \\
\texttt{gaia\_pmra}              & Gaia proper motion in R.A. direction (absolute)                                                                             & mas~yr$^{-1}$                 \\
\texttt{gaia\_pmra\_err}          & Gaia proper motion error in R.A. direction                                                                                  & mas~yr$^{-1}$                 \\
\texttt{gaia\_pmdec}             & Gaia proper motion in Declination direction (absolute)                                                                      &  mas~yr$^{-1}$                \\
\texttt{gaia\_pmdec\_err}         & Gaia proper motion error in Dec. direction                                                                                  &  mas~yr$^{-1}$                \\
\texttt{gaia\_phot\_g\_mean\_mag}   & Gaia G band mean magnitude                                                                                                  & mag                 \\
\texttt{gaia\_hq\_flag}           & Flag indicating whether a star was considered reliable in both Gaia and HST and used for rotation measurements                                            & -                 \\

\texttt{hst\_pm\_hq\_flag}         & Flag indicating whether a star passed the exemplary combined quality criterion                                            & -                 \\ \hline

\end{tabular}
\end{table*}

\begin{table*}[h]
\caption{Content of each photometric catalog.}
\label{tab:cat_photometry}
\scriptsize
\begin{tabular}{p{3cm}p{8cm}l}

\hline
Column   & Description        & Unit \\ \hline
\texttt{ID}                     & oMEGACat II Identifier                                            & -                \\
\texttt{corrected\_mag}                      & Photometry with empirical local corrections            & mag            \\
\texttt{corrected\_mag\_err}                  & $Red. \chi^2$ scaled error including error on corrections             & mag            \\
\texttt{m1\_weighted\_mean}                   & Weighted mean of the calibrated method 1 photometry            & mag            \\
\texttt{m1\_weighted\_mean\_error}             & Standard error of the weighted mean of the method 1 photometry             & mag            \\
\texttt{m1\_weighted\_rms}                    &  Weighted RMS of the calibrated method 1 photometry             & mag            \\
\texttt{m1\_median}                          & Median of the calibrated method 1 photometry             &  mag           \\
\texttt{m1\_mad}                             & Median absolute deviation of the calibrated method 1 photometry            & mag            \\
\texttt{m1\_mean}                            & Standard mean of the calibrated method 1 photometry             & mag            \\
\texttt{m1\_rms}                             & RMS of the calibrated method 1 photometry             & mag            \\
\texttt{n\_measurements}                     & Number of measurements used to determine the combined photometric results for this filter            & -            \\
\texttt{chi2}                               & $\chi^2$ value of the combined calibrated magnitude            & -            \\
\texttt{chi2\_red}                           & reduced $\chi^2$ value of the combined calibrated magnitude              & -            \\
\texttt{qfit\_weighted\_mean}                 & Weighted mean of the QFIT parameter of all individual measurements              & -            \\
\texttt{o\_weighted\_mean}                    & Weighted mean of the $o$ value (ratio $f_{Source}/f_{Neighbors}$) of all individual measurements              & -            \\
\texttt{rx\_weighted\_mean}                   &  Weighted mean of the radial excess parameter of all individual measurements             & -            \\
\texttt{m2\_weighted\_mean}                   & Weighted mean of the calibrated method 2 photometry             & mag            \\
\texttt{m3\_weighted\_mean}                   & Weighted mean of the calibrated method 3 photometry              & mag            \\
\texttt{iter\_00\_flag}                       & Flag indicating whether initial (non PM) crossmatch was used             & -            \\
\texttt{brightlist\_flag}                    & Flag indicating whether photometry had to substituted from \texttt{hst1pass} due to saturation             & -            \\
\texttt{phot\_hq\_flag}                       & Flag indicating whether star passed exemplary photometric criteria             & -            \\ \hline
\end{tabular}
\end{table*}

\clearpage
\section{Numerical values of rotation profile}
\label{sec:appendixrotationtable}
Table \ref{tab:rotationprofile} shows the numerical values for the plane-of-sky rotation profile determined in Section \ref{subsec:rotation_sky}. It is also available in machine-readable form in the Zenodo Repository: \href{https://zenodo.org/doi/10.5281/zenodo.11104046}{doi:10.5281/zenodo.11104046}

\begin{table}[h]
\caption{Numerical values of the rotation profile from Fig.~\ref{fig:rotation}}
\label{tab:rotationprofile}

\begin{tabular}{llllll}
\hline
Lower limit  & Median radius   & Upper limit  & Number of Stars & Median tangential    & Inferred           \\
of bin             & of stars in bin & of bin             &                 & proper motion        & Rotation Velocity  \\
{[}arcsec{]} & {[}arcsec{]}    & {[}arcsec{]} & -               & {[}mas\,yr$^{-1}${]} & {[}km\,s$^{-1}${]} \\ \hline
0.0          & 20.66           & 30.0         & 270             & 0.0548$\pm$0.0324    & 1.41$\pm$0.83      \\
30.0         & 47.6            & 60.0         & 1020            & 0.086$\pm$0.0181     & 2.21$\pm$0.47      \\
60.0         & 76.65           & 90.0         & 1620            & 0.1478$\pm$0.0131    & 3.8$\pm$0.34       \\
90.0         & 105.91          & 120.0        & 2423            & 0.2088$\pm$0.0105    & 5.38$\pm$0.27      \\
120.0        & 135.99          & 150.0        & 3031            & 0.2556$\pm$0.0093    & 6.58$\pm$0.24      \\
150.0        & 165.6           & 180.0        & 3882            & 0.2562$\pm$0.0083    & 6.59$\pm$0.21      \\
180.0        & 195.11          & 210.0        & 4361            & 0.2678$\pm$0.008     & 6.89$\pm$0.21      \\
210.0        & 225.38          & 240.0        & 4470            & 0.2648$\pm$0.0078    & 6.82$\pm$0.20       \\
240.0        & 254.06          & 270.0        & 4186            & 0.2797$\pm$0.0083    & 7.2$\pm$0.21       \\
270.0        & 282.86          & 300.0        & 3020            & 0.2619$\pm$0.0103    & 6.74$\pm$0.27      \\
300.0        & 309.92          & 330.0        & 985             & 0.2658$\pm$0.0176    & 6.84$\pm$0.45      \\
330.0        & 340.91          & 360.0        & 246             & 0.2455$\pm$0.0421    & 6.32$\pm$1.08      \\ \hline
\end{tabular}
\end{table}

\clearpage

\clearpage
\bibliography{main_bibliography,additional_bibliography}
\bibliographystyle{aasjournal}



\end{document}